\documentclass[12pt]{article}
\usepackage{epsfig}
\newlength{\dinwidth}                                                          
\newlength{\dinmargin}                                                         
\newcommand{\qsd}       {\mbox{${Q^2}$}} 
\newcommand{\x}         {\mbox{${\it x}$}}

\newcommand {\pom}  {I\hspace{-0.2em}P}
\newcommand {\alphapom} {\mbox{$\alpha_{_{\pom}}$}}
\newcommand {\xpom} {\mbox{$x_{_{\pom}}$}}

\newcommand{\qf}       {\mbox{$Q^{4}$}}

\newcommand{\qq}       {\mbox{$Q^{2}$}}
\newcommand{\ncp}       {^{\rm NC}}
\newcommand{\born}      {_{\rm Born}}
\newcommand{\sigxqq}    {\mbox{$d^{2}\sigma\ncp\born/dxd\qq$}}
\newcommand{\chiz}     {\mbox{$\chi_{_{\rm Z}}$}}

\newcommand{\Mzq}      {\mbox{$M_{\rm Z}^{2}$}}
\newcommand{\thw}      {\mbox{$\theta_{\scriptscriptstyle \rm W}$}}
\newcommand{\sthw}     {\mbox{$\sin^2\!\thw$}}
\newcommand{\cthw}     {\mbox{$\cos^2\!\thw$}}
\def\be{\begin{equation}}
\def\ee{\end{equation}}
\def\bea{\begin{eqnarray}}
\def\eea{\end{eqnarray}}

\begin{document}
\title{
{DESY 01-058}\hfill{May 2001}\\
\vspace{1cm}
{\bf Deep Inelastic Scattering at\\
 Large Energy and Momentum Transfers:\\
 Recent Results from the HERA Collider\footnote{This report is based on lectures presented at the IX Mexican School on Particles and Fields, Metepec, Puebla, Mexico, 9 - 19 August, 2000, and at the International School of Subnuclear Physics, Erice, Italy, 27 August - 5 September 2000.}}
\author{G\"{u}nter Wolf\\
{Deutsches Elektronen Synchrotron DESY  }\\
 }}
\date{}
\maketitle
\begin{abstract}
Data from H1 and ZEUS on the structure and the quark and gluon densities of the proton are discussed. A brief excursion is made into the field of inclusive diffraction by deep inelastic scattering. The comparison of $e^-p$ and $e^+p$ scattering at large momentum transfers demonstrates clearly the presence of weak contributions in neutral current interactions. The comparison with the corresponding charged current results shows at $Q^2$ values above the masses squared of the heavy vector bosons the unification of electromagnetic and weak interactions. The new data are testing the validity of the Standard Model down to spatial resolutions of the order of $10^{-16}$ cm. Intensive searches have been performed for a manifestation of instantons, for signs of compositeness of quarks and leptons and for the effect of extra dimensions.   
\end{abstract}

\section{Introduction}

Ever since pointlike constituents, quarks, have been found in the nucleon~\cite{Taylor69} the question has been raised whether quarks and leptons also have substructure. Obviously, the detection of quarks and leptons as extended objects would produce a revolution in particle physics, e.g. by the possibility of constructing quarks and leptons from common subconstituents. Deep inelastic lepton nucleon scattering is particularly well suited for the study of the spatial structure of quarks and leptons. The sensitivity to substructure depends on the virtuality $Q^2$ of the exchanged current and, therefore, on the energy of the scattering partners. The highest energies are provided by HERA where by now a spatial resolution of $10^{-16}$ cm or one thousandth of the proton radius has been achieved. The well defined initial state and the large energies and virtualities make HERA an excellent place to probe the Standard Model also in other areas such as new currents and new particles. 

The first surprise encountered at HERA by H1~\cite{Hepf92} and ZEUS~\cite{Zepf92} in deep inelastic electron proton scattering has been a rapid rise of the proton structure function $F_2$ as Bjorken-x tends to zero. The data are consistent with a power like rise, $F_2 \propto x^{-\lambda}$, a behaviour which for a number of reasons~\cite{GLR1983}, such as overlap of partons or a Froissart like bound for the total virtual photon proton cross section or the requirement of unitarity, should not persist down to infinitely small values of $x$. Instead, the parton densities should saturate at some $x < x_{lim}$. Saturation or not is a hot research topic at HERA. The discovery of large rapidity gap events in deep inelastic scattering~\cite{Zeplrg93,Heplrg94}, which result from diffraction, has opened another avenue on this issue. The link provided by the optical theorem between the total cross section (which is proportional to $F_2$) and the diffractive cross section suggests that the rise should be even faster in diffraction.

This report presents new data from the two collider experiments H1 and ZEUS at HERA with particular emphasis on the results for structure functions, diffraction and neutral and charged current processes at very large $Q^2$. For a recent and comprehensive introduction to the physics at HERA see~\cite{AbramCald98}.

\section{The HERA Collider and the Experiments}
\label{s:HERA}

\begin{figure}[ht]
\begin{center}
\epsfig{file=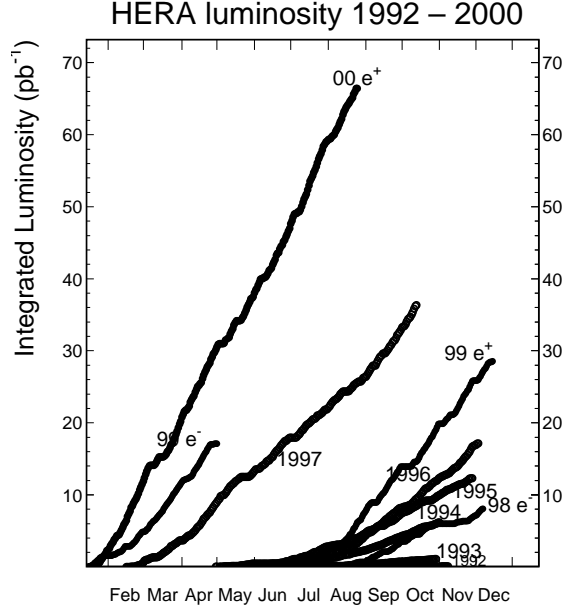,bbllx=-120pt,bblly=20pt,bburx=610pt,bbury=701pt,height=12cm}
\end{center}
\caption{Integrated yearly luminosity delivered by HERA per experiment during 1992 - 2000.}
\label{f:heralumi9200bw}
\end{figure}

The HERA collider can store electrons (positrons) of 30 GeV and protons of 920 GeV in two rings of 6.3 km circumference~\cite{HERA1981}. Table~\ref{t:HERApar} lists some of the salient parameters of the machine~\cite{Willeke00}. In order to maximize the luminosity up to 210 bunches of particles can be stored for each beam. The time interval between consecutive bunches is 96 ns. The circulating $e^-$ ($e^+$) beam becomes transversly polarized by the Sokolov-Ternov effect~\cite{Sokolov}. The measured specific luminosity is almost a factor of two larger compared to the design value. The maximum peak luminosity achieved so far is about 30$\%$ above the design value. The integrated luminosity per year provided by HERA for e-p collisions since 1992 is diplayed in Fig.~\ref{f:heralumi9200bw}. The total yearly luminosity increased by about a factor of two every year reaching 70 pb$^{-1}$ in 2000. An upgrade program has been started which promises a factor of 3 -  5 increase in luminosity by the insertion of additional magnets close to the interaction point. Operation in the new configuration will start in the year 2001.

\begin{table}[hbt]
\centering
\caption{HERA machine parameters.} 
\label{t:HERApar}
\begin{tabular}{|l|rcr|}
\hline
parameter                           & electron ring &          & proton ring \\
\hline
circumference (m)                       &           &  6336  &          \\
beam energies, design (GeV)             &  30       &        &  820     \\
beam energies, achieved (GeV)           &  27.6     &        &  920     \\
e p c.m. energy squared (GeV$^2$)       &           &  10$^5$   &          \\
magn. bending field (T)                 &  0.15     &        &  5.25   \\
dipole bend. radius  (m)                &  610      &        &  584     \\
max. circ. curr., design (mA)           &  60       &        &  160     \\
max. circ. curr., achiev.(mA)           &  50       &        &  110     \\
n. bunch buckets                        &  220      &        &  220     \\
n. bunches (typical)                    &  180      &        &  180    \\
time betw. cross. (ns)                  &           &  96    &          \\
max. lumi. design /achiev. $10^{31}cm^{-2}s^{-1}$     &      & 1.5/2.0    &          \\
max. yearly lumi. deliv. per expt.  ($pb^{-1}$)  &           &   70   &          \\
electron polarization                   &  50 - 70$\%$  &    &          \\
\hline
\end{tabular}
\end{table}

The data collected by H1 and ZEUS between 1992 and 2000 for different beam conditions correspond to integrated luminosities of about 17 $pb^{-1}$ for $e^-p$ and 116 $pb^{-1}$ for $e^+p$ collisions, per experiment.

\section{Structure Functions of the Proton}
\subsection{Kinematics}
\begin{figure}[ht]
\centerline{\epsfig{file=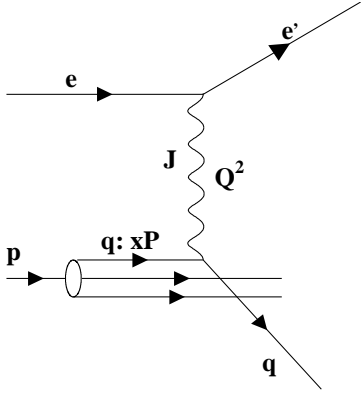,bbllx=0pt,bblly=45pt,bburx=510pt,bbury=480pt,height=6cm,clip=}}
\caption{Diagram for deep inelastic $ep$ scattering.}
\label{f:diagdisgeneric}
\end{figure}

Inclusive deep inelastic scattering (DIS) by neutral current (NC) exchange, 
\begin{displaymath}
e(k)+p(P) \to e^\prime (k^\prime)+anything
\end{displaymath} 
can be described as a function of Bjorken-$x$ and $Q^2$ (see Fig.~\ref{f:diagdisgeneric}). The basic quantities,
in the absence of QED radiation, are:   
\begin{eqnarray}
  s & = & 4 \cdot E_e \cdot E_p \\
\qsd & = & -q^{2}  =  -(e - e^{\prime})^{2}  \\
  x &  = & \frac{\qsd}{2 P \cdot q} \\
  \nu & = & (q \cdot P)/M_p \\
  y &  = & \frac{ q \cdot P}{e \cdot P} \\
\qsd & = & x \cdot y \cdot s \\
  W^2 & = &\frac{Q^2(1-x)}{x}+M_p^2 \approx \frac{\qsd}{x}   
        \;  {\rm for}\;  x \ll 1 
\end{eqnarray} 
where $e$ and $e^{\prime}$ are the four-momenta of the initial and
final state electrons, $P$ is the initial state proton four-momentum,
$M_p$ is the proton mass, $s$ is the square of the $ep$ c.m. energy, $-Q^2$ is the mass squared of the exchanged current, $\nu$ is the energy transfer and $y$  the fractional energy transfer from the incident electron to the
proton as measured in the proton rest frame, $y = \nu/\nu_{max}$, $\nu_{max} = s/(2M_p)$, $x$ is the fractional momentum of the proton carried by the struck quark and $W$ is the $\gamma^* p$ c.m. energy.

\subsection{Definition of the structure functions}
The differential cross section for deep inelastic scattering DIS can be expressed in terms of three structure functions, ${\cal F}_2,{\cal F}_L,{\cal F}_3$~\cite{IngelmanRueckl}:
\begin{eqnarray}
\frac{d^2\sigma}{dxdQ^2} & = & \frac{2\pi\alpha^2}{xQ^4}[(1+(1-y)^2){\cal F}_2 -y^2{\cal F}_L \pm (1 -(1-y)^2)x{\cal F}_3]
\label{eq:sigmanc}
\end{eqnarray}
where $\alpha$ is the fine structure constant and the upper (lower) sign applies to $e^-$~($e^+$)~p scattering. Since the contributions from ${\cal F}_L$ and $x{\cal F}_3$ are expected to be small in the measured region, the radiatively corrected NC cross section can be written as:
\begin{eqnarray}
\frac{d^2\sigma}{dxdQ^2} & = & \frac{2\pi\alpha^2}{xQ^4}[(1+(1-y)^2){\cal F}_2](1-\delta_L - \delta_3).
\label{eq:f2corr}
\end{eqnarray}
The structure function ${\cal F}_2$ receives contributions from photon and $Z^o$ exchange and can be written as
\begin{eqnarray}
{\cal F}_2 & = & F^{em}_2 (1 + \delta_Z).
\end{eqnarray}
where $F^{em}_2$ denotes the contribution from photon exchange alone.

The corrections $\delta_Z,_L,_3$ are functions of $x$ and $Q^2$ but are, to a good approximation, independent of ${\cal F}_2$, i.e. they are insensitive to the parton density distributions. They were calculated from theory using structure functions which gave a good representation of the data. In the measured region $\delta_L$ is small except when $y \ge 0.7$ where $\delta_L \approx 0.12$. The contributions from $\delta_Z,_3$ are negligible for $Q^2 <$ 1000 GeV$^2$ and small up to $Q^2 \approx 5000$ GeV$^2$.
Whenever, in the following, $Q^2$ is below 1000 GeV$^2$ the notations ${\cal F}_2$, $F^{em}_2$ and $F_2$ are used interchangeably.

In QCD, ignoring $Z^0$ exchange, ${\cal F}_2$ can be expressed in terms of the quark densities $q(x,Q^2)$ of the proton:
\begin{eqnarray} 
{\cal F}_2  & = & \sum_q e^2_q x q(x,Q^2),
\end{eqnarray}
where $e_q$ is the electric charge of quark q and the summation is performed over all quarks and antiquarks.  

A recent in-depth discussion of the structure functions of the nucleon has been presented in~\cite{Devenish}.

\subsection{Structure function $F_2$ in the DIS regime} 
\begin{figure}[ht]
\begin{center}
\epsfig{file=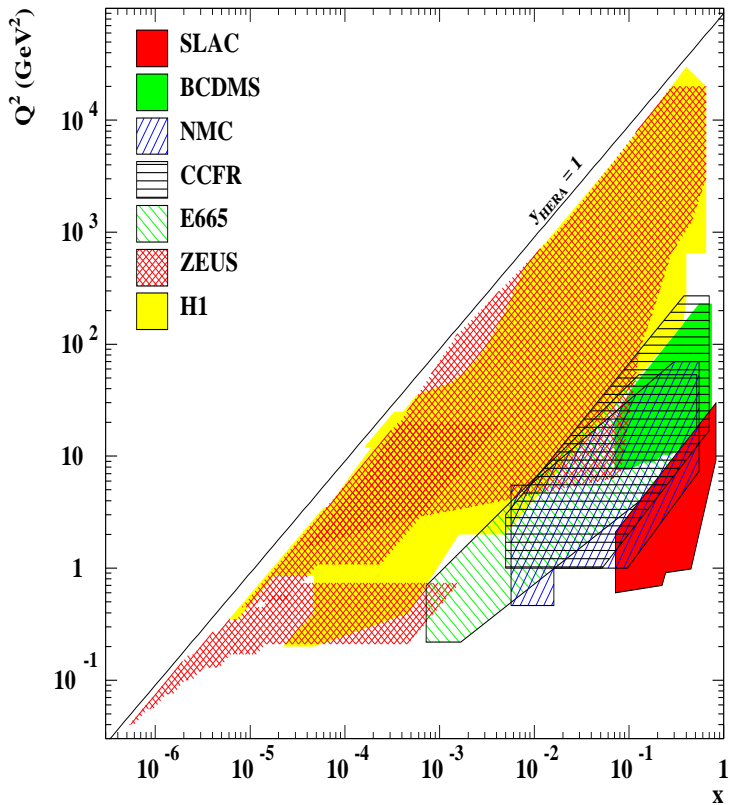,bbllx=50pt,bblly=150pt,bburx=610pt,bbury=701pt,width=10cm,height=12cm,clip=}
\end{center}
\caption{\label{f:HZ9200xvsq2}
        { The $x\, - \, Q^2$ plane: the regions covered by H1 and ZEUS
          and by fixed target experiments.
        }}
\end{figure}

The proton structure function ${\cal F}_2$ has been measured at HERA over a wide range in $x$ and $Q^2$ as shown in Fig.~\ref{f:HZ9200xvsq2}. At large $x$ the HERA data overlap with those obtained by fixed target experiments. Until recently, most of the HERA results had come from data taken up to 1994, corresponding to about 2 - 3 pb$^{-1}$ per experiment. H1 and ZEUS have now presented preliminary analyses of data from 1996-7 which are based on an order of magnitude higher integrated luminosity. The region covered in $x,Q^2$ has been enlarged substantially. At low $Q^2$ ZEUS and H1 have started to map out the transition region from photoproduction to deep inelastic scattering for $x$ between $ 10^{-4}$ and $10^{-6}$. At the high $Q^2$ end, the structure function measurements have been extended up to $Q^2 \approx 30000$ GeV$^2$. Progress has also been made in the measurement of the charm contribution to $F_2$ which provides for a direct test of the gluon density $g(x,Q^2)$ extracted with QCD fits from the $F_2$ data.

\begin{figure}
\begin{center}
\epsfig{file=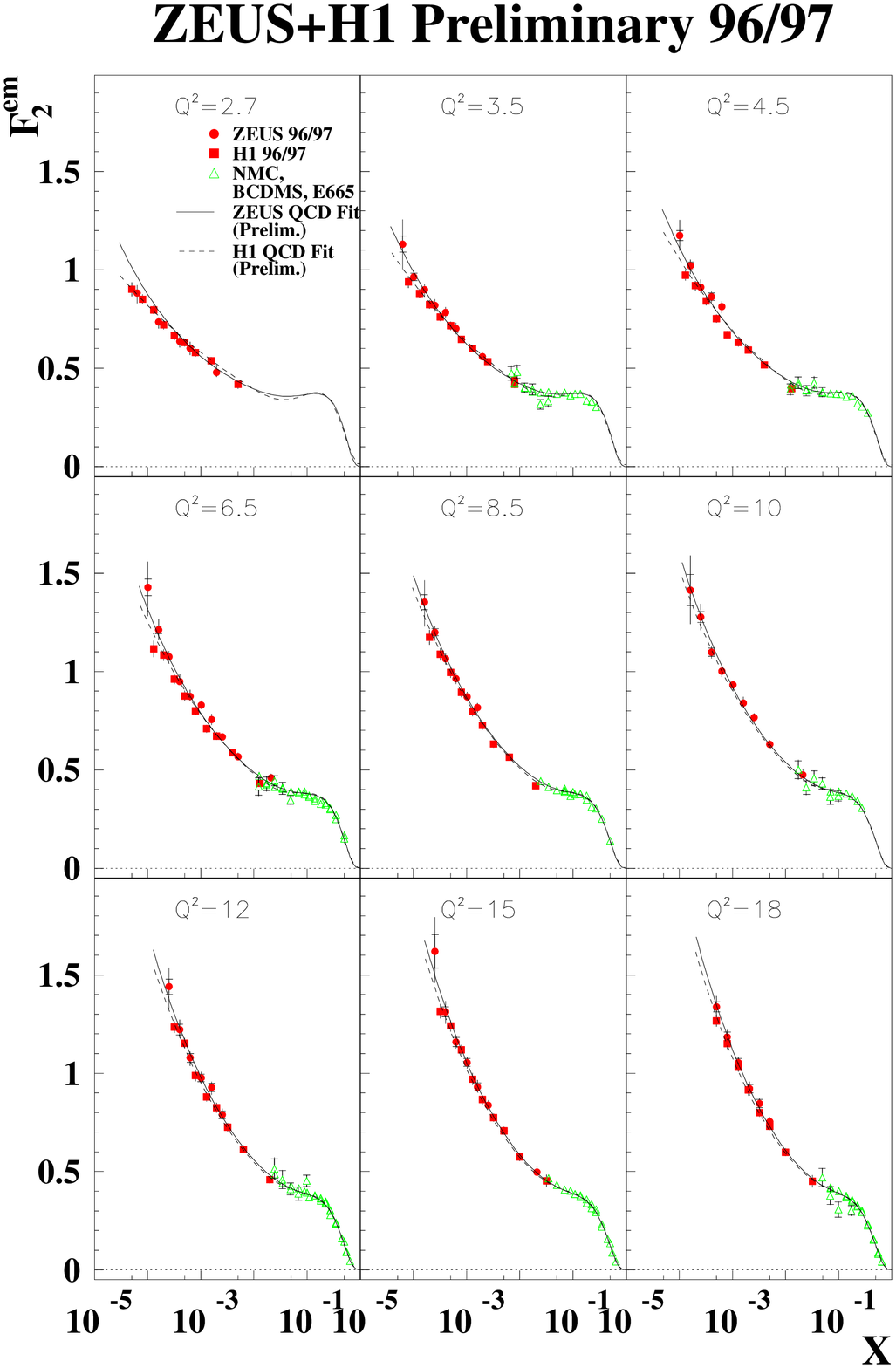,bbllx=100pt,bblly=0pt,bburx=410pt,bbury=701pt,width=8cm,height=14cm}
\end{center}
\caption{\label{f:HZ967f2vsxa}
        { Structure function ${\cal F}_2$ from NC scattering as a function of
          $x$ for fixed values of $Q^2$ between 2.7 and 18 GeV$^2$ as 
          measured by H1, ZEUS. Also shown 
          are the data from the fixed target experiments BCDMS, E665 and NMC.
          The lines indicate QCD NLO fits to the data by H1 and ZEUS.
         }}
\end{figure}

\begin{figure}
\begin{center}
\epsfig{file=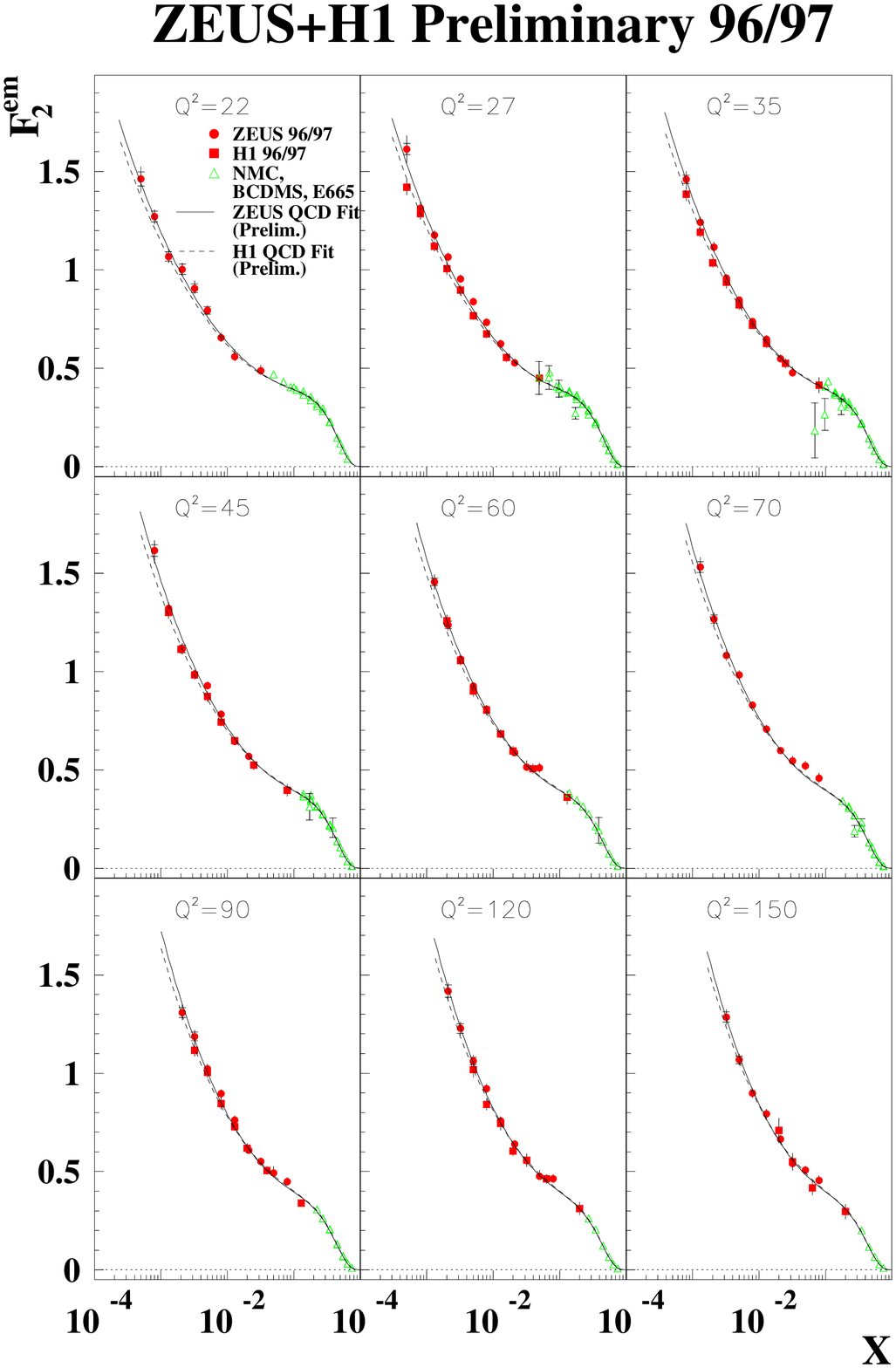,bbllx=100pt,bblly=0pt,bburx=410pt,bbury=701pt,width=8cm,height=14cm}
\end{center}
\caption{\label{f:HZ967f2vsxb}
        { Structure function ${\cal F}_2$ from NC scattering as a function of
          $x$ for fixed values of $Q^2$ between 22 and 150 GeV$^2$ as 
          measured by H1, ZEUS. Also shown 
          are the data from the fixed target experiments BCDMS, E665 and NMC.
          The lines indicate QCD NLO fits to the data by H1 and ZEUS.
         }}
\end{figure}

The results on the $x$ dependence of ${\cal F}_2$ from H1~\cite{Hepf2967} and ZEUS~\cite{Zepf2967} are presented in Figs.~\ref{f:HZ967f2vsxa},~\ref{f:HZ967f2vsxb} for different $Q^2$ intervals. The error bars show the statistical and systematic uncertainties added in quadrature. For $Q^2 <$ 100 GeV$^2$ the typical statistical errors are 2\% at low $Q^2$ rising to 6\% at $Q^2 \approx 100$ GeV$^2$. There is good agreement between the two HERA experiments. Also shown are the data from the fixed target experiments: BCDMS~\cite{Bcdms}, E665~\cite{E665}, NMC~\cite{Nmc} and SLAC~\cite{Slac} which cover the region of $`large` \, x$. In the region, where the HERA and the fixed target data overlap, good agreement is observed.

\begin{figure}
\begin{center}
\epsfig{file=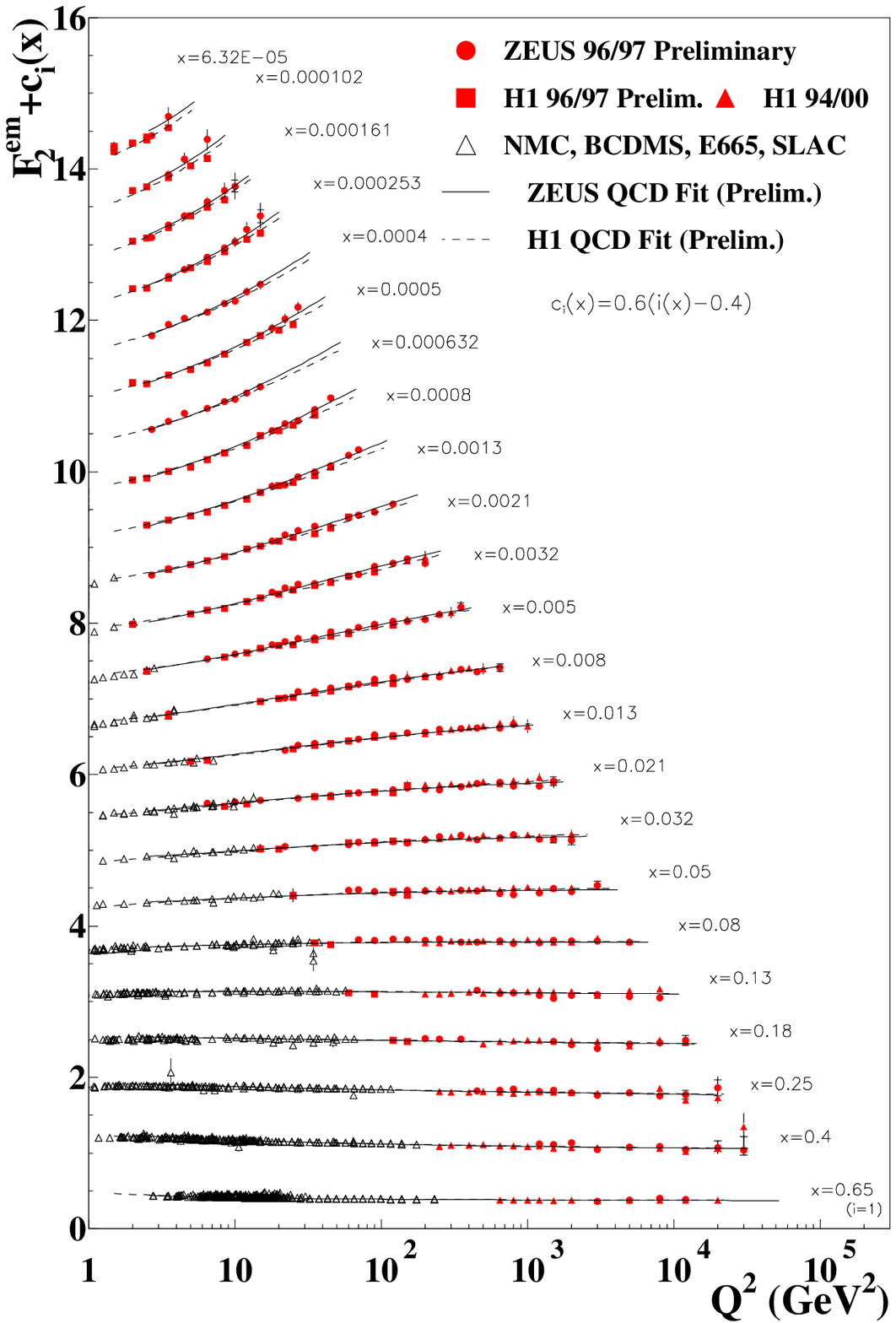,bbllx=-20pt,bblly=0pt,bburx=560pt,bbury=501pt,width=21cm,height=16cm}
\end{center}
\caption{\label{f:HZ967f2vsq}
        { Structure function ${\cal F}_2$ from NC scattering as a function of
          $Q^2$ for fixed values of $x$ as measured by H1, ZEUS. Also shown 
          are the data from the fixed target experiments BCDMS, E665, NMC and
          SLAC. The dashed (solid) lines indicate QCD NLO fits to the data by 
          H1 and ZEUS.
        }}
\end{figure}
  
Figure~\ref{f:HZ967f2vsq} shows the ${\cal F}_2$ values as a function of $Q^2$ for fixed $x$. For $x >$ 0.1 the data now span almost four decades in $Q^2$. Scaling violations proportional to $\ln Q^2$ are observed which decrease as $x$ increases.

\begin{figure}[ht]
\begin{center}
\epsfig{file=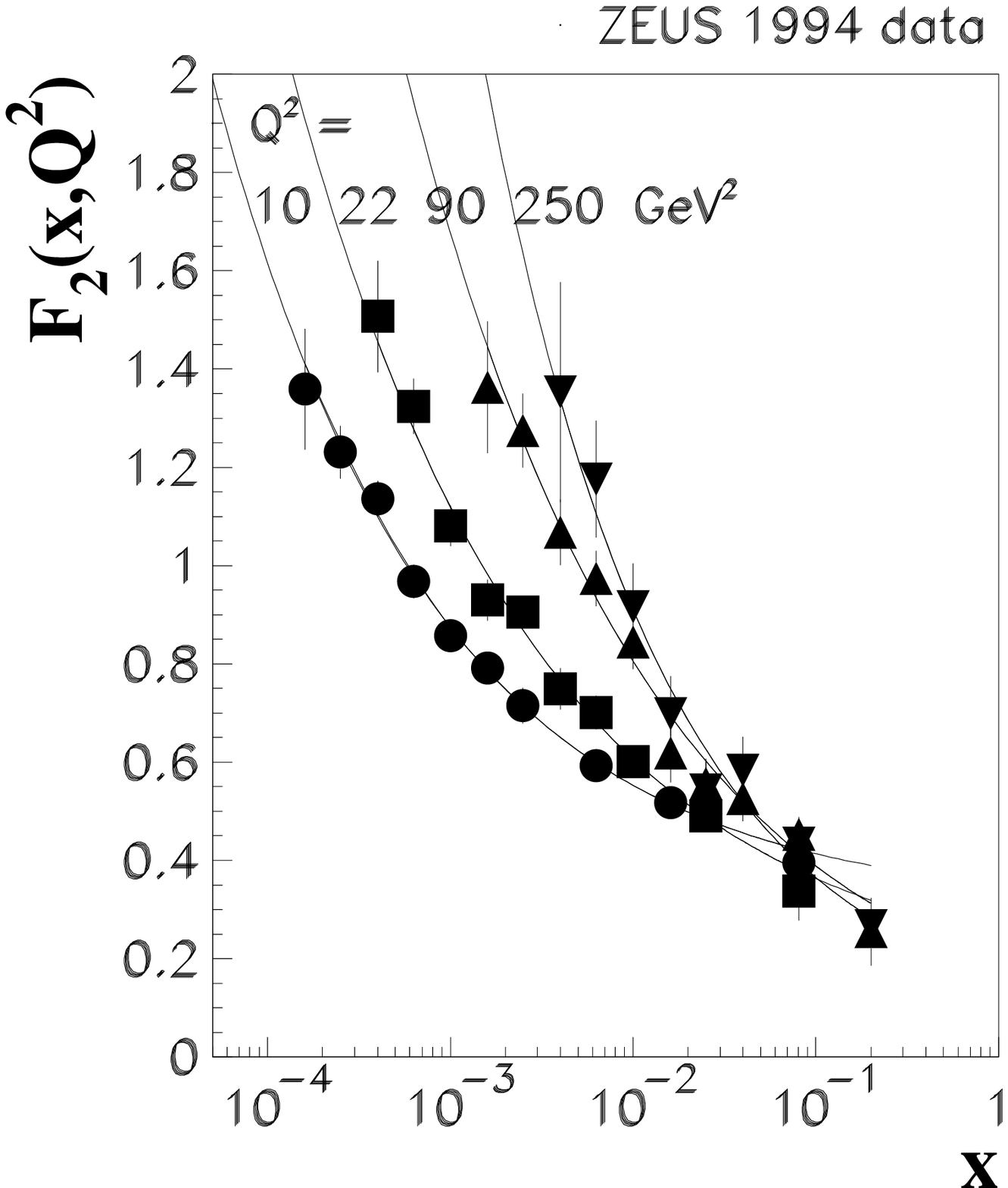,bbllx=100pt,bblly=0pt,bburx=510pt,bbury=701pt,height=12cm}
\end{center}
\caption{\label{f:zf2vsx94fitcon}
        { Structure function ${\cal F}_2$ from NC scattering as a function of
          $x$ for fixed values of $Q^2 =$ 10, 22, 90 and 250 GeV$^2$ as 
          measured by ZEUS. For the curves see text. 
        }}
\end{figure}

The most striking feature of the HERA data is the rapid rise of ${\cal F}_2$ as $x \to 0$ which is seen to persist down to $Q^2$ values as small as 1.5 GeV$^2$. This rise accelerates with increasing $Q^2$ as shown in Fig.~\ref{f:zf2vsx94fitcon} where ZEUS data from $Q^2 = 10,22,90$ and $250$ GeV$^2$ have been overlaid. Fitting ${\cal F}_2$ data to the form ${\cal F}_2 = a + b x^{-\lambda}$ leads to the results given in Table~\ref{t:f2fita}~\cite{Wolf96}. The constant term $a$ advocated in~\cite{Wolf94} can be thought of representing the nonperturbative contributions such as expected from the aligned jet configuration, see~\cite{Bjorken}. The contribution from $a$ decreases with $Q^2$: for $Q^2 \ge 90$ GeV$^2$ it is zero within the errors.  The power $\lambda$ of the $x$ dependent term is rather constant with $Q^2$ while the coefficient $b$ rises for $Q^2$ between 10 and 90 GeV$^2$ and appears to be driving the rapid rise of ${\cal F}_2$. 

If the constant term is arising from a nonperturbative, hadron-like contribution then one may want to modify its contribution as follows: ${\cal F}_2 = a_{had} (\frac{W^2}{W^2_0})^{\alpha_{pom}(0) -1} + b x^{-\lambda}$, where $W^2\approx \frac{Q^2}{x}$ is the square of the c.m. energy of the virtual photon proton system,  $W_0 = 140$ GeV is the average $W$ value of the data and $\alpha_{pom}(0)$ is the intercept of the pomeron trajectory obtained by fitting the cross section data from hadron-hadron scattering~\cite{Donlan84,Cudell}. The fit results for this form are presented in Table~\ref{t:f2fitb}. The conclusions are the same as before.

\begin{table}[hbt]
\begin{center}
\caption{Parameters from the fit of ZEUS 1994 ${\cal F}_2$ data to the form ${\cal F}_2 = a + bx^{-\lambda}$.}
\vspace{0.5cm}
\label{t:f2fita}
\begin{tabular}{|c|c|c|c|c|} 
\hline
parameter & $Q^2$=10GeV$^2$ & 22GeV$^2$     & 90GeV$^2$     & 250GeV$^2$ \\
\hline
a         & 0.31$\pm$0.03   & 0.14$\pm$0.05 & 0.01$\pm$0.11 & 0.06$\pm$0.11\\
b         & 0.05$\pm$0.01   & 0.10$\pm$0.02 & 0.18$\pm$0.07 & 0.11$\pm$0.06\\
$\lambda$ & 0.36$\pm$0.02   & 0.32$\pm$0.02 & 0.32$\pm$0.05 & 0.44$\pm$0.10\\
\hline
\end{tabular}
\end{center}
\end{table}

\begin{table}[hbt]
\begin{center}
\caption{Parameters from the fit of ZEUS 1994 ${\cal F}_2$ data to the form 
${\cal F}_2 = \rm{a}_{had} (\frac{W^2}{W^2_0})^{\alpha_{pom}(0) -1} +
\rm{b} x^{-\lambda}$ with $W_0 = 140$ GeV and ${\alpha_{pom}(0) = 1.09}$.}

\vspace{0.5cm}
\label{t:f2fitb}
\begin{tabular}{|c|c|c|c|c|} 
\hline
parameter & $Q^2$=10GeV$^2$ & 22GeV$^2$     & 90GeV$^2$     & 250GeV$^2$ \\
\hline
$\rm{a}_{had}$ & 0.58$\pm$0.03   & 0.32$\pm$0.06 & 0.01$\pm$0.17 & 0.11$\pm$0.15\\
b         & 0.013$\pm$0.003  & 0.063$\pm$0.016 & 0.18$\pm$0.08 & 0.09$\pm$0.06\\
$\lambda$ & 0.47$\pm$0.03   & 0.37$\pm$0.03 & 0.32$\pm$0.05 & 0.47$\pm$0.11\\
\hline

\end{tabular}
\end{center}
\end{table}

\begin{figure}[ht]
\begin{center}
\epsfig{file=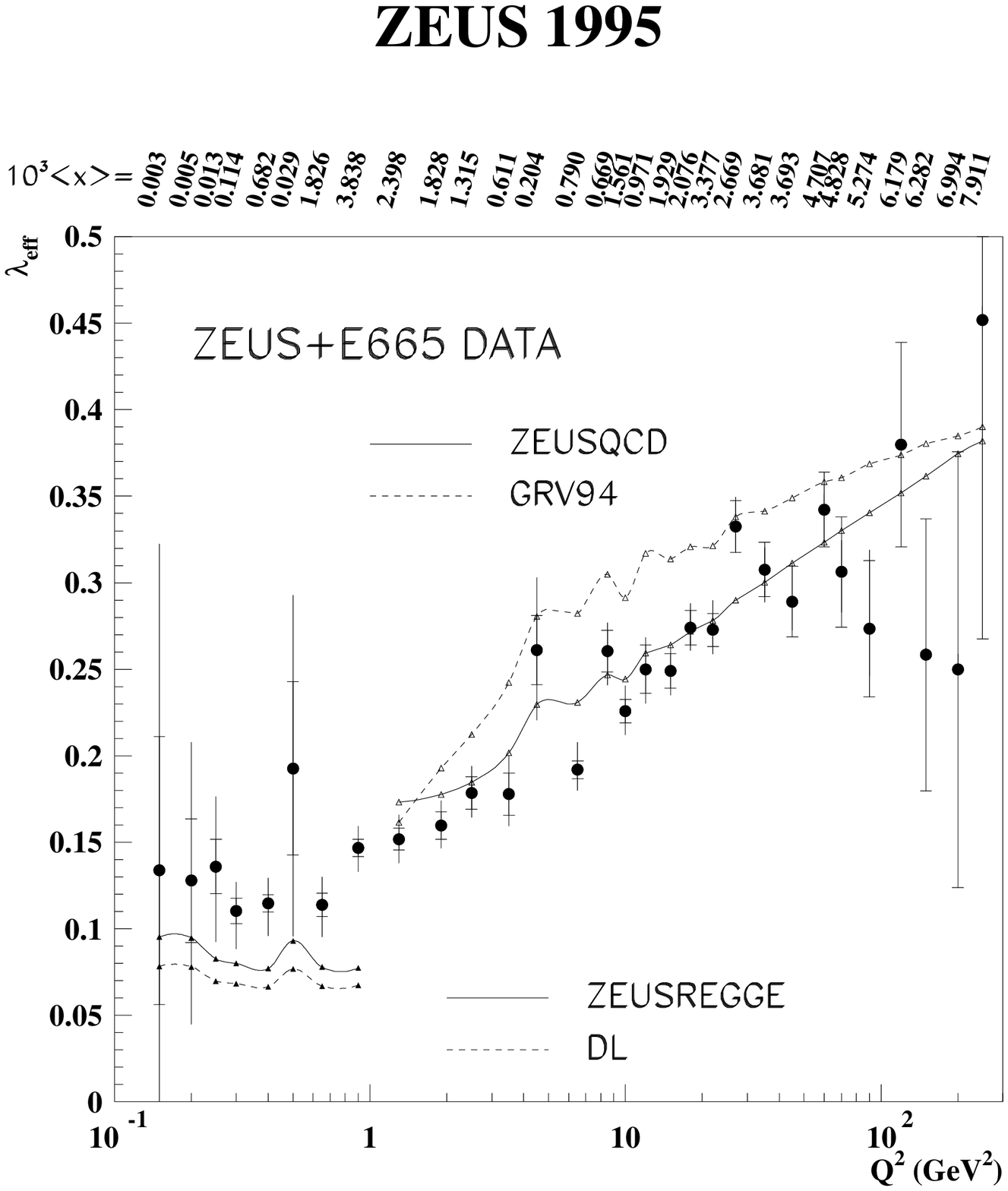,height=12cm,clip=}
\end{center}
\caption{\label{f:Z95lambdavsq}
        { The power $\lambda$ as a function of $Q^2$  obtained from fitting
          ${\cal F}_2$ to the form ${\cal F}_2 = b x^{-\lambda}$. From
          \protect\cite{Zepf295}.
        }}
\end{figure}

Fits performed without a constant term ($a \equiv 0$) result in considerably larger values of $\chi^2/ndf$ for $Q^2 < 40$ GeV$^2$. Thus, in this range of $Q^2$ the data indicate the presence of a soft term $a \neq 0$ as has been noticed before~\cite{Wolf94}. If $a$ is set to zero, $\lambda$ is found to rise with $Q^2$ as shown in Fig.~\ref{f:Z95lambdavsq}, see e.g.~\cite{Hepf295,Zepf295}.

The curves shown in Figs.~\ref{f:HZ967f2vsxa},~\ref{f:HZ967f2vsxb} and~\ref{f:HZ967f2vsq} are the result of QCD NLO fits by H1~\cite{Hepf2967}, ZEUS~\cite{Zepf2967} and ~\cite{Botje,MRST1,CTEQ4M,MRSd0prime} based on DGLAP evolution~\cite{DGLAP}. The fits show that NLO DGLAP evolution can give a consistent description of the data over the full $Q^2$ range.

\subsection{Structure function ${\cal F}_L$}
In the Quark Parton Model (QPM = zero order QCD), ${\cal F}_L$ vanishes for spin 1/2 partons. In LO QCD ${\cal F}_L$ acquires a nonzero value due to the contribution from gluon radiation which is proportional to the strong coupling constant $\alpha_s$. A direct determination of ${\cal F}_L$ requires the measurement of the DIS cross section at fixed $x,Q^2$ for different values of $y$ which can be accomplished e.g. by varying the $ep$ c.m. energy squared $s$. 
 
H1 has shown that for a limited region of high $y$, ${\cal F}_L$ can be extracted from the ${\cal F}_2$ measurements at a single value of $s$ if these measurements are combined with a rather weak assumption on the validity of the DGLAP evolution~\cite{Hepf2967,Hepfl94}.  At high $y$ the factors   $1+(1-y)^2$ and   $y^2$ which multiply ${\cal F}_2$ and ${\cal F}_L$, respectively,  in the expression for the DIS cross section (see Eq.~\ref{eq:sigmanc}), are of comparable magnitude. With this in mind, the following procedure was chosen. The ${\cal F}_2$ values measured by H1 for $y <$ 0.35 and by BCDMS at larger values of $x$ are used to extract the parton distribution functions. The DGLAP equations allow to evolve the parton distribution functions in $Q^2$ for fixed $x$ and to predict ${\cal F}_2$ at high $y$. Subtraction of the ${\cal F}_2$ contribution yields then ${\cal F}_L$. Note, as shown above, NLO DGLAP gives a good description of the ${\cal F}_2$ data over four orders of magnitude in $x$ and $Q^2$, while for the determination of ${\cal F}_L$ the evolution extends the maximum $Q^2$ at fixed $x$ by only a factor of two. Nevertheless, this analysis cannot strictly exclude the possibility that ${\cal F}_2$ behaves differently than assumed.

\begin{figure}[ht]
\begin{center}
\epsfig{file=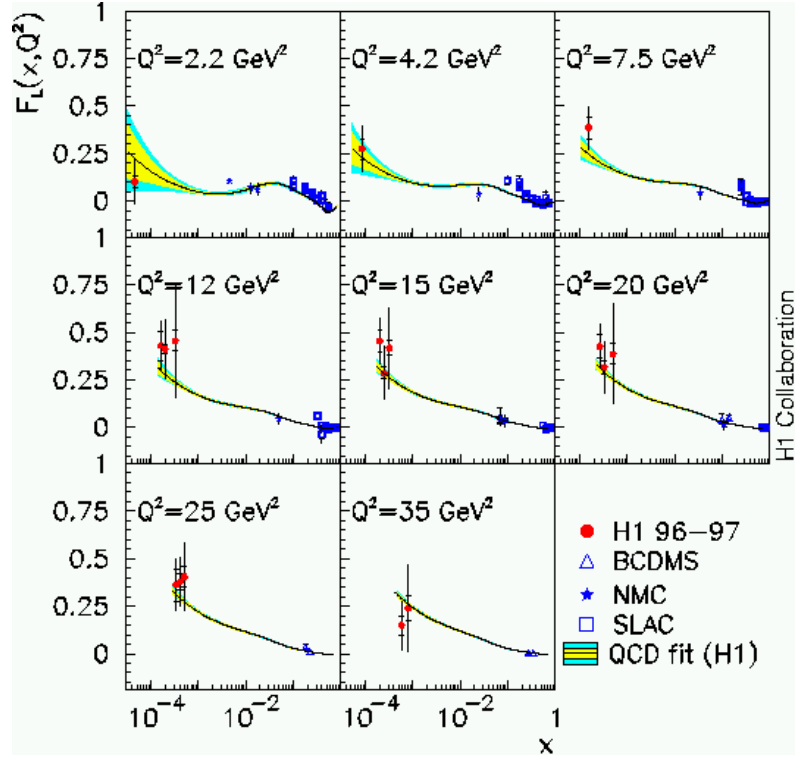,height=10cm}
\end{center}
\caption{\label{f:Hep967flvsx}
        { Structure function ${\cal F}_L$ from NC scattering as a function
          of $x$ as determined by H1. 
          The dashed band shows ${\cal F}_L$ as expected from
          the QCD NLO analysis.
        }}
\end{figure}

The longitudinal structure function ${\cal F}_L$ extracted by H1 is shown in Fig.~\ref{f:Hep967flvsx} as a function of $x$. The full error bars represent the statistical and systematic uncertainties added in quadrature. The ${\cal F}_L$ values are significantly above zero and a factor of 2 - 3 below those of ${\cal F}_2$. The dashed bands, which show ${\cal F}_L$ as expected from the QCD NLO analysis, are consistent with the extracted ${\cal F}_L$ values. 
 
This result lends also support to the procedure applied in the extraction of ${\cal F}_2$ from the DIS cross ection where the (small) contribution from ${\cal F}_L$ was taken from a QCD analysis, see above.

\subsection{Gluon density of the proton}
\begin{figure}[ht]
\centerline{\epsfig{file=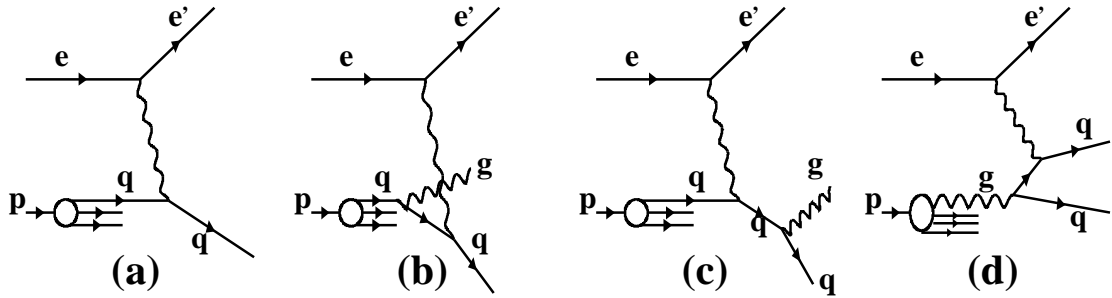,bbllx=50pt,bblly=45pt,bburx=510pt,bbury=180pt,height=4.5cm,clip=}}
\caption{DIS diagrams: (a) no QCD radiation, (b)-(d) lowest order QCD processes.}
\label{f:diagdislo}
\end{figure}

In the QPM (diagram(a) in Fig.~\ref{f:diagdislo}) the structure functions scale. Violations of scaling arise from QCD radiative effects which in LO (diagrams (b)-(d) in Fig.~\ref{f:diagdislo}) have a simple mathematical expression:

\begin{eqnarray}
\frac{d{\cal F}_2}{d\ln Q^2} = \sum_q e^2_q \frac{\alpha_s(Q^2)}{2 \pi} \int^1_x \frac{dy}{y} [P_{qq}(\frac{x}{y}) q(y,Q^2) + P_{qg}(\frac{x}{y}) g(x,Q^2)].
\label{eq:df2dlnq2}
\end{eqnarray}

Here, $P_{qq}$ and $P_{qg}$ are the quark and gluon splitting functions and $g(x,Q^2)$ is the gluon density of the proton. At small $x$, $x < 10^{-2}$, the dominant contribution is expected to come from quark-pair creation by gluons (diagram (d) in Fig.~\ref{f:diagdislo} and second term in Eq.~\ref{eq:df2dlnq2}) which offers the possibility to determine the density of gluons $g$ in the proton rather directly. It is instructive to look first at the approximate relation derived by ~\cite{Prytz93} who considers only the contribution from diagram(d):

\begin{eqnarray}
x \cdot g(x,Q^2) \approx \frac{27 \pi}{10 \alpha_s(Q^2)} \frac{dF_2(x,Q^2)}{d\ln Q^2}.
\end{eqnarray}
The scaling violations of $F_2$, $\frac{d{\cal F}_2}{d\ln Q^2}$, measure directly the gluon momentum density $xg(x,Q^2)$. 

H1 and ZEUS determined the gluon density by performing a global DGLAP type QCD fit to the ${\cal F}_2$ data such as shown by the curves in Figs.~\ref{f:HZ967f2vsxa}-~\ref{f:HZ967f2vsq}. This method takes the contributions from the quark densities automatically into account but requires assumptions on the $x$ dependence of the quark and gluon densities at the evolution scale $Q^2_0$ while for the approximate method these are not needed. 

\begin{figure}[ht]
\begin{center}
\epsfig{file=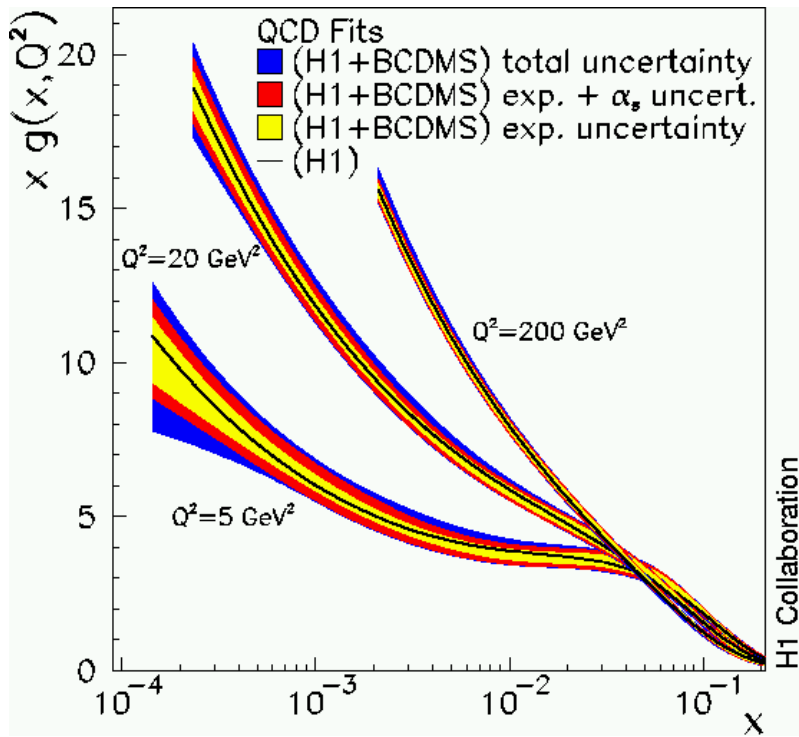,bbllx=90pt,bblly=130pt,bburx=540pt,bbury=710pt,height=11cm}
\end{center}
\caption{\label{f:Hxg967vsx.eps}
      { The gluon momentum density of the proton as a function of $x$ 
        at $Q^2 =$ 5, 20, 200 GeV$^2$ as determined by H1.
      }}
\end{figure}

The gluon momentum density $xg(x,Q^2)$ of the proton as determined by H1~\cite{Hepf2967} at $Q^2$ = 5, 20 and 200 GeV$^2$ is shown in Fig.~\ref{f:Hxg967vsx.eps}. Similar results have been reported by ZEUS~\cite{Zepf295,Botje}. The precision now achieved for $xg(x,Q^2)$ is around 8\% at $x = 5.10^{-4}$ and $Q^2$ = 20 GeV$^2$.

It is instructive to convert the HERA data for ${\cal F}_2$ and $xg(x,Q^2)$ which give the quark plus antiquark and gluon momentum densities, respectively, into numbers of partons and compare these with those obtained at large $x$ by~\cite{MRSd0prime} from the parton density set MRSD0'. The result is shown in Table~\ref{t:npartons} for $Q^2 = 20$ GeV$^2$ which corresponds to a spatial resolution of about $5.10^{-15}$ cm. For high $x$, $x>0.06$, one finds about 2.4 quarks which is close to the canonical number of 3 quarks; in addition there are roughly 1.8 gluons. The parton numbers increase rapidly towards small $x$: for $5.10^{-4} < x < 5.10^{-3}$ there are 4 times as many $q,\overline{q}$ and 15 times as many gluons as compared to the high-$x$ regime.

\begin{table}[hbt]
\centering
\caption{Equivalent number of partons of the proton for $Q^2 =20$ GeV$^2$ determined  at low $x$ from the H1 and ZEUS data and at large $x$ from the MRSD0' set.}
\vspace{0.5cm}
\label{t:npartons}
\begin{tabular}{|l|c|c|} \hline
                     & $x > 0.06$    & $5\cdot 10^{-4}<x< 5\cdot 10^{-3}$\\
\hline
$N_{q,\overline{q}}$ & $\approx 2.4$ & 9 $\pm$ 1\\
$N_g$                & $\approx 1.8$ & 27 $\pm$ 2\\
\hline
\end{tabular}
\end{table}

\subsection{Charm contribution to $F_2$}
The structure function $F_2$ measures the momentum densities of quarks in the proton summed over all quark flavours. Charm quarks are expected to contribute via fusion of the virtual photon with a gluon from the proton (boson-gluon fusion) (diagram (d) of Fig.~\ref{f:diagdislo}). Therefore  pair production of charm quarks in DIS offers another way to measure the gluon density of the proton.

\begin{figure}[ht]
\begin{center}
\epsfig{file=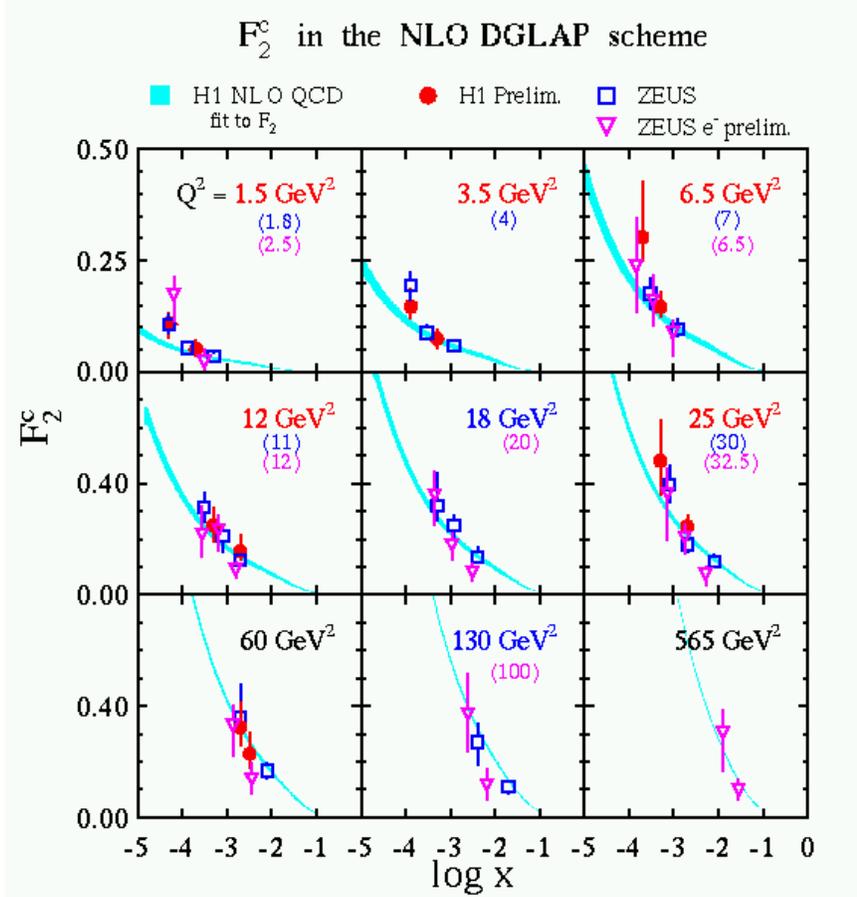,height=12.cm}
\end{center}
\caption{\label{f:HZ967f2cvsx}
     {  The charm contribution $F^c_2(x,Q^2)$ to the proton structure
        function $F_2$ for fixed $Q^2$ as a function
        of $x$ as measured by H1 and ZEUS. The curves show the 
        predictions from the H1 NLO QCD fit to the 1996-7 $F_2$ data.
     }}
\end{figure}

\begin{figure}[ht]
\begin{center}
\epsfig{file=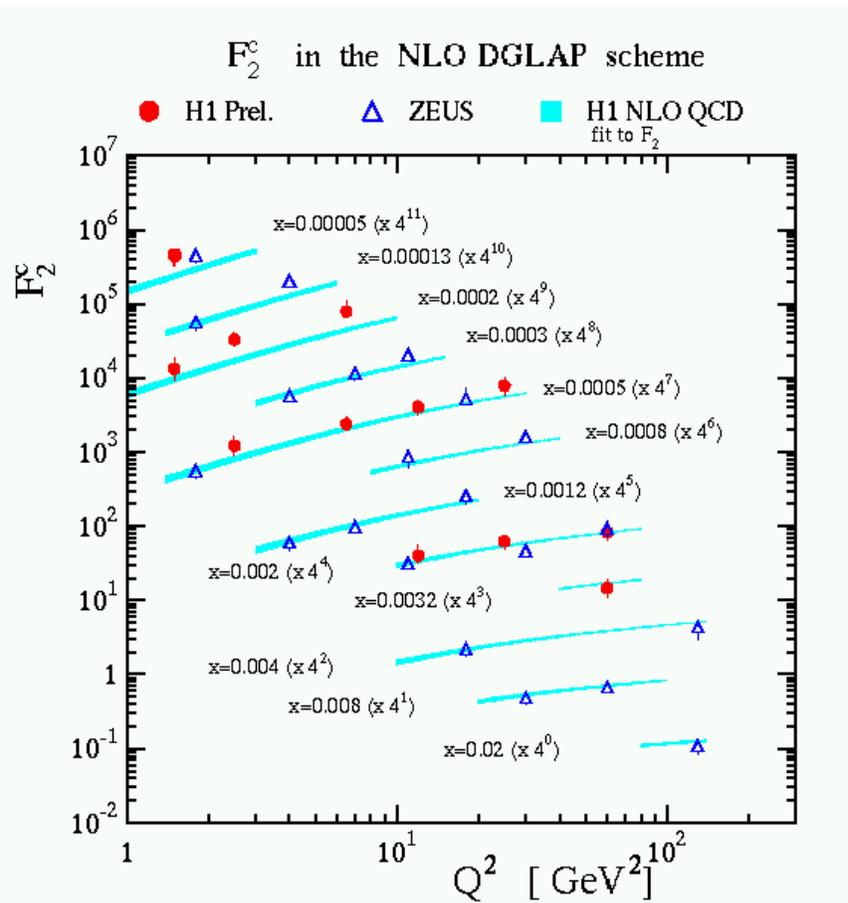,bburx=610pt,bbury=601pt,height=12cm}
\vspace{-1.0cm}
\end{center}
\caption{\label{f:HZ967f2cvsq}
     {  The charm contribution $F^c_2(x,Q^2)$ to the proton structure
        function $F_2$ for fixed $x$ as a function
        of $Q^2$ as measured by H1 and ZEUS. The curves show the 
        predictions from the H1 NLO QCD fit to the 1996-7 $F_2$ data.
     }}
\end{figure}

The charm contribution to DIS was determined by detecting  $D^*$ production in a limited $\eta^{D^*},p^{D^*}_{\perp}$ region~\cite{H1epc94,Zepc94,Zepc967,Zepc967Os,Hepc967}. The measured cross section $\sigma(ep \to eD^* X)$ was extrapolated to the full range in $\eta^{D^*},p^{D^*}_{\perp}$ with the help of a QCD model~\cite{Harrissmith}. Using the branching ratio for $c \to D^*$ measured at LEP the cross section $\sigma(ep \to e c\overline{c} X)$ and from this the charm contribution $F^c_2(x,Q^2)$ were determined. In Figs.~\ref{f:HZ967f2cvsx},~\ref{f:HZ967f2cvsq} $F^c_2$ is displayed as a function of $x$ for fixed $Q^2$ values, and for fixed $x$ as a function of $Q^2$. A comparison with the $F_2$ data shows that for $Q^2 \ge 6$ GeV$^2$ charm contributes about 20 - 30\% of $F_2$. This is in broad agreement with the ratio of 4/10 expected for a democratic sea assuming massless quarks and neglecting the $b$ quark contribution.

The curves in Fig.~\ref{f:HZ967f2cvsx},~\ref{f:HZ967f2cvsq} show the prediction for $F^c_2$  using the gluon density as obtained from the QCD fit to $F_2$. They are in good agreement with the data providing an important test for the QCD analyses and fits performed by H1 and ZEUS on $F_2$. 

\begin{figure}[ht]
\begin{center}
\epsfig{file=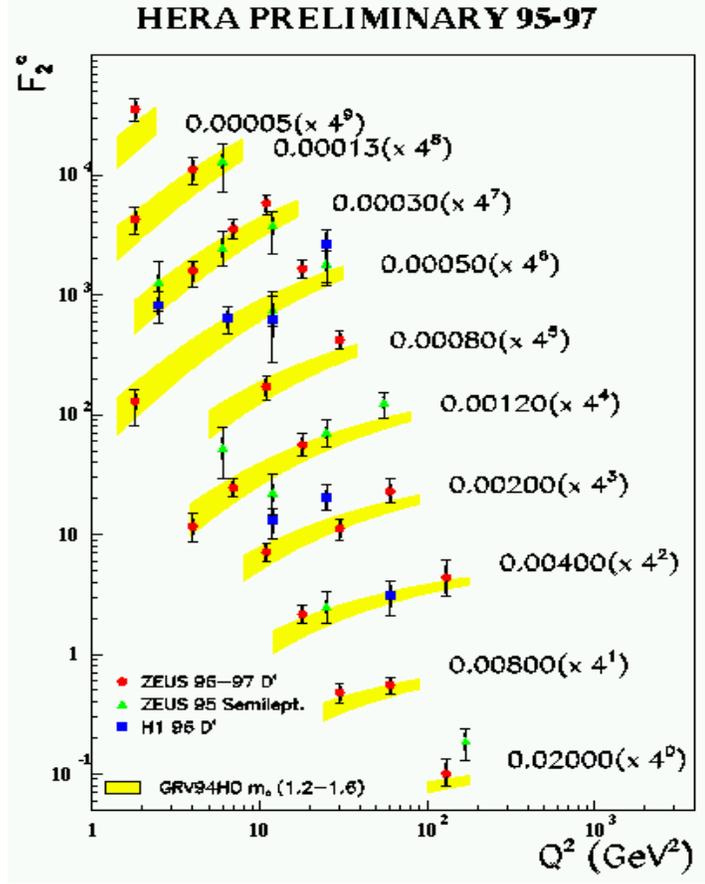,bbllx=-100pt,bblly=0pt,bburx=510pt,bbury=601pt,width=14cm,height=12cm}
\vspace{-1.0cm}
\end{center}
\caption{\label{f:f2cvsq2withsemih1.onlygrv}
     {  The charm contribution $F^c_2(x,Q^2)$ to the proton structure
        function $F_2$ for fixed $x$ as a function
        of $Q^2$ as measured by H1 and ZEUS. The bands show the 
        GRV94HO predictions for masses of the charm quark between 1.2 and 1.6
        GeV. 
     }}
\end{figure}

In Fig.~\ref{f:f2cvsq2withsemih1.onlygrv} the $Q^2$ dependence of $F^c_2$ is compared for $x$ values between $5.10^{-5}$ and 0.02. The bands show the predictions of GRV94HO~\cite{GRV94c} for a range of charm quark masses between 1.2 and 1.6 GeV. The agreement is impressive.

\subsection{The rise of ${\cal F}_2$ as $x\to 0$ in the light of theory and models}

\subsubsection{The rise of ${\cal F}_2$ as $x \to 0$ and the $W$ dependence of $\sigma^{tot}_{\gamma^*p}$}

Neglecting contributions from $Z^0$ exchange, the DIS cross section can be expressed in terms of the flux of virtual photons times the total cross section for virtual - photon proton scattering, $\sigma^{tot}_{\gamma^*p}$ ~\cite{Hand}, which is written in terms of the cross sections for the scattering of transverse and longitudinal photons,
\begin{eqnarray}
\sigma^{tot}_{\gamma^*p}(x,Q^2) = \sigma_T(x,Q^2) + \sigma_L(x,Q^2).
\end{eqnarray}
The cross section defined in this manner can be interpreted in a way similar to the case of the interaction of real photons provided the lifetime of the virtual photon is large compared to the interaction time (in the proton rest frame), which means $x \ll 1/(2M_p \cdot R_p)$, where $R_p$ is the proton radius, $R_p = 4$ GeV$^{-1}$~\cite{Ioffe}. This requirement is well satisfied if $x \ll 0.1$. The expression for ${\cal F}_2$ in terms of $\sigma_T$ and $\sigma_L$ is
\begin{eqnarray}
{\cal F}_2(x,Q^2) = \frac{Q^2 (1 - x)}{4 \pi^2 \alpha} \sigma^{tot}_{\gamma^*p}.
\end{eqnarray}

\begin{figure}[ht]
\begin{center}
\epsfig{file=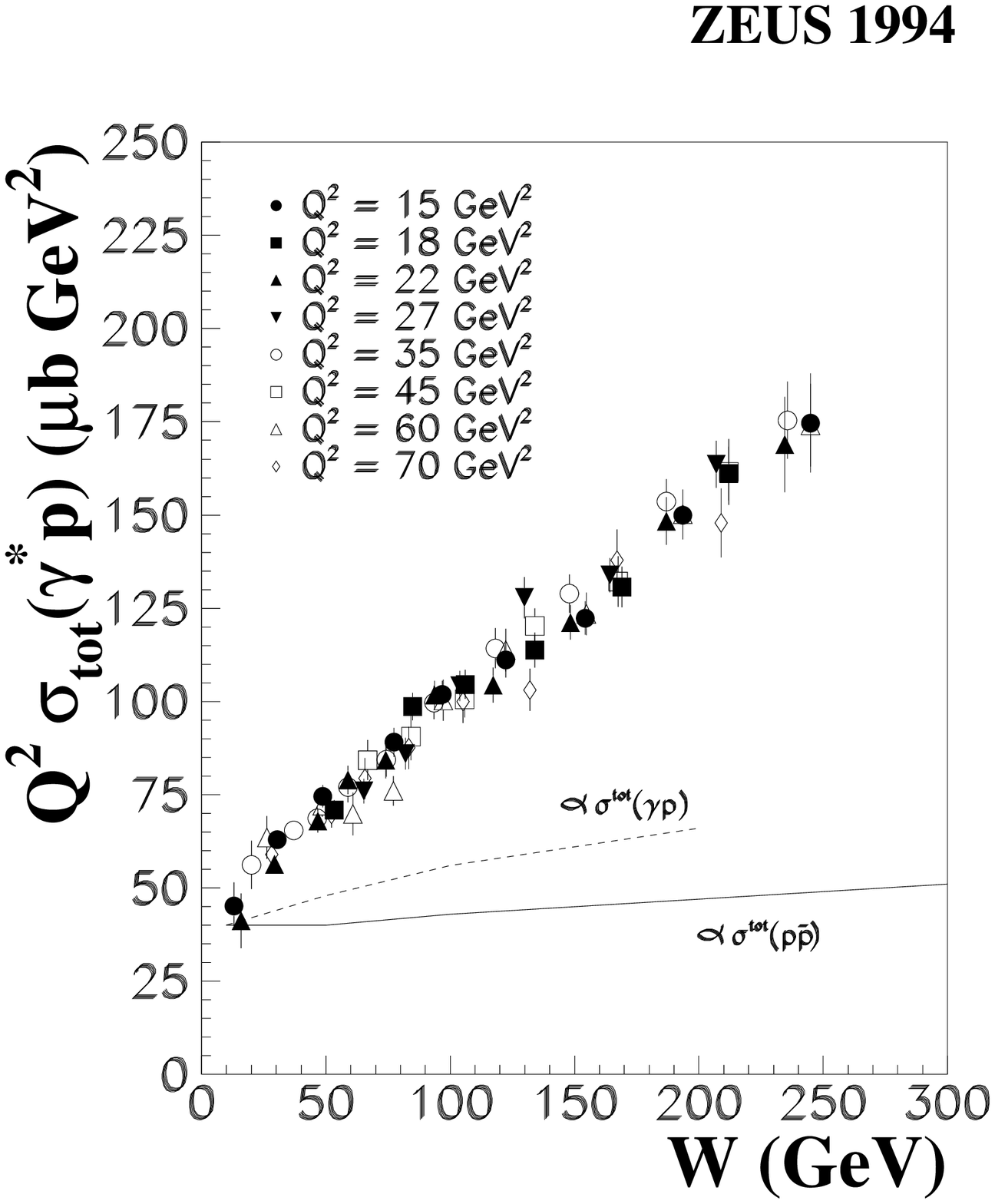,bbllx=30pt,bblly=140pt,bburx=510pt,bbury=671pt,height=10cm,clip=}
\end{center}
\caption{\label{f:Zepq2sigvsw}
     {  $Q^2 \cdot \sigma^{tot}_{\gamma^*p}$ as a function of $W$ for fixed
        $Q^2$ between 15 and 70 GeV$^2$ as determined from the ZEUS data. The
        curves are proportional to the total cross sections 
        $\sigma^{tot}_{\gamma p}$ and 
        $\sigma^{tot}_{p \overline{p}}$.
     }}
\end{figure}

At small $x$ the expression can be rewritten in terms of the virtual-photon proton c.m. energy $W$, $W^2 \approx Q^2/x$ leading to
\begin{eqnarray}
\label{sigtot}
\sigma^{tot}_{\gamma*p} \approx \frac{4 \pi^2 \alpha}{Q^2} {\cal F}_2(W,Q^2)
\end{eqnarray}
Equation~\ref{sigtot} was used by ZEUS~\cite{Zepf293} to determine from the 1993 ${\cal F}_2$ data  $\sigma^{tot}_{\gamma^*p}$. Figure~\ref{f:Zepq2sigvsw} shows  for the 1994 data $Q^2 \sigma^{tot}_{\gamma^*p}$ as a function of $W$ from 20 to 260 GeV for fixed $Q^2$ between 15 and 70 GeV$^2$~\cite{Wolf97}.  The data cluster around a narrow band rising almost linearly with $W$. A fit of the form $Q^2 \sigma^{tot}_{\gamma^*p} = a + b\cdot W^{\epsilon}$ gave the value of $\epsilon \approx 0.9$ (a fit with a=0 yielded a smaller value, $\epsilon_{a=0}  \approx 0.5$, though with considerable larger $\chi^2/ndf$). If $\sigma^{tot}_{\gamma^*p}$ is described in terms of a pomeron trajectory, $\sigma^{tot}_{\gamma^*p} \propto (W^2)^{\alphapom(0) -1}$, the intercept $\alphapom(0) = 1 +\frac{\epsilon}{2} \approx 1.45$.  

The observed rise of  $\sigma^{tot}_{\gamma^*p}$ is in marked contrast to the behaviour of the total cross section for antiproton-proton scattering and for $real$ photon-proton scattering which are compatible with $\sigma^{tot}_{\gamma^*p} \approx W^{0.2}$, see curves in Fig.~\ref{f:Zepq2sigvsw}. Thus, the rise of ${\cal F}_2$ as $x \to 0$ or that of $\sigma^{tot}_{\gamma^*p}$ as $W \to \infty$ signals the presence of a new phenomenon. In QCD, the rise is the result of the strong increase of the number of partons at small $x$ which blacken the proton and reduce its transparency for virtual photons of high $Q^2$.

\subsubsection{DGLAP evolution}
The DGLAP equations describe the evolution of the parton densities with $Q^2$. In order to solve these equations one must provide the parton densities as a function of $x$ at some reference scale $Q^2_o$ which should be large enough for perturbative QCD to be applicable~\cite{Kwiecinski}. Assuming a Regge-type behavior~\cite{Leader80,Lopez80}, the small $x$ dependence of the valence quark (v), sea quark (s)  and gluon densities is of the form:
\begin{eqnarray}
xq_v(x,Q^2)  \propto x^{1-\alpha_R} \\
xq_s(x,Q^2)  \propto x^{1-\alphapom} \\
xg(x,Q^2)      \propto x^{1-\alphapom}
\end{eqnarray}
where $\alpha_R$ and $\alphapom$ denote the intercepts of the reggeon and pomeron trajectories. For $\alpha_R \approx 0.5$ and $\alphapom \approx 1$ one obtains $xq_v(x,Q^2_o) \propto x^{0.5}$ and $xq_s(x,Q^2_o) \propto x g(x,Q^2_o) \propto const$. 

In the leading log $Q^2$ approximation the $x^{0.5}$ behavior of the valence distribution remains unchanged by the $Q^2$ evolution while the sea quark and gluon distributions at small $x$ become steeper. In fact,
in perturbative QCD, ${\cal F}_2$ is expected to grow faster than any power of $\ln (1/x)$ as $x \to 0$~\cite{Gribov72,Derujula74}:

\begin{eqnarray}
{\cal F}_2(x,Q^2) &  \approx & C_o [\frac{33-2n_f}{576\pi^2 \ln \frac{1}{x} \ln\frac{\alpha_s(Q^2_o)}{\alpha_s(Q^2)}}]^{1/4} \cdot exp\sqrt{\frac{144 \ln \frac{1}{x}}{33-2n_f} \ln \frac{\alpha_s(Q^2_o)}{\alpha_s(Q^2)}.}
\label{eq:f2qcd}
\end{eqnarray} 
where $n_f$ is the number of quark flavors. The rise of ${\cal F}_2$ as $x \to 0$ can be accelerated by decreasing the reference scale $Q^2_o$.
This can be seen by applying Eq.~\ref{eq:f2qcd} for a specific set of parameters, e.g. $n_f =3$ and $\alpha_s(Q^2) = \frac{4\pi}{(11-\frac{2}{3}n_f)\ln Q^2/\Lambda^2}$ with $\Lambda = 0.2$ GeV. Starting with $x$-independent parton distributions at $Q^2_o$ and parametrizing the ${\cal F}_2(x,Q^2)$ values obtained from Eq.~\ref{eq:f2qcd} as ${\cal F}_2(x,Q^2) = b(Q^2)(1/x)^{\lambda (Q^2)}$ for $10^{-4}<x<10^{-2}$ yields 
\begin{tabbing}
\hspace{1cm}\=for $Q^2_o=4$ GeV$^2$ at $Q^2$ = 10 (20,\,100)GeV$^2$: $\lambda \approx$ 0.15 (0.21, 0.29)\\
\>for $Q^2_o=1$ GeV$^2$ at $Q^2$ = 10 (20,\,100)GeV$^2$: $\lambda \approx$ 0.29 (0.32, 0.38).
\end{tabbing}

Hence, theory predicted that $F_2$ would rise as $x \to 0$ but it could {\it not} predict the value of $\lambda$ and therefore the speed of the rise since the starting parton distributions at $Q^2_0$ were unknown. We will return to this point below in a discussion of the GRV model~\cite{GRV94,GRV98}.

\subsubsection{BFKL evolution} 
The DGLAP scheme requires angular ordering and neglects terms proportional to $\ln \frac{1}{x}$, an approximation, which may run in difficulties as $x \to 0$. The BFKL formalism~\cite{Lipatov} does not impose angular ordering and resums terms proportional to $\ln \frac{1}{x}$. Based on the BFKL formalism which performs QCD evolution for fixed $Q^2$ as function of $x$ the gluon density in the proton was predicted to rise as $g(x,Q^2) \propto x^{-(1+ \lambda)}$ as $x \to 0$, where $\lambda \approx \alpha_s (12/\ln 2)/\pi \approx 0.5$ for $Q^2 = 20$ GeV$^2$. BFKL-type calculations in NLO predict a considerably smaller value for $\lambda$, $\lambda \approx 0.15$~\cite{Bottazzi,Scorletti,Catani}. BFKL inspired fits to the data have been performed by~\cite{Kwiecinski97}.

\subsubsection{Lopez-Yndurain model}
The authors of~\cite{Lopez80} presented in 1980 - i.e. long before HERA came into operation - a NLO QCD model which predicted the rise of ${\cal F}_2$ at small $x$ observed at HERA with remarkable accuracy. According to the model, ${\cal F}_2$ should behave at small $x$ as a power in $x$, ${\cal F}_2 \propto x^{-\lambda_s}$ where $\lambda_s$ is independent of $Q^2$, except for heavy flavor thresholds. Extending the scanty data available then down to $x \approx 0.05$ led to the prediction $\lambda_s = 0.37 \pm 0.07$. Adding a constant term to ${\cal F}_2$ a fit of the model to the new data from H1 and ZEUS provided a good description of the measurements and yielded $\lambda_s = 0.355 \pm 0.01$~\cite{Adel97}. Dividing the data into different $Q^2$ intervals indicated a possible but small rise of $\lambda_s$ with $Q^2$ from $0.325\pm 0.01$ at $Q^2 <$ 10 GeV$^2$ to $0.355 \pm 0.01$ at $Q^2 >$ 100 GeV$^2$.

\subsubsection{GRV model}
The rapid rise of ${\cal F}_2$ at small $x$ observed by the HERA experiments was anticipated in the GRV model~\cite{GRV94}. In this model, the steep rise of ${\cal F}_2$ at fixed $Q^2$ and small $x$ is generated dynamically by QCD evolution starting from an input scale $Q_0 = \mu$ through bremsstrahlung of gluons from quarks and by gluon annihilation into quark-antiquark pairs. At the evolution scale $Q_0 = \mu$ the input parton densities were chosen to be valence-like. 

\begin{figure}[ht]
\begin{center}
\epsfig{file=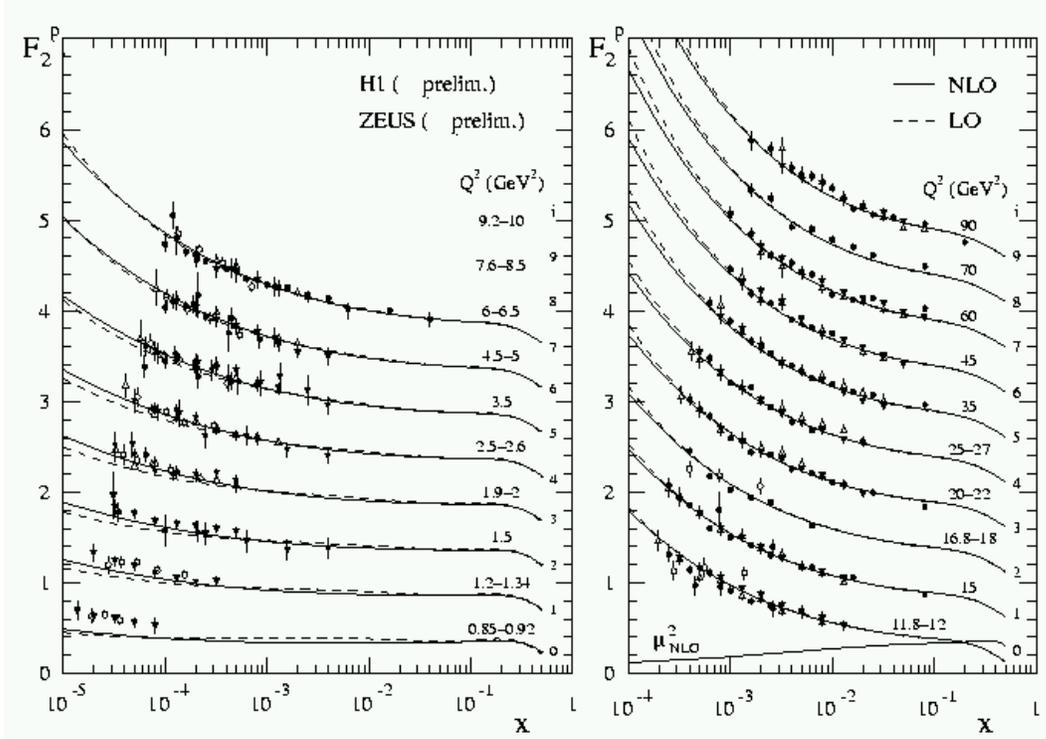,height=14cm,angle=-90}
\end{center}
\caption{\label{f:f2vsxgrv98}
     {  Structure function $F_2$ from NC scattering as a function of
        $x$ for fixed values of $Q^2$ as measured by H1 (1994/5 data) and 
        ZEUS (1994 data). The $F_2$ values are plotted as 
        $F_2(x,Q^2) + i(Q^2) \times 0.5$ where for each $Q^2$ set 
        the value of $i$ is indicated in the figure.
        The solid (dashed) lines show the GRV98 predictions calculated in 
        NLO (LO). 
     }}
\end{figure}

The steepness of the rise depends on the input scale for which originally a value of $\mu^2 = 0.34$ GeV$^2$ was chosen (GRV94). The GRV94 predictions gave a good account of the 1993 ${\cal F}_2$ data from HERA but with the advent of the more precise ${\cal F}_2$ values from the 1994/5 and now from the 1996/7 data it became apparent that the rise predicted by GRV94 was somewhat steeper than measured. Recently, it was shown by the same authors~\cite{GRV98} that agreement with the high precision data from HERA can be obtained by raising slightly the starting scale to $\mu^2 = 0.4$ GeV$^2$ (GRV98). The predictions of GRV98 are shown in Fig.\ref{f:f2vsxgrv98} together with 1994 and 1995 data from H1 and ZEUS. For $Q^2 \ge 1.5$ GeV$^2$ the predictions follow closely the measurements. For lower $Q^2$ the predictions fall below the data which is not surprising since there also nonperturbative contributions are expected.  

\subsubsection{Haidt parametrisation}
Double-logarithmic scaling of ${\cal F}_2$ with respect to $x$ and $Q^2$ has been investigated in~\cite{Ball,Buchmuller-3}. The rise observed in the HERA data at small $x$ is consistent with a logarithmic rise in $x$ as well as in $Q^2$.  An economical parametrization which decribes the ${\cal F}_2$ data of H1 and ZEUS for $x < 10^{-3}$, $Q^2 \ge 0.11$ GeV$^2$ has been obtained in~\cite{Haidt}:
\begin{eqnarray}
{\cal F}_2(x,Q^2) = m \, \log_{10}(1+\frac{Q^2}{Q^2_0}) \, \log_{10}(\frac{x_0}{x})
\end{eqnarray}
with $Q^2_0 = 0.5$ GeV$^2$, $x_0 = 0.04$ and $m= 0.41$.

Besides its simplicity, this parametrisation has the property that for fixed $Q^2$ the total cross section for virtual photon proton scattering, $\sigma_{tot}(\gamma^*p) \approx \frac{1}{Q^2}F_2(x,Q^2)$, does not violate the Froissart bound for hadronic total cross sections.

\subsection{$F_2$ in the transition region between photoproduction and DIS}

Virtual photon proton scattering at large $Q^2$ with its rapidly rising cross section as $W \to \infty$ behaves markedly different from ordinary hadron hadron or real photon proton collisions. The rise is a sign for hard scattering as described by perturbative QCD (pQCD). This leads to the question where in $Q^2$ does the transition from hadronic-type to hard scattering occur. When $\sqrt{Q^2}$ is below the typical hadronic transverse momenta of 200 - 500 MeV one expects confinement effects to dominate the interaction. On the other hand, the success of the GRV model may indicate that already at $Q^2 \approx 1$ GeV$^2$ hard scattering is the dominant type of interaction. In order to gain further insight, a precise mapping of the proton structure functions in the region from photoproduction up to $Q^2$ of the order of a few GeV$^2$ is essential.

\begin{figure}[ht]
\begin{center}
\epsfig{file=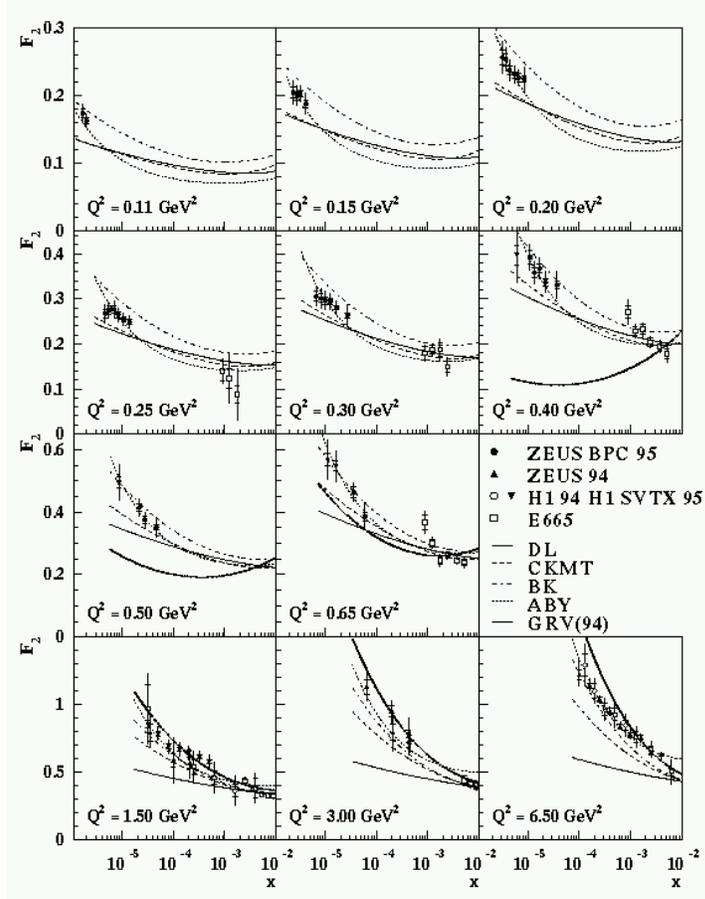,height=12.0cm}
\end{center}
\caption{\label{f:Zepf2loq95vsq}
     {  $F_2$ as a function of $Q^2$ for fixed values of $x$ as measured by ZEUS, H1 and E665. Also shown are predictions of various models, see text.
     }}
\end{figure}

The structure function $F_2(x,Q^2)$ vanishes as $Q^2 \to 0$. This can be seen from its relation with the total virtual photon proton cross section, which is nonzero for $Q^2 = 0$ (Eq.~\ref{sigtot}).
Measurements of $F_2$ at small $Q^2$ and small $x < 10^{-4}$ were published by H1~\cite{H1epf2loq95} and ZEUS~\cite{Zepf295,Zepf2loq45}. Figure~\ref{f:Zepf2loq95vsq} shows $F_2$ for fixed $x$ as a function of $Q^2$. 
A number of phenomenological models~\cite{GRV94,Donlanvdm,Badelek,Capella,Adel,Donlan92,AL97} have been put forward to describe the behaviour at low $x$ and low $Q^2$. The curves in Fig.\ref{f:Zepf2loq95vsq} show predictions of some of these models.  The Vector Dominance type model DL~\cite{Donlanvdm} fails to reproduce the rise of ${\cal F}_2$ for $Q^2$ above $\approx 0.5$ GeV$^2$. The BK model~\cite{Badelek} assumes a VDM-like component which dominates the region of low $Q^2$ plus a hard QCD-like component for the high $Q^2$ regime. The predictions are somwhat above the data at low $Q^2$. The model CKMT~\cite{Capella} assumes at high $Q^2$ the dominance of a bare pomeron with an intercept of $\alphapom(0) \approx 1.24$. At low $Q^2$ the pomeron intercept is assumed to decrease due to rescattering corrections leading to  $\alphapom(0) = 1.08$, the value obtained from hadron-hadron scattering.  The CKMT predictions are found to be below the data for $Q^2 < 0.6 - 1$ GeV$^2$. The GRV model~\cite{GRV94} considers only the hard scattering contribution. The GRV94 predictions for ${\cal F}_2$ are close to zero for $Q^2$ near the evolution scale $Q^2 =$ 0.34 GeV$^2$. At $Q^2 =$ 0.44 GeV$^2$ GRV94 accounts for about 40$\%$ of the measured ${\cal F}_2$ and for about 80$\%$ at $Q^2 =$ 0.57 GeV$^2$. At $Q^2 = 0.9$ GeV$^2$ basically all of the DIS cross section is attributed to hard scattering. The model ABY~\cite{Adel}, which assumes a hard plus a soft component evolved in NLO-QCD, gives a rather good descripton of the full set of data. The comparison suggests that the transition from soft to hard scattering occurs at $Q^2$ values somewhere between 0.8 and several GeV$^2$.

\begin{figure}[ht]
\begin{center}
\epsfig{file=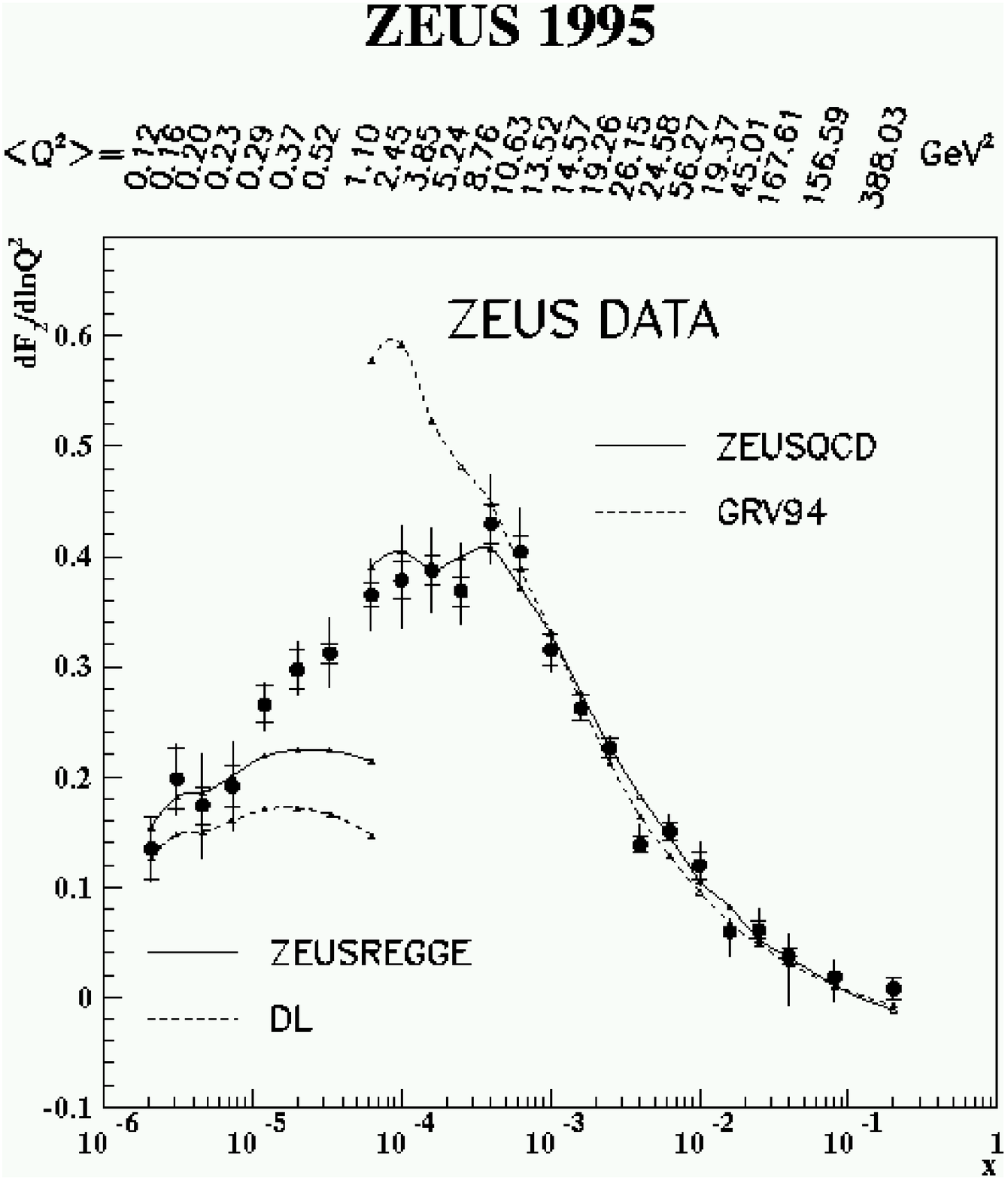,height=7cm,width=9cm}
\end{center}
\caption{\label{f:Zeplnf2dq95vsx}
     {  $dF_2/d\ln Q^2$ as a function of $x$ calculated by fitting
        ZEUS $F_2$ data in bins of $x$ to the functional form 
        $a + b \ln Q^2$. For each $x$ bin the average $Q^2$ value is 
        indicated on top of the figure. The linked points labelled DL
        and GRV94 are from the Donnachie-Landshoff Regge fit and the GRV94
        NLO QCD fit. In both cases, the points are obtained using the same 
        $Q^2$ range as for the experimental data. From ZEUS.
     }}
\end{figure}

ZEUS studied also the scaling violations, $dF_2/d\ln Q^2$, in the transition region~\cite{Caldwell97,Zepf295,Zepf2loq45}. The logarithmic slope $dF_2/d\ln Q^2$ was derived from the data by fitting $F_2 = a + b \ln Q^2$ in bins of fixed $x$ for $W^2 \simeq Q^2/x > 10$ GeV$^2$. Figure~\ref{f:Zeplnf2dq95vsx} shows $dF_2/d\ln Q^2$ as a function of $x$ (Caldwell plot). Also shown (on the top of the figure) for each $x$ bin is the weighted mean of $Q^2$ ($<Q^2>$) which increases as $x$ increases due to kinematics and detector acceptance. For $x$ values down to $3 \times 10^{-4}$ the slope $dF_2/d\ln Q^2$ increases as $x$ increases. At lower $x$ (equivalent to lower $Q^2$) the slope decreases. The prediction of the GRV94 model, for which $dF_2/d\ln Q^2$ was determined in the same manner as for the data, reproduces the data for $x > 3 \times 10^{-4}$ ($Q^2 > 8$ GeV$^2$). For smaller $x$ the GRV94 slope keeps on rising while in the data it decreases. 

\begin{figure}[ht]
\begin{center}
\epsfig{file=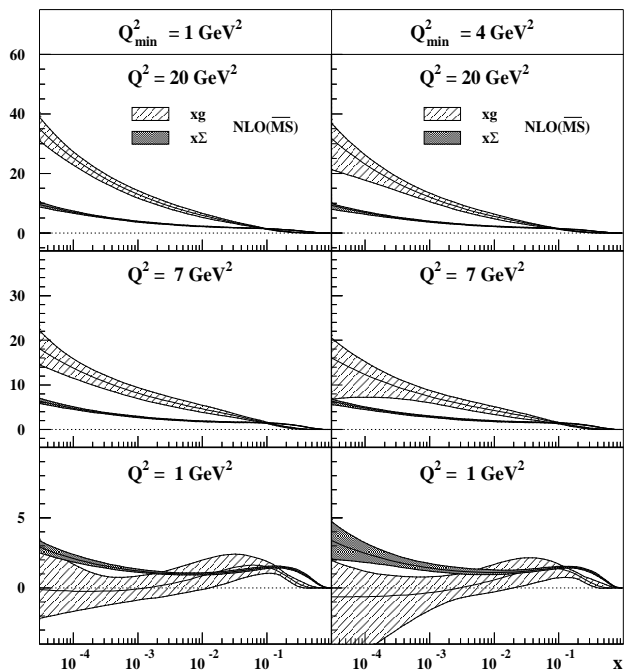,bbllx=50pt,bblly=150pt,bburx=510pt,bbury=701pt,width=8cm,height=10cm,clip=}
\end{center}
\caption{\label{f:Z945xqxgvsx}
     { The quark singlet momentum distribution, $x\Sigma (x,Q^2)$ (shaded), 
       and the gluon momentum density, $xg(x,Q^2)$ (hatched), as a function
       of $x$ at fixed values of $Q^2 = 1,7,20$ GeV$^2$ for two 
       different evolution scales $Q^2_{min}$. From ZEUS.
     }}
\end{figure}

In order to gain further insight into the scaling violations at low $x$ and $Q^2$ ZEUS performed QCD NLO fits using all their data with $3 \times 10^{-5} < x < 0.7$ and $Q^2 > 1$ GeV$^2$ together with those from NMC~\cite{NMC97} and BCDMS~\cite{BCDMS90}. A reasonable description of the data was achieved by the fits. Figure~\ref{f:Z945xqxgvsx} shows the singlet quark momentum density ($x\Sigma \equiv x\sum_q [q(x) + \overline{q}(x)]$) and the momentum density of the gluon as a function of $x$ for $Q^2 = 1,\; 7$ and 20 GeV$^2$. For $Q^2 \ge 7$ GeV$^2$ the gluon density is much larger than the singlet quark density while at $Q^2 = 1$ GeV$^2$ the gluon density has become equal to or lies below the singlet quark density. Also, $x\Sigma$ is seen to rise as $x \to 0$ for all three $Q^2$ values; $xg$ , on the other hand, rises at $Q^2 = 7$ and $20$ GeV$^2$ but may become constant at $Q^2 = 1$ GeV$^2$, or even zero. Such a behaviour was also found by~\cite{MRST98}. 

One may tentatively conclude that at low $Q^2$, of the order of 1 GeV$^2$, the quark sea drives the gluon density while at higher $Q^2$ the gluon drives the density of the sea quarks. However, this conclusion may be premature since the fits did not allow for a contribution from soft scattering which could still be substantial at $Q^2 \ge 1$ GeV$^2$ (see e.g. Fig.~\ref{f:Zepf2loq95vsq}).

\section{Diffraction in deep inelastic scattering}

Diffraction has been studied extensively in hadron-hadron scattering at small momentum transfers~\cite{Diffractionover}. An elegant parametrization of the data has been provided by the Regge formalism through the introduction of a pomeron trajectory~\cite{Gribovetal}. The hypothesis that diffraction may have a partonic component~\cite{Ingelmanschlein} has been substantiated by the observation of high transverse energy jets produced in $p\overline{p}$ scattering~\cite{UA8}. However, in hadron-hadron scattering both collision partners are extended objects which makes extraction of the underlying partonic process(es) difficult.
In DIS, on the other hand, the virtual photon has a pointlike coupling to quarks. Hence, HERA offers a unique opportunity to study the partonic structure of diffraction since it gives, in a well defined manner, access to the regime of large photon virtualities and large energy transfers between the virtual photon and the target proton in its rest system, $\nu = Q^2/(2m_px)=2 - 20$ TeV.

Diffraction in virtual photon proton scattering has been studied at HERA in the quasielastic processes of vector meson production, $\gamma^* p \to Vp$, where $V\;=\; \rho^0, \omega, \phi$, $J/\Psi, \Upsilon$. While low mass $V$ production ($V = \rho^0,\omega,\phi$) contributes more than 10\% of the total cross section at $Q^2 = 0$~\cite{Vphoto} it becomes negligible at large $Q^2$~\cite{Vdis}. However, diffractive dissociation of the virtual photon, $\gamma^* p \to XN$ (N= proton or a low mass nucleon system), into a large mass $M_X$, first recognized by the presence of a class of events with a large rapidity gap~\cite{Zeplrg93,Heplrg94} remains a substantial fraction of the total DIS cross section also at large $Q^2$~\cite{Zepdiff956}. This has opened a window for a systematic study of diffraction in reactions initiated by a hard probe~\cite{Zepdiff956,ZEUSdisdiff,H1epf2d393,Zepdiff93,H1epf2d394,Zeplpsf2d394,Zepdiff94}.

\subsection{t-dependence of the diffractive cross section}
\begin{figure}[ht]
\begin{center}
\epsfig{file=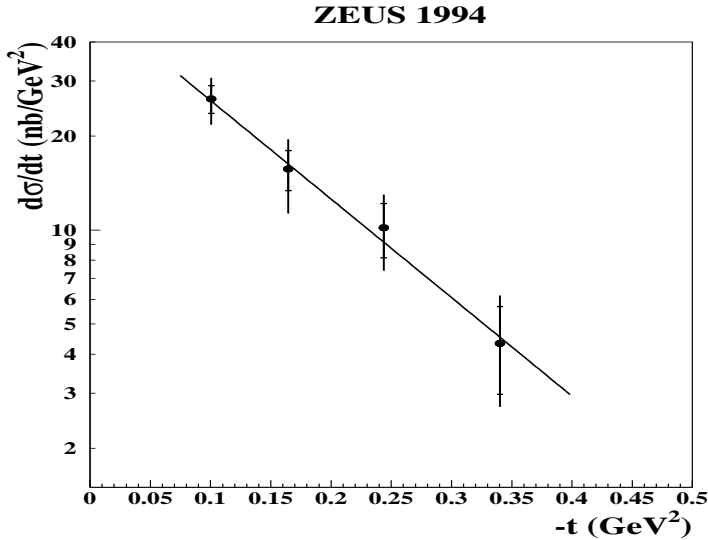,bbllx=120pt,bblly=200pt,bburx=510pt,bbury=701pt,width=8cm,height=7cm}
\end{center} 
\caption{\label{f:Zeplps94sigvst}
     {  The diffractive cross section $d\sigma_{\gamma^* p \to Xp}/d|t|$
        for events with a leading proton carrying more than 97\% of the 
        incoming proton momentum for $5 < Q^2 < 20$ GeV$^2$, $50 < W < 270$ GeV
        and $0.015 < \beta \approx Q^2/(M^2_X+Q^2) < 0.5$. From ZEUS. 
     }}
\end{figure}

The dependence of the diffractive cross section $d\sigma_{\gamma^* p \to Xp}/dM_X$ on the square of the four-momentum transfer $t$ between the incoming and outgoing proton was measured by ZEUS by detecting the scattered proton in the leading proton spectrometer (LPS) and the system X in the central detector~\cite{Zeplpsf2d394}. The cross section is steeply falling with $-t$ as shown in Fig.~\ref{f:Zeplps94sigvst}; a fit of the form $d\sigma_{\gamma^* p \to Xp}/dM_X \propto exp(bt)$ yielded $b=7.2\pm 1.1(stat)^{+0.7}_{-0.9}(syst)$ GeV$^{-2}$. This shows that small momentum transfers between incoming and outgoing proton dominate, as expected for diffractive scattering.

\subsection{Diffractive structure function and cross section}

Isolation of diffractive events with the LPS is rather straightforward: detection of a proton scattered under very small angles and carrying a large fraction of the momentum of the incoming proton, $x_L = p^{LPS}/p_{p_{beam}}> 0.95$ ensures a large rapidity gap between the outgoing proton and the system $X$. However, the event rate is limited by the acceptance of the LPS. 

\begin{figure}[ht]
\centerline{\epsfig{file=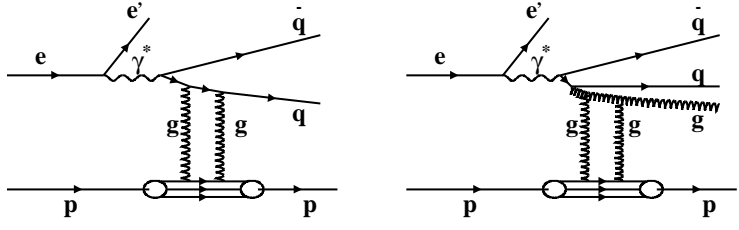,bbllx=-130pt,bblly=0pt,bburx=510pt,bbury=150pt,height=4cm,clip=}}
\caption{QCD diagrams for diffraction in DIS.}
\label{f:diagdiffqcd}
\end{figure}

In QCD, diffraction is characterized by the exchange of a colourless object, e.g. a colour singlet two-gluon system, between the incoming virtual photon and proton, producing a massive photon-like state and a nucleon, see Fig.~\ref{f:diagdiffqcd}, or a low mass excited nucleonic system. In comparison with nondiffractive scattering, the exchange of a colourless system suppresses QCD radiation and therefore also the production of additional hadrons between the massive photon-like state and the nucleon or nucleonic system. Large acceptance for diffractive events has been achieved by requiring either a large rapidity gap between the nucleonic system $N$ produced in the forward direction and the system $X$ detected in the central detector, or by using the fact that in diffractive events most of the hadronic energy is carried away by the system $N$ which escapes detection leaving behind, in the region of the central detector, a low mass system. Therefore, by measuring the distribution of the mass of the hadronic system observed in the central detector the diffractive contribution can be separated from the nondiffractive one ($M_X$ method). Analyses based on the first method were performed by H1~\cite{H1epf2d393,H1epf2d394} and based on the $M_X$ method by ZEUS~\cite{Zepdiff93,Zepdiff94}.

The results were presented in terms of the diffractive cross section 
\newline $d\sigma_{\gamma^* p \to Xp}/dM_X$ and the diffractive structure function $F^{D(3)}_2(x_{\pom},\beta,Q^2)$~\cite{Ingelmanschlein}. The cross section for the process $ep \to e X N$ is expressed in terms
of the transverse (T) and longitudinal (L) cross sections,
$\sigma_T^{di\!f\!f}$ and $\sigma_L^{di\!f\!f}$, for $\gamma^* p \to X N$ as:
\begin{eqnarray}
\frac{d\sigma^{di\!f\!f}_{\gamma^*p \to XN}(M_X,W,Q^2)}{dM_X} & 
\equiv & \frac{d(\sigma_T^{di\!f\!f}+\sigma_L^{di\!f\!f})}{dM_X} \nonumber \\
 &\approx &
\frac{2\pi}{\alpha}\,\frac{Q^2}{ (1-y)^2 +1}
 \,\frac{d\sigma^{di\!f\!f}_{ep \to eXN}(M_X,W,Q^2)}{dM_X d\ln W^2 dQ^2} \; .
\label{eq:sgp} 
\end{eqnarray}
The diffractive structure function of the proton can be related to the diffractive cross section in terms of the scaling variables $x_{\pom} \approx (M^2_X+Q^2)/(W^2+Q^2)$ and $\beta \approx Q^2/(M^2_X+Q^2)$. In models where diffraction is described by the $t$-channel exchange of a system, for example the pomeron, $x_{\pom}$ is the momentum fraction of the proton carried by the pomeron and $\beta$ is the momentum fraction of the struck quark within the pomeron. One obtains~\cite{Ingelmanprytz}:
\begin{eqnarray}
\frac{1}{2M_X}\frac{d\sigma^{di\!f\!f}_{\gamma^*p \to XN}(M_X,W,Q^2)}{dM_X} =   4\pi^2\alpha \frac{W^2}{(Q^2 +W^2)^2 Q^2} F^{D(3)}_2(\beta,\xpom,Q^2).
\label{eq:f2d3sdiff}
\end{eqnarray}
For $W^2 \gg Q^2$, Eq.~\ref{eq:f2d3sdiff} can be written as:
\begin{eqnarray}
\frac{1}{2M_X}\frac{d\sigma^{di\!f\!f}_{\gamma^*p \to XN}(M_X,W,Q^2)}{dM_X} \approx   \frac{4\pi^2\alpha}{Q^2(Q^2 +M^2_X)} \xpom F^{D(3)}_2(\beta,\xpom,Q^2).
\label{eq:f2d3sdiffapprox}
\end{eqnarray}
If $F^{D(3)}_2$ is interpreted in terms of quark densities then it specifies for a diffractive process the probability to find a quark carrying a momentum fraction $x = \beta \xpom$ of the proton momentum.

\subsection{Diffractive structure function measurement by H1}

\begin{figure}[ht]
\begin{center}
\epsfig{file=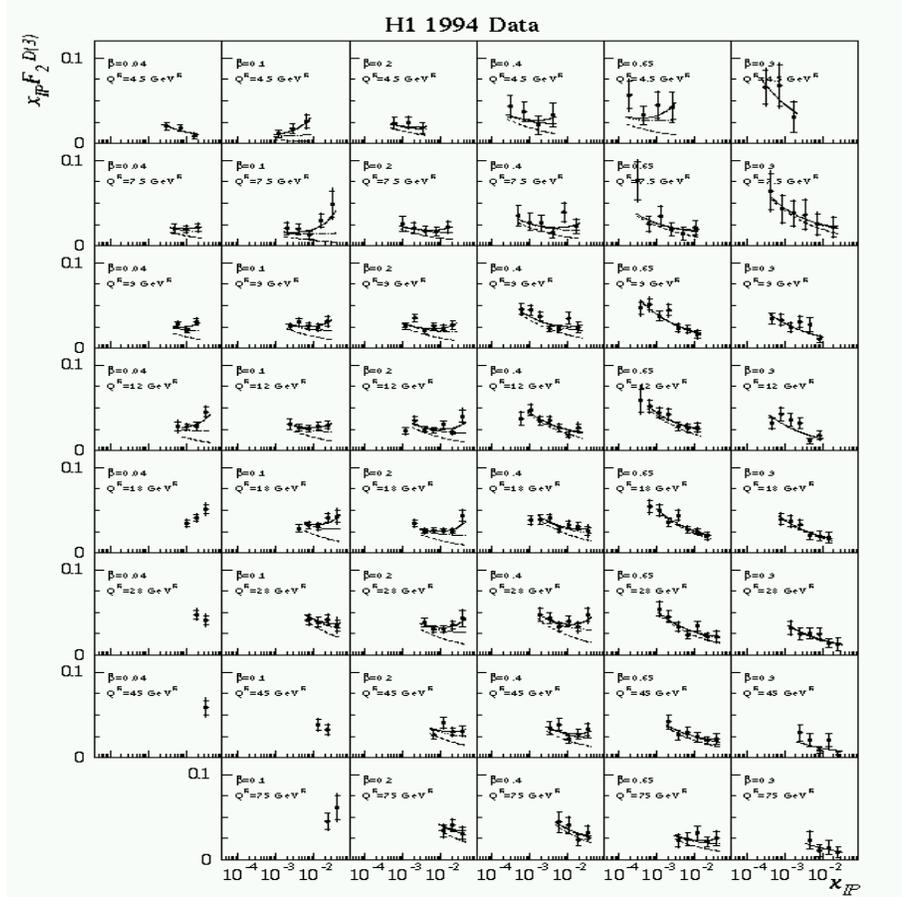,width=12cm,height=12cm}
\end{center}
\caption{\label{f:H1epf2d394fig5}
     {  The diffractive structure function $x_{\pom}F^{D(3)}_2$ as a function
        of $x_{\pom}$ for various $\beta$ and $Q^2$ values. The solid curves
        show the results from the two-component Regge fit. The dashed curves
        show the pomeron contribution alone. From H1.
     }}
\end{figure}

H1 presented their results for $\gamma^*p \to XN$ in terms of the diffractive structure function. The mass of the nucleon system $N$ was restricted to $M_N < 1.6$ GeV. In Fig.~\ref{f:H1epf2d394fig5}  $x_{\pom}F^{D(3)}_2$ is shown as a function of $x_{\pom}$ for fixed $\beta$ values and fixed $Q^2$ between 4.5 and 75 GeV$^2$. The variation of $x_{\pom}F^{D(3)}_2$ with $\beta$ and $Q^2$ is rather modest, indicating moderate scaling violations. In general, $x_{\pom}F^{D(3)}_2$ is falling in the region $x_{\pom} \le 10^{-2}$ followed sometimes by an increase at large $x_{\pom}$ values. 

The $x_{\pom}$ dependence of $F^{D(3)}_2$ is related to the $W$ dependence of the diffractive cross section and, if analyzed in a Regge approach, to the Regge trajectories exchanged in the $t$-channel. By writing $\xpom F^{D(3)}_2(\xpom,\beta,Q^2) =(C/\xpom)\cdot (x_0/\xpom)^n  F^{D(2)}_2(\beta,Q^2)$ one obtains $n=2(\overline{\alphapom}-1)$ if only the pomeron trajectory $\alphapom$ (here averaged over $t$) is contributing. Because of the rise of $x_{\pom}F^{D(3)}_2$ seen at large $x_{\pom}$ H1 concluded that in addition to the pomeron a lower lying trajectory R is also contributing. The solid curves in Fig.~\ref{f:H1epf2d394fig5} show the result of a two-component fit to the data. The dashed curves show the pomeron contribution alone as obtained from the fit. The fit yielded for the intercept of the pomeron trajectory $\alphapom(0) = 1.203 \pm 0.020(stat) \pm (0.013(syst)^{+0.030}_{-0.035}(model)$, a value which is above the results deduced from (soft) hadron-hadron scattering where $\alphapom(0) = 1.08$~\cite{Donlan84} and $1.096^{+0.012}_{-0.009}$~\cite{Cudell} was found.

\begin{figure}[ht]
\begin{center}
\epsfig{file=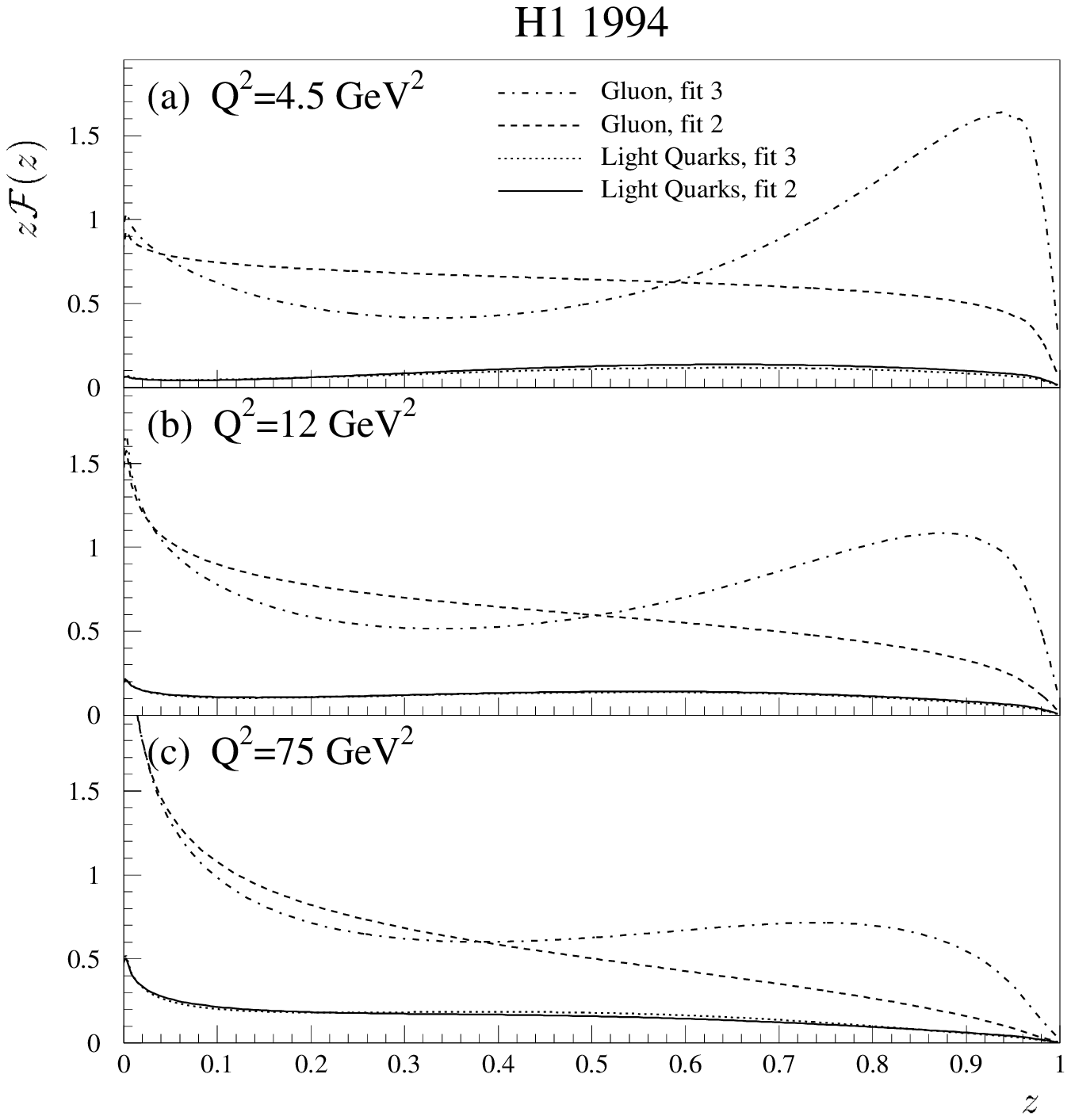,bbllx=150pt,bblly=400pt,bburx=410pt,bbury=820pt,width=6cm,height=8cm}
\end{center}
\caption{\label{f:H1epf2d394fig8}
     {  The sum of the light quark and gluon distributions as a function of
        the momentum fraction ${\it z}$ of the pomeron carried by the parton
        at different values of $Q^2$. From H1.
     }}
\end{figure}

H1 fitted their data with a QCD motivated model, in which parton distributions are assigned to the leading and subleading exchanges. Figure~\ref{f:H1epf2d394fig8} shows the resulting contributions to the parton densities of the pomeron as a function of the fraction ${\it z}$ of the pomeron momentum carried by the parton. Within this model the majority of the momentum of the pomeron is found to be carried by gluons. 

\subsection{Diffractive cross section and structure function measured by ZEUS}

\begin{figure}[ht]
\begin{center}
\epsfig{file=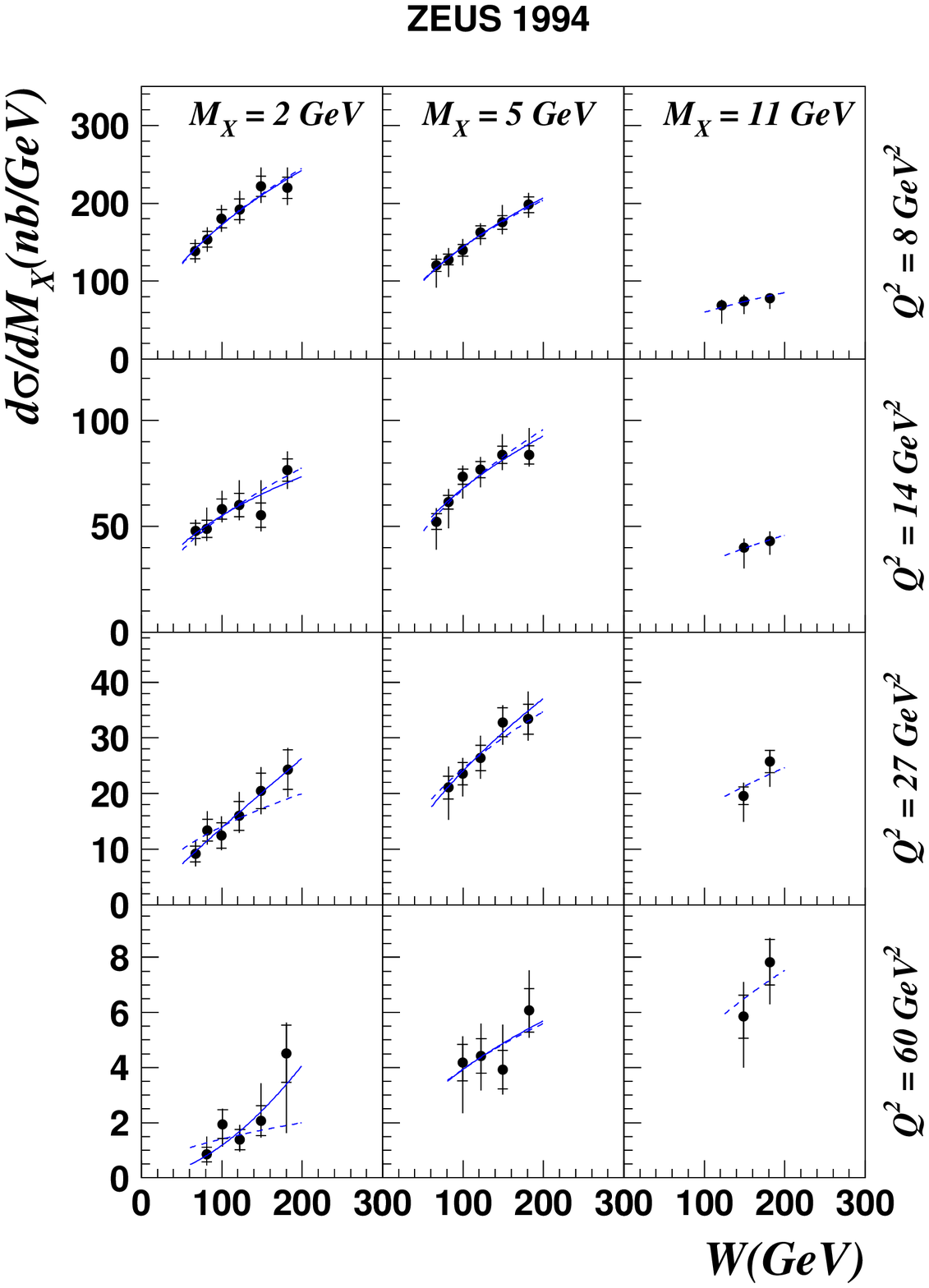,bbllx=-150pt,bblly=10pt,bburx=700pt,bbury=601pt,height=11cm,clip=}
\end{center}
\caption{\label{f:Zepdiff94fig5}
     {  The diffractive cross section $d\sigma^{di\!f\!f}
             _{\gamma^* p \to X N}/dM_X, \; M_N < 5.5$ GeV, as a function
             of $W$ for different $M_X$ and $Q^2$ values. The solid curves show             the result from fitting the diffractive 
             cross section for each ($W,Q^2$) bin separately using 
             the form  $d\sigma^{di\!f\!f}_{\gamma^* p \to X N}/dM_X 
             \propto W^{a^{diff}}$ where $a^{diff}$ and the normalization
             constants were treated as free parameters. The dashed curves show
             the result from the fit where $a^{diff}$ was assumed to be the
             same for all ($W,Q^2$) bins. From ZEUS.
     }}
\end{figure}

ZEUS determined the diffractive cross section and structure function for $\gamma^* p \to XN$ where $M_N < 5.5$ GeV~\cite{Zepdiff94}. The diffractive cross section is presented in Fig.~\ref{f:Zepdiff94fig5} as a function of $W$ for various $M_X$ and $Q^2$ values. The diffractive cross section rises rapidly with $W$ at all $Q^2$ values for $M_X$ up to 7.5 GeV. A fit to the form
\begin{eqnarray}
\frac {d\sigma^{di\!f\!f}_{\gamma^*p \to XN}(M_X,W,Q^2)}{ dM_X } & = h \cdot W^{a^{diff}} \; \; \; , 
\label{eq:gxpint}
\end{eqnarray}
where $a^{diff}$ and the normalization constants $h$ were treated as free parameters, yielded $a^{diff} = 0.507 \pm 0.034 (stat)^{+0.155}_{-0.046} (syst)$ which corresponds to a $t$-averaged $\overline{\alphapom} = 1+a^{diff}/4 = 1.127 \pm 0.009 (stat)^{+0.039}_{-0.012}(syst)$. This value is consistent with the H1 result since averaging over the $t$-distribution gives approximately $\overline{\alphapom} = \alphapom (0) - 0.03$.

\begin{figure}[ht]
\begin{center}
\epsfig{file=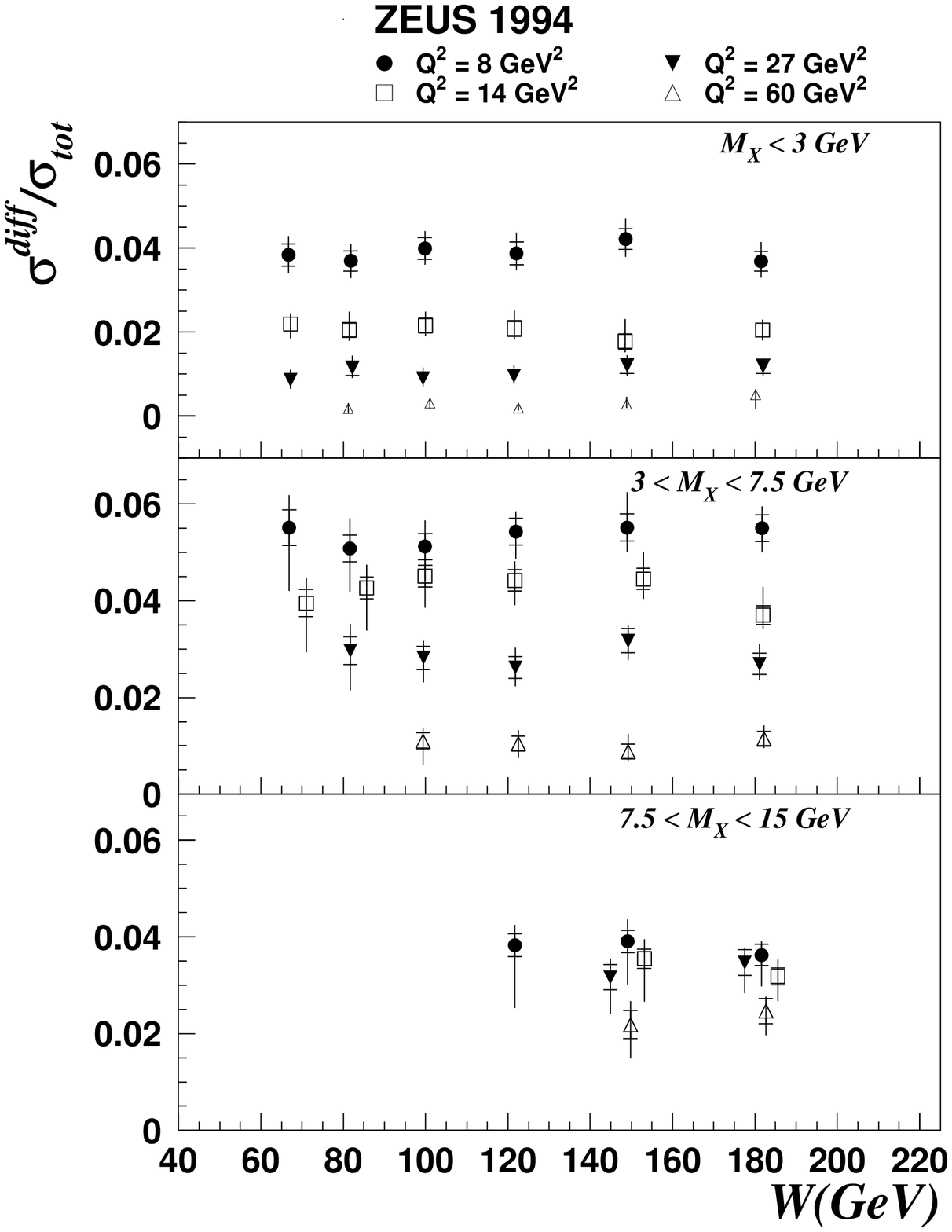,height=12cm}
\end{center}
\caption{\label{f:Zepdiff94fig8}
     {       The ratio of the diffractive cross section, 
             integrated over the $M_X$ intervals indicated, 
             $\sigma^{diff} =  
             \int^{M_b}_{M_a} dM_X\sigma^{diff}_{\gamma^* p\to XN}$,
             for $M_N < 5.5$ GeV, to the total cross section for virtual
             photon proton scattering,
             $r^{diff}_{tot} = \sigma^{diff}/\sigma_{\gamma^* p}^{tot}$,
             as a function of $W$
             for the  $M_X$ intervals and $Q^2$ values indicated. From ZEUS.
     }}
\end{figure}

The diffractive cross section was compared with the measured total virtual-photon proton cross section. The ratio of the two cross sections,
\begin{eqnarray}
r^{diff}_{tot}\;=\; \frac{{\bf\int^{M_b}_{M_a}} dM_X d\sigma^{di\!f\!f}_{\gamma^* p \to XN}/dM_X}{\sigma_{\gamma^* p}^{tot} },
\end{eqnarray}
 is displayed in Fig.~\ref{f:Zepdiff94fig8} as a function of $W$ for the different $M_X$ bins and $Q^2$ values. The data show that, for fixed $M_X$, contrary to naive expectations, the diffractive cross section possesses the same $W$ dependence as the total cross section. The rapid rise of $\sigma_{tot}$ with $W$, which is equivalent to the rapid rise of $F_2$ as $x \to 0$, in QCD is attributed to the evolution of partonic processes. The observation of similar $W$ dependences for the total and diffractive cross sections suggests, therefore, that diffraction in DIS receives sizeable contributions from hard processes. The same $W$ dependence for the diffractive and total cross sections was predicted in~\cite{Buchmuller95}.  

\begin{figure}[ht]
\begin{center}
\epsfig{file=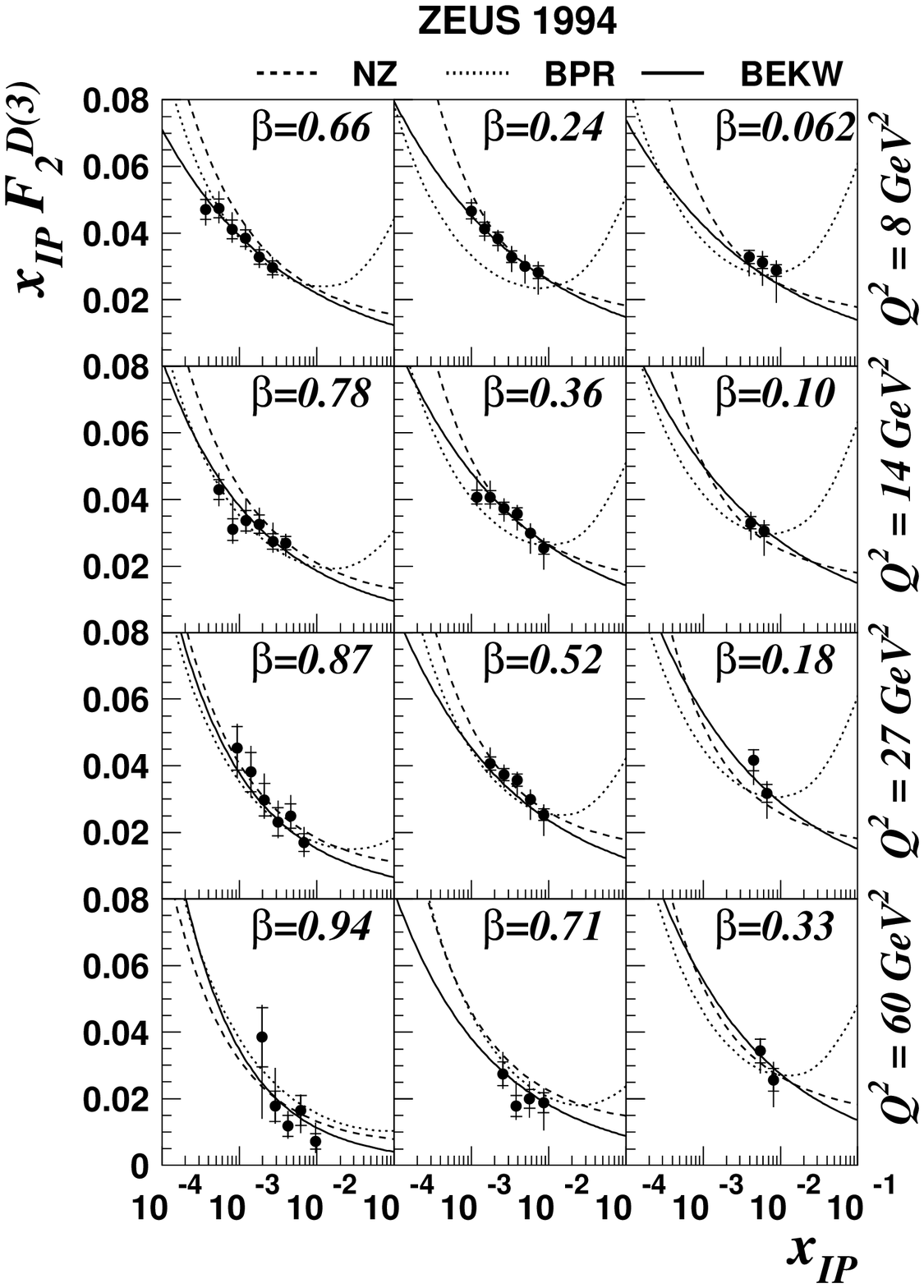,bbllx=-150pt,bblly=10pt,bburx=700pt,bbury=601pt,height=12cm,clip=}
\end{center}
\caption{\label{f:Zepdiff94fig9}
     {   The diffractive structure function of the proton multiplied
         by $\xpom$, $\xpom F^{D(3)}_2$, as a function of $\xpom$.
         The curves show the results from the models of Nikolaev and 
         Zhakarov (NZ), Bialas, Peschanski and Royon (BPR) and Bartels,
         Ellis, Kowalski and W\"{u}sthoff (BEKW).      
     }}
\end{figure}

The diffractive structure function, multiplied by $x_{\pom}$ is shown in Fig.~\ref{f:Zepdiff94fig9} as a function of $x_{\pom}$ for different values of $\beta$ and $Q^2$. $\xpom F^{D(3)}_2(\xpom,\beta,Q^2)$ decreases with increasing $\xpom$, which reflects the rapid increase of the diffractive cross section with rising $W$. The data are consistent with the assumption that the diffractive structure function $F^{D(3)}_2$ factorizes into a term depending only on $\xpom$ and a structure function $F^{D(2)}_2$ which depends on ($\beta,Q^2$). The rise of $\xpom F^{D(3)}_2$ with $\xpom$ can be described as $\xpom F^{D(3)}_2 \propto (1/\xpom)^n$ with $n = 0.253 \pm 0.017 (stat)^{+0.077}_{-0.023}(syst)$. The data are also consistent with models which break factorization. 

\begin{figure}[ht]
\begin{center}
\epsfig{file=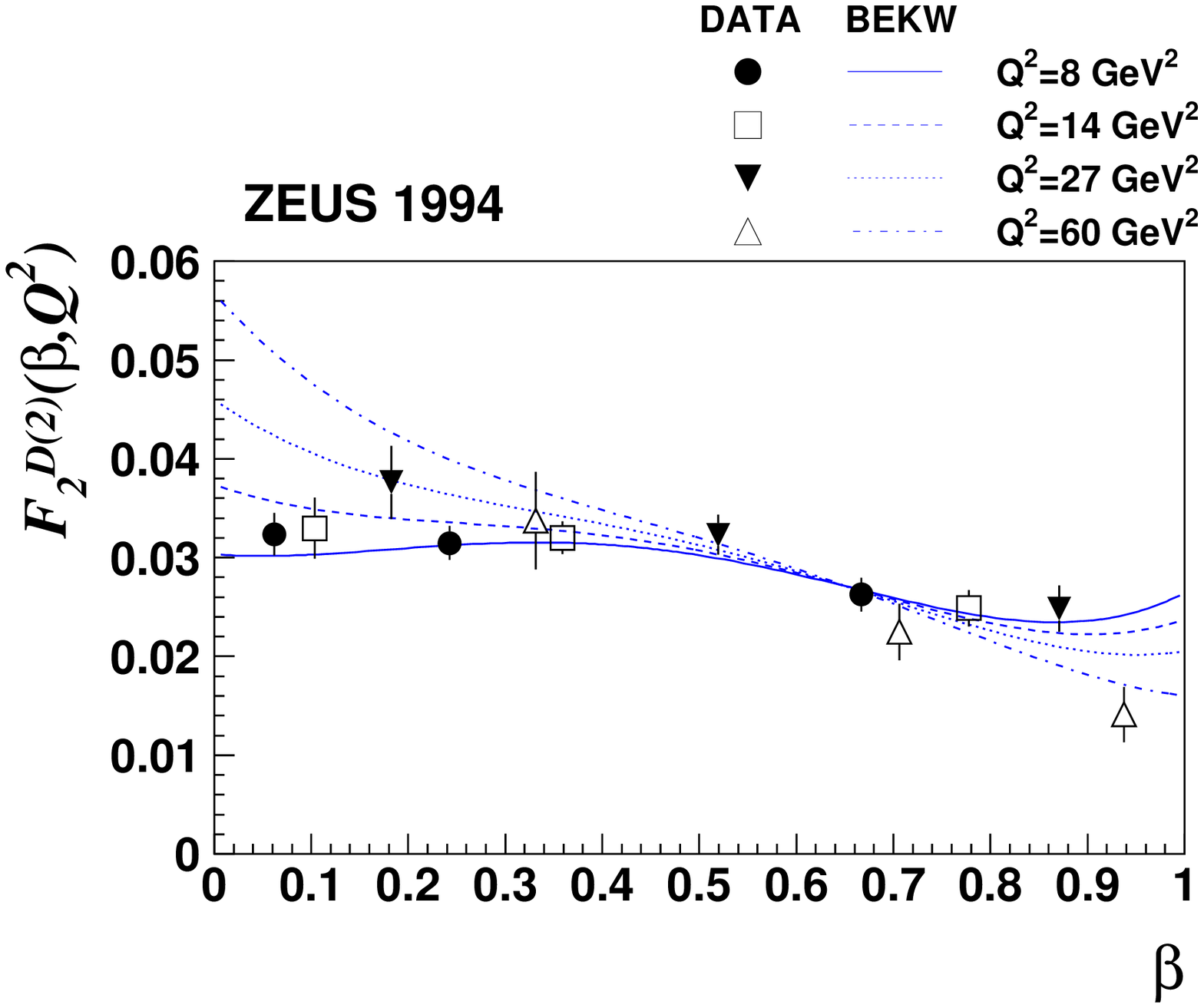,bbllx=-100pt,bblly=10pt,bburx=700pt,bbury=501pt,height=10cm,clip=}
\end{center}
\caption{\label{f:Zepdiff94fig11}
     {    The structure function $F^{D(2)}_2(\beta,Q^2)$
          for $\gamma^* p \to XN, M_N < 5.5$ GeV, for the $Q^2$
          values indicated, as a function of $\beta$ as extracted from
          a fit to the measured $\xpom F^{D(3)}_2$ values.
          The curves show the fit results obtained with the BEKW model.
          From ZEUS.
     }}
\end{figure}

Figure~\ref{f:Zepdiff94fig11} shows $F^{D(2)}_2(\beta,Q^2) = x_0 F^{D(3)}_2(x_0,\beta,Q^2)$ where $F^{D(3)}_2$ was evaluated at $x_{\pom} = x_0 = 0.0042$. The data show that $F^{D(2)}_2$ has a simple behaviour. For $\beta < 0.6$ and $Q^2 < 14$ GeV$^2$, $F^{D(2)}_2$ is approximately independent of $\beta$.  For $\beta < 0.8$ also the data from different $Q^2$ values are rather similar suggesting a leading twist behaviour characterized by a slow $\ln Q^2$ type rescaling. For $\beta > 0.9$ the data show a decrease with $\beta$ or $Q^2$. The approximate constancy of $F^{D(2)}_2$ for $\beta < 0.9$ combined with the rapid rise of $F^{D(3)}_2$ as $\xpom$ decreases can be interpreted as evidence for a substantial partonic component in DIS diffraction dissociation.

The $Q^2$ behaviour of $\xpom F^{D(3)}_2(\xpom,\beta,Q^2)$ is different from that of the proton structure function $F_2(x,Q^2)$, taken at $x = \xpom$, which rises gradually with $Q^2$. It is in broad agreement with the conjecture~\cite{Buchmuller95} that \newline
 $\xpom F^{D(3)}_2(\xpom,\beta,Q^2) \propto F_2(x=\xpom,Q^2)/\log_{10}(Q^2 / Q^2_0)$ where $Q^2_0 = 0.55$ GeV$^2$. 

\subsection{Comparison with partonic models}
The data were compared with several partonic models of diffraction, NZ~\cite{NZ}, BPR~\cite{BPR} and BEKW~\cite{BEKW}. Good agreement with the data can be achieved. The models provide a first glimpse of how the different components may build up the diffractive structure function.

In the BEKW model basically three components build up the diffractive structure function, $\xpom F^{D(3)}_2(\beta,\xpom,Q^2) = c_T \cdot F^T_{q\overline{q}} + c_L \cdot F^L_{q\overline{q}} + c_g \cdot F^T_{q\overline{q}g}$. The three terms represent the contributions from transverse photons fluctuating into a $q\overline{q}$ or a $q\overline{q}g$ system and from longitudinal photons fluctuating into a $q\overline{q}$ system, and have the following functional dependence on $\xpom$, $\beta$ and $Q^2$:
\begin{eqnarray}
F^T_{q\overline{q}} = (\frac{x_0}{\xpom})^{n_T(Q^2)} \cdot \beta (1-\beta) \\
F^L_{q\overline{q}} = (\frac{x_0}{\xpom})^{n_L(Q^2)} \cdot \frac{Q^2_0}{Q^2} \cdot [\ln(\frac{7}{4}+\frac{Q^2}{4\beta Q^2_0})]^2 \cdot \beta^3 (1-2\beta)^2 \\
F^T_{q\overline{q} g} = (\frac{x_0}{\xpom})^{n_T(Q^2)} \cdot \ln(1+\frac{Q^2}{Q^2_0}) \cdot (1-\beta)^3 \\
n_{T,L}(Q^2) = 0.1 + n^0_{T,L} \cdot \ln[1+ \ln(\frac{Q^2}{Q^2_0})]
\end{eqnarray}

In the model $x_0, Q^2_0, n^0_{T,L}$ are free parameters which have to be determined from the data. The three terms $F^T_{q\overline{q}},\; F^L_{q\overline{q}},\; F^T_{q\overline{q} g}$ behave differently as a function of $Q^2$. Except for a possible $Q^2$ dependence of the power $n_T$, $F^T_{q\overline{q}}$ does not depend on $Q^2$. The term $F^L_{q\overline{q}}$ is of higher twist but the power $1/Q^2$ is softened by a logarithmic $Q^2$ factor; $F^T_{q\overline{q}g}$ grows logarithmically with $Q^2$.

\begin{figure}[ht]
\begin{center}
\epsfig{file=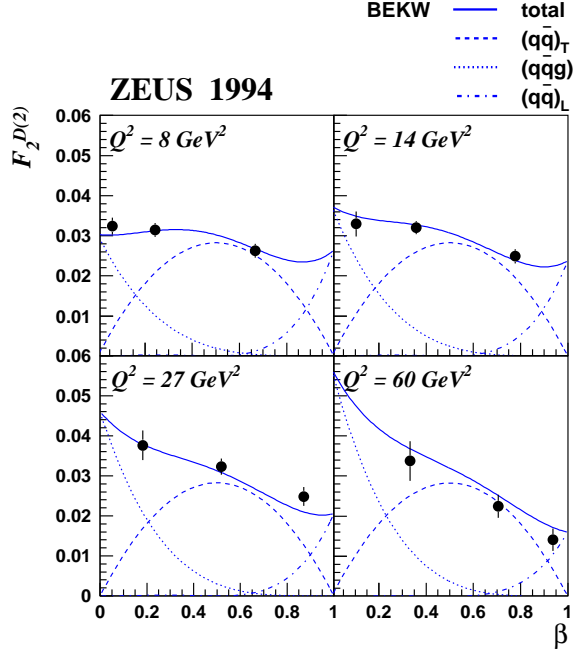,bbllx=-150pt,bblly=10pt,bburx=700pt,bbury=500pt,height=10cm,clip=}
\end{center}
\vspace{-0.6cm}
\caption{\label{f:Zepdiff94fig14a}
        {The three components $(q\overline{q})_T$, $(q\overline{q}g)$
         and $(q\overline{q})_L$ of the BEKW model building 
         up the diffractive structure function of the proton and their sum
         $F^{D(2)}_2(\beta,Q^2)$ 
         as a function of $\beta$ for $Q^2 = 8, 14, 27$ and $60$ GeV$^2$, 
         as obtained from a fit of the model to the data. From ZEUS.
         }}
\end{figure}

\begin{figure}[ht]
\begin{center}
\epsfig{file=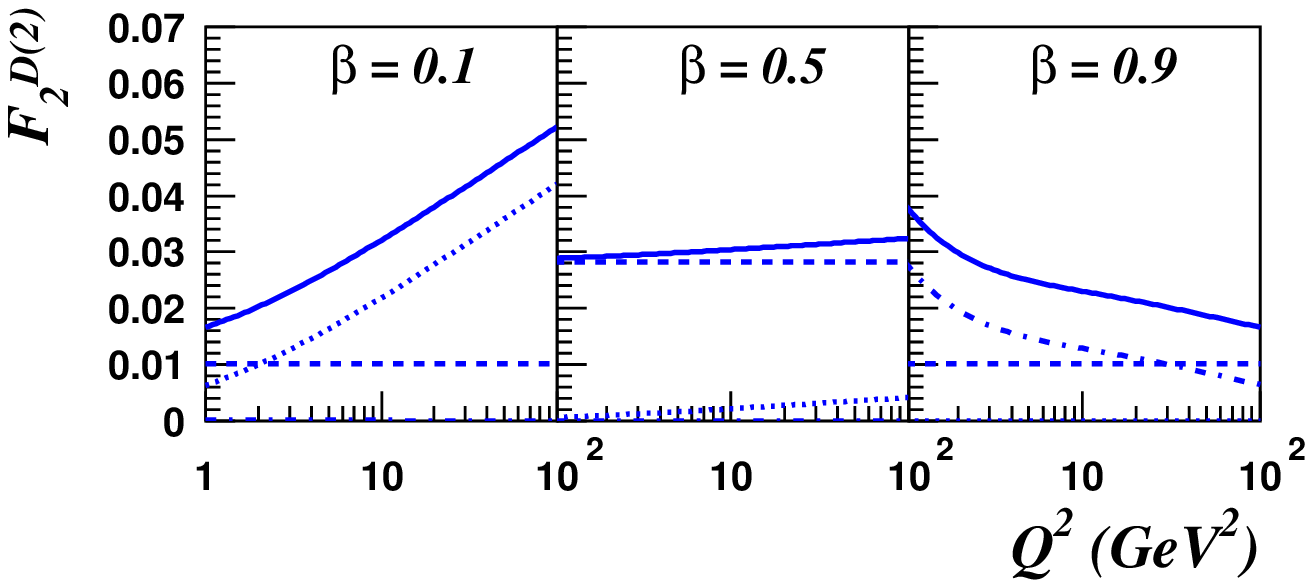,bbllx=-30pt,bblly=-20pt,bburx=700pt,bbury=320pt,height=10cm,clip=}
\end{center}
\vspace{-0.6cm}
\caption{\label{f:Zepdiff94fig14b}
        {The three components $(q\overline{q})_T$, $(q\overline{q}g)$
         and $(q\overline{q})_L$ of the BEKW model building 
         up the diffractive structure function of the proton and their sum
         $F^{D(2)}_2(\beta,Q^2)$ as a function of $Q^2$ for
         $\beta = 0.1, 0.5$ and $0.9$ as obtained from a fit of the model
         to the data. The notation is the same as in the 
         previous figure. From ZEUS.
         }}
\end{figure}

It is instructive to compare the $\beta$ and $Q^2$ dependences of the three components. Figure~\ref{f:Zepdiff94fig14a} shows $c_T F^T_{q\overline{q}}$ (dashed), $c_L F^L_{q\overline{q}}$ (dashed-dotted), $c_g F^T_{q\overline{q}g}$ (dotted) and their sum $\xpom F^{D(3)}_2(\xpom,\beta, Q^2)$ at $\xpom = x_0$ (solid curves) as a function of $\beta$ for $Q^2 = 8, 14, 27, 60$ GeV$^2$. For $\beta > 0.2$ the colourless system couples predominantly to the quarks in the virtual photon. The region $\beta \ge 0.8$ is dominated by the contributions from longitudinal photons. The contribution from coupling of the colourless system to a $q \overline{q} g$ final state becomes important for $\beta < 0.3$. The last result is in contrast to the H1 observation (see above) that the large $\beta$ region is dominated by the gluon contribution. Figure~\ref{f:Zepdiff94fig14b} shows the same quantities as a function of $Q^2$ for $\beta = 0.1, 0.5, 0.9$. The gluon term, which dominates at $\beta = 0.1$ rises with $Q^2$ while the quark term, which is important at $\beta = 0.5$ shows no evolution with $Q^2$, i.e. is of leading twist. The contribution from longitudinal photons, which is higher twist and dominates at $\beta = 0.9$, decreases with $Q^2$.

In the BEKW model the $\xpom$-dependence of the quark and gluon contributions for transverse photons is expected to be dominated by the aligned jet configuration~\cite{Bjorken} and, therefore, to be close to that given by the soft pomeron. Writing $F^T_{q\overline{q}} \propto (x_0/x_{\pom})^{n_T}$ this implies $n_T \approx 2(\overline{\alphapom}^{soft} - 1)$. However, perturbative admixtures in the diffractive final state are expected to have a somewhat stronger energy dependence, leading to an effective  $n_T > 2(\overline{\alphapom}^{soft} - 1)$. The $\xpom$ dependence of the longitudinal contribution is driven by the square of the proton's gluon momentum density leading to $n_L > n_T$. The fit of the BEKW model to the data indicates that transverse (longitudinal) photons dominate the region $\beta < 0.8$ ($\beta > 0.8$).

In ~\cite{GolecWuest} a novel model (GBW) has been developed which links inclusive diffraction with the total cross section. In a frame where photon and proton are collinear, the total $\gamma^*p$ and the diffractive cross sections can be written as~\cite{Nikolaev90,Forshaw97}:
\begin{eqnarray}
\sigma_{T,L}(x,Q^2) = \int d^2{\bf r} \int dz \vert\Psi_{T,L}(z,{\bf r})\vert^2 \hat{\sigma }(x,r^2) \\
\frac{d\sigma^{diff}_{T,L}(t=0)}{dt} = \frac{1}{16\pi}\int d^2{\bf r}\int dz \vert\Psi_{T,L}(z,{\bf r})\vert^2 \vert \hat{\sigma}(x,r^2) \vert^2
\end{eqnarray}
where $\Psi_{T,L}(z,{\bf r})$ denotes the wave function for transverse ($T$) and longitudinal ($L$) photons, $\hat{\sigma}(x,r^2)$ the dipole cross section for the $q \overline{q}$ pair with the proton, $z$ the momentum fraction of the photon carried by the quark and $r$ the relative transverse separation between the quarks.

\begin{figure}[ht]
\begin{center}
\epsfig{file=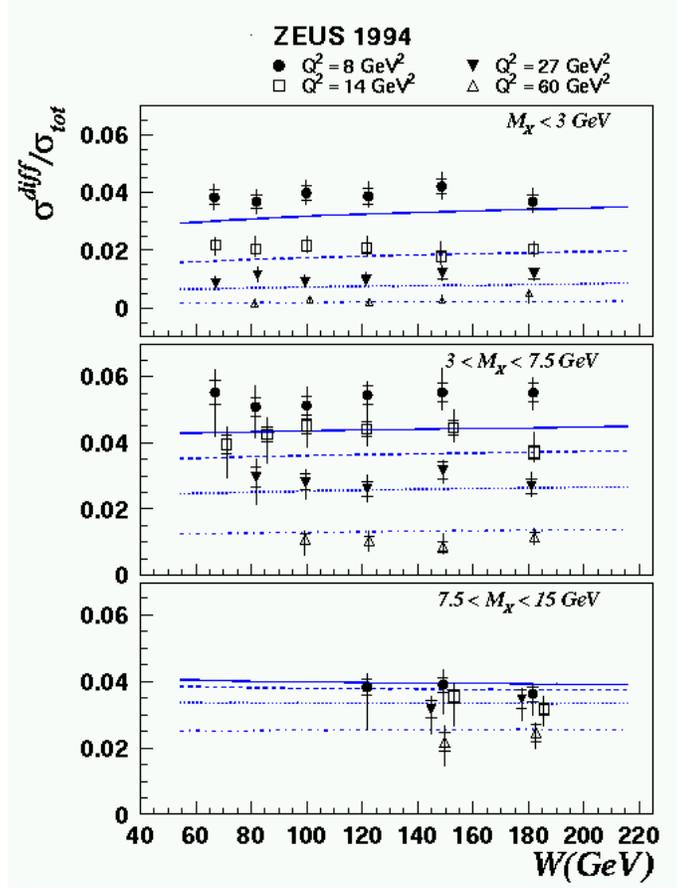,bbllx=-120pt,bblly=-20pt,bburx=650pt,bbury=660pt,height=12cm}
\end{center}
\vspace{-0.6cm}
\caption{\label{f:ZrdifftotGBW}
        {Comparison of the predictions by the GBW model with the ZEUS data for the ratio of the diffractive to total cross section as a function of $W$, for the $M_X$ intervals and $Q^2$ values indicated. Figure taken from \protect\cite{Kowalski99}. 
         }}
\end{figure}

The wave functions are determined by the photon-$q \overline{q}$ coupling and are known in QCD~\cite{Forshaw97}. The novelty of the model is an ansatz for the dipole cross section whose free parameters are determined by a comparison with the data for $F_2 \approx Q^2\sigma^{tot}_{\gamma^* p}$. Given $\hat{\sigma}(x,r^2)$, an absolute prediction can now be made for the diffractive cross section. These predictions are found to give an almost quantitative representation of the ratio of diffractive cross section to total cross section as shown by Fig.~\ref{f:ZrdifftotGBW}. This is a further step towards a quantitative description of DIS diffraction within QCD.

An interesting discussion of DIS diffraction in terms of QCD radiation and the connection with the problem of confinement has recently been presented in~\cite{BartelsKow00}.

\section{Weak Interactions become strong: NC and CC Scattering at High $Q^2$}
In the standard model, electron proton scattering at low and medium $Q^2$ ($Q^2 \le $ 1000 GeV$^2$) proceeds almost exclusively through photon exchange. At higher values of $Q^2$ substantial contributions are expected also from the exchange of the heavy vector bosons $W^{\pm}$ and $Z^{o}$. The interference between photon and $Z^0$ exchange contributes to $e^-p$ and $e^+p$ scattering with opposite sign (see $xF_3$ component in the expression for the cross section below) which allows for a direct detection of the weak contribution in NC scattering.

The structure functions can be expressed as sums over quark flavors of the proton's quark densities $q(x,Q^2)$ weighted according to the gauge structure of the scattering amplitudes~\cite{IngelmanRueckl}, see also~\cite{Devenish}. 

\subsection{NC cross section}
For the neutral current (NC) reaction, $e^{\pm}p \to e^{\pm} X$, mediated by $\gamma$ and $Z^o$ exchange, the electroweak Born--level NC DIS differential cross-section can be written as
\begin{equation}
  \frac{d^2\sigma\ncp\born(e^{\pm}p)}{dxd\qq} = \frac{2\pi \alpha ^2}
    {x\qf}  \left[Y_+\,F_2\ncp(x,\qq)
    {\mp}  Y_-\,xF_3\ncp(x,\qq)  -y^2\,F_L\ncp(x,\qq)\right] ,
    \label{eq:born} 
\end{equation}
where $Y_{\pm} = 1 \pm (1-y)^2$. The structure functions $F_2\ncp$ and $xF_3\ncp$ for longitudinally unpolarized beams may be described in leading order QCD 
as sums over the quark flavor $f=u,...,b$ of the product of electroweak
quark couplings and quark momentum distributions in the proton
\begin{eqnarray}
  F_2\ncp & = & \; \frac{1}{2}  \sum\limits_{f}\, {xq^+_f} \left[(V_f^L)^2+(V_f^R)^2+(A_f^L)^2+(A_f^R)^2 \right],\nonumber \\
  xF_3\ncp & = & \sum\limits_{f}\, {xq^-_f} [V_f^LA_f^L-V_f^RA_f^R]
\end{eqnarray}
where $xq^\pm_f=x q_f(x,\qq)\pm x {\bar q}_f(x,\qq)$ and
$x q_f$ ($x {\bar q}_f$) are the quark (anti-quark) momentum
distributions. In leading order QCD, we have $F_L\ncp = 0$.
The functions $V_f$ and $A_f$ can be written as
\begin{eqnarray}
  V_{f}^{L,R} & = &  \; e_{f}  -(v_{e} \pm a_{e})\,v_{f} \nonumber \\
  A_{f}^{L,R} & = & -(v_{e} \pm a_{e})\,a_{f}\chiz(\qq),
\end{eqnarray}
where the weak couplings, $a_i=T^3_i$ and $v_i=T^3_i - 2e_i\sthw$, are
functions of the weak isospin, $T^3_i = \frac{1}{2}$ ($-\frac{1}{2}$) for
$u,\nu$ ($d,e$), and the weak mixing angle, \thw; 
$e_i$ is the electric charge in units of the positron charge; and
\chiz\ is proportional to the ratio of $Z^0$-boson 
and photon propagators
\begin{equation}
\chiz =\frac{1}{4 \sthw \cthw}\frac{\qq}{\qq + \Mzq} .
\label{eq:chi} 
\end{equation}

The contribution of $F_L\ncp$ to \sigxqq\ is 
predicted to be approximately $1.5\%$
averaged over the kinematic range considered below.  However, 
in the region of small $x$ at the lower end of the \qq\ range of the data shown below the $F_L\ncp$ contribution to the cross-sections can be as large as $12\%$.

\subsection{CC cross section}
\begin{figure}[ht]
\centerline{\epsfig{file=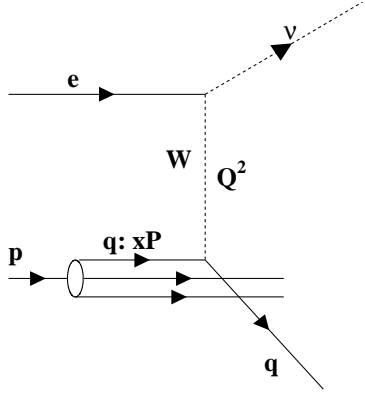,bbllx=0pt,bblly=45pt,bburx=510pt,bbury=480pt,height=6cm,clip=}}
\caption{Diagram for deep inelastic $ep$ scattering by charged current exchange.}
\label{f:diagdiscc}
\end{figure}

   The electroweak Born cross section for the charged current reactions (see Fig.~\ref{f:diagdiscc}): 
\begin{equation}
e^- p \rightarrow \nu_e X \; \; \; {\rm and} \; \; \;  
e^+ p \rightarrow \bar{\nu}_e X
\end{equation}
can be written as
\begin{equation}
  {{d^2\sigma^{\rm CC}_{\rm Born}(e^{\pm}p)} \over {dx \, dQ^2}}  = 
   {G^2_F \over 4 \pi x } \Biggl( {M^2_W \over M^2_W + Q^2}\Biggr)^2
  \Biggl[
Y_+ { F}^{\rm CC}_2(x, Q^2) \mp Y_- x{ F}^{\rm CC}_3(x, Q^2) - y^2{ F}^{\rm CC}_{ L}(x, Q^2)
  \Biggr],
  \label{e:Born}
\end{equation}
where $G_F$ is the Fermi constant and $M_W$ is the mass of the $W$ boson.

The structure functions ${ F}^{\rm CC}_2$ and 
$x{ F}^{\rm CC}_3$ , in leading-order (LO) QCD, measure
sums and differences of quark and antiquark parton momentum
distributions.  For longitudinally unpolarized
beams,
$e^-p$:
\begin{eqnarray}
{ F}^{\rm CC}_2 =  x[u(x,Q^2)+c(x,Q^2)+ \bar{d}(x,Q^2) + \bar{s}(x,Q^2)] \\
x{ F}^{\rm CC}_3 = x[u(x,Q^2)+c(x,Q^2)- \bar{d}(x,Q^2) - \bar{s}(x,Q^2)] 
\end{eqnarray}
$e^+p$:
\begin{eqnarray}
{ F}^{\rm CC}_2 =  x[d(x,Q^2)+s(x,Q^2)+ \bar{u}(x,Q^2) + \bar{c}(x,Q^2)] \\
x{ F}^{\rm CC}_3 = x[d(x,Q^2)+s(x,Q^2)- \bar{u}(x,Q^2) - \bar{c}(x,Q^2)] 
\label{e:F3}
\end{eqnarray}
where $u(x,Q^2)$ is, for example, the number density
of an up quark with momentum fraction $x$ in the proton.
Since the top quark mass is large and the off-diagonal elements of
the CKM matrix are small, the contribution from the third generation
quarks to the structure functions may be safely ignored~\cite{Katz}.
The chirality of the CC interaction is reflected by the factors $Y_\pm$
multiplying the structure functions.
The longitudinal structure function, $F^{\rm CC}_{L}$,
is zero at leading order but is finite at next-to-leading-order (NLO) QCD.
It gives a negligible contribution to the cross section except at
$y$ values close to 1, where it can be as large as 10\%.

\subsection{Experimental results: NC scattering}

Both the H1~\cite{HepCC93sig} and ZEUS~\cite{ZepNCCC93sig} experiments have previously reported cross section measurements, based on data collected in 1993, which established that the $Q^2$ dependence of the CC cross section is consistent with the expectations from the $W$ propagator. The data from ZEUS demonstrated also that the CC and NC cross sections are of similar magnitude for $Q^2 \ge M^2_W$. 

\begin{figure}[ht]
\begin{center}
\epsfig{file=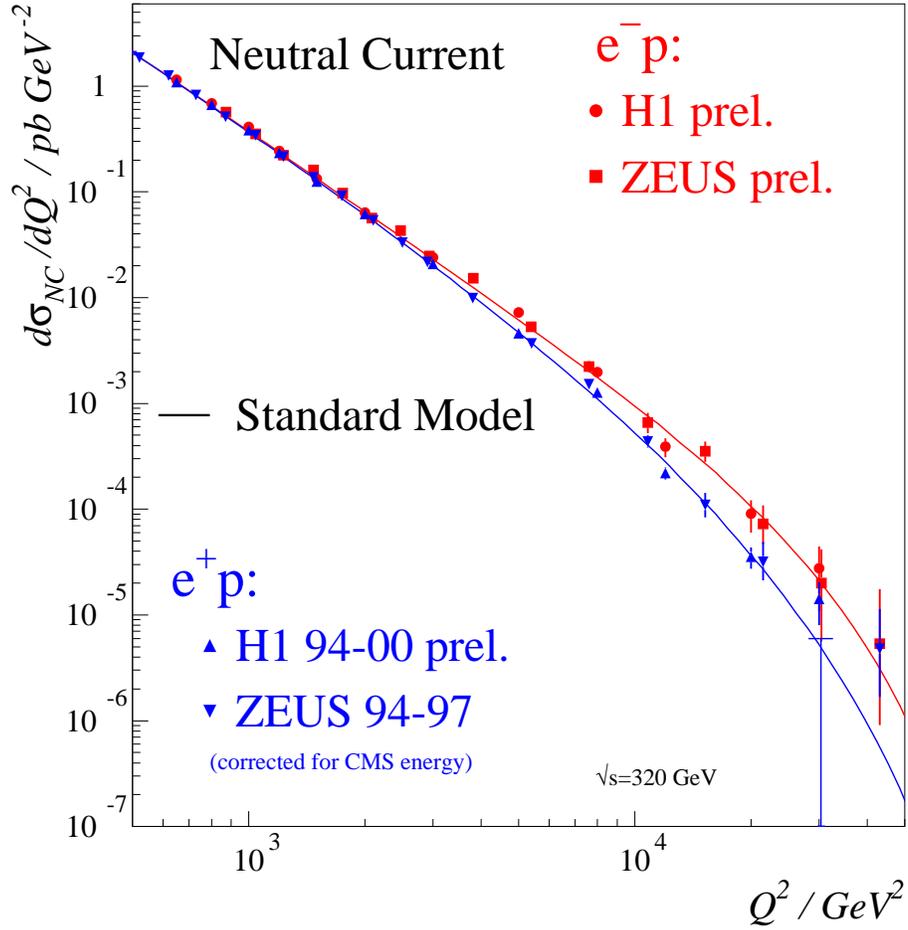,bbllx=50pt,bblly=150pt,bburx=510pt,bbury=680pt,height=13cm}
\end{center}
\caption{\label{f:HZncvsq40}
      { The NC cross sections for $e^-p$ and $e^+p$ as a function of $Q^2$ 
        as measured by H1 and ZEUS. The curves show QCD-NLO predictions.
      }}
\end{figure}

The high precision NC data presented recently by H1~\cite{H1NCCC89,HNCe+47} ($e^-p$ data from 1998-9 with 15 pb$^{-1}$; $e^+ p$ data from 1994-2000 with 46 pb$^{-1}$) and ZEUS~\cite{ZNCe+47,ZNCe-89} ($e^-p$ data from 1998-9 with 16 pb$^{-1}$; $e^+ p$ data from 1996 with 30 pb$^{-1}$) allow, for the first time, to see the $Z^0$ contribution to the NC cross section. In Fig.~\ref{f:HZncvsq40}  the $e^-p$ and $e^+p$ cross sections  are shown in terms of $d\sigma/dQ^2$ as a function of $Q^2$. Both cross sections decrease by about six orders of magnitude between $Q^2$ = 500 and 40000 GeV$^2$, mainly governed by the photon propagator which leads to a behaviour of the form $d\sigma/dQ^2 \propto 1/Q^4$. For $Q^2$ values above 3000 GeV$^2$ there is a clear difference between the two charge states: the cross section for $e^-p$ scattering is larger than for $e^+p$, which demonstrates the presence of a weak contribution. 

\begin{figure}[ht]
\begin{center}
\epsfig{file=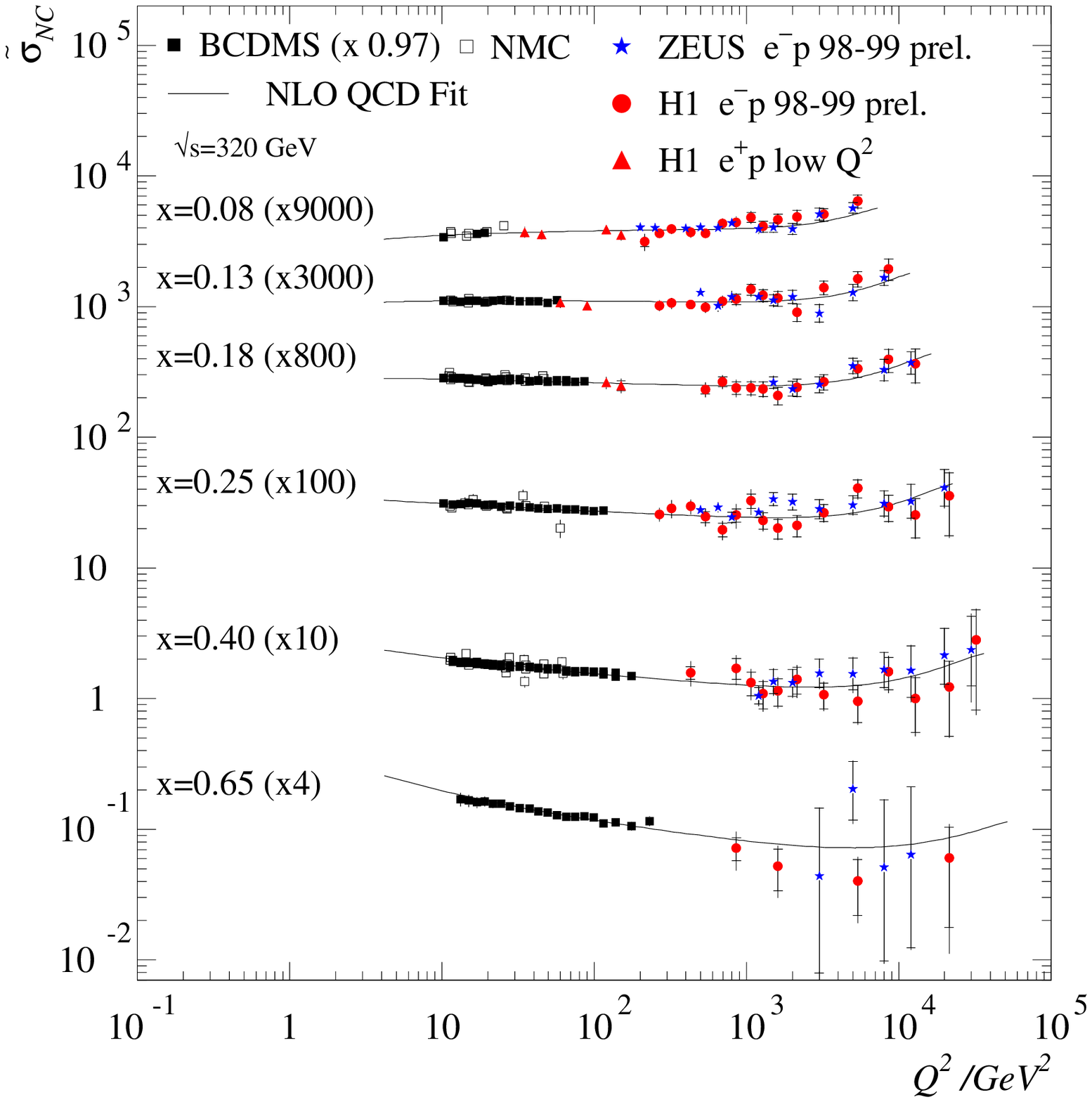,bbllx=50pt,bblly=150pt,bburx=510pt,bbury=701pt,width=11cm,height=12cm}
\end{center}
\caption{\label{f:HZe-ncrvsq89}
      { The $e^-p$ NC reduced cross section for different values of $x$ 
        as a function of $Q^2$ 
        as measured by H1 and ZEUS. The curves show QCD-NLO fits
        for $\gamma + Z^0$ exchange.
      }}
\end{figure}

\begin{figure}[ht]
\begin{center}
\epsfig{file=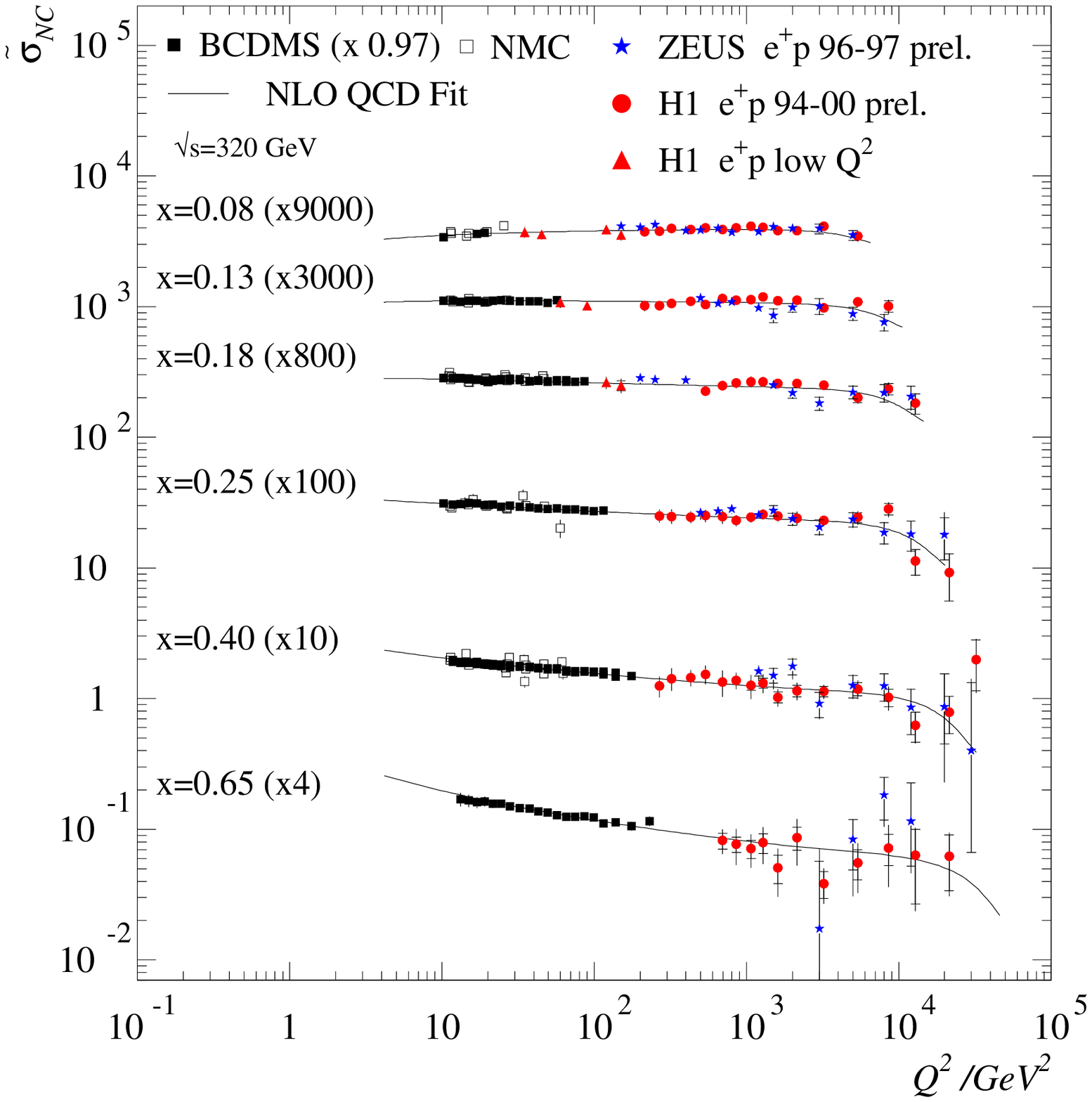,bbllx=50pt,bblly=150pt,bburx=510pt,bbury=701pt,height=12cm,clip=}
\end{center}
\caption{\label{f:HZe+ncrvsq40}
      { The $e^+p$ NC reduced cross section for different values of $x$ 
        as a function of $Q^2$ 
        as measured by H1 and ZEUS. The curves show QCD-NLO fits
        for $\gamma + Z^0$ exchange.
      }}
\end{figure}

The predictions of the Standard Model (solid curves) give a good description of the data. This is also true when the model is compared with the data for different values of $x$. This is shown in Figs.~\ref{f:HZe-ncrvsq89},~\ref{f:HZe+ncrvsq40} where the reduced cross sections,

\begin{eqnarray}
\tilde{\sigma}^{\mp}_{NC} = \frac{xQ^4}{2\pi\alpha^2}\frac{1}{Y_+}\frac{d^2\sigma_{NC}}{dxdQ^2} = F^{NC}_2 \pm \frac{Y_-}{Y_+}xF^{NC}_3
\end{eqnarray}

are given for fixed $x$ as a function of $Q^2$.

\begin{figure}[ht]
\begin{center}
\epsfig{file=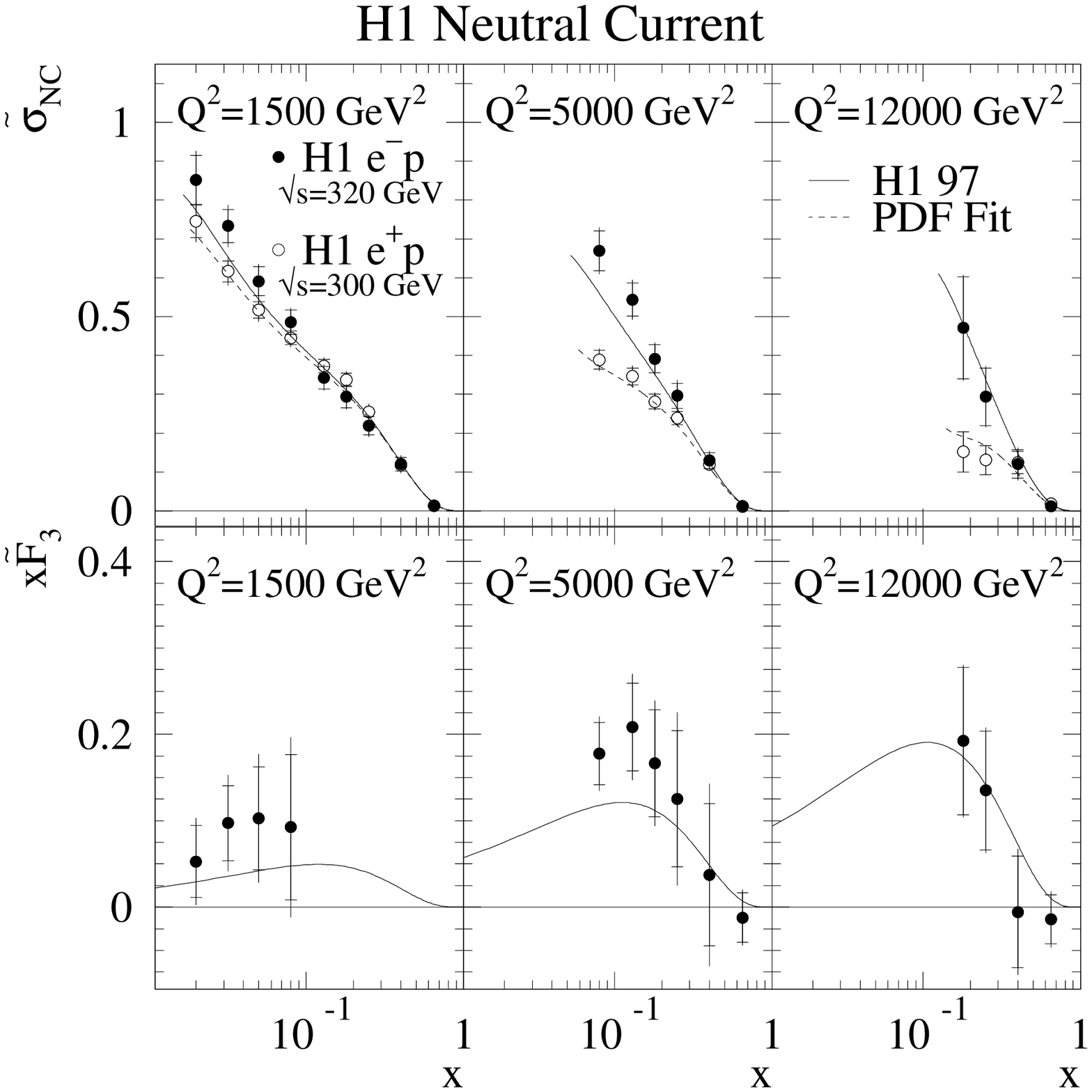,bbllx=50pt,bblly=0pt,bburx=510pt,bbury=601pt,height=10cm}
\end{center}
\caption{\label{f:Hncxf3vsxab}
      {  The $e^+p$ NC reduced cross section and $xF_3$ for different values 
        of $Q^2$ as a function of $x$ as measured by H1. 
        The curves show the predictions from the H1 97 PDF fit.
      }}
\end{figure}

From the comparison of the $e^-p$ and $e^+p$ data the structure function $xF^{NC}_3$
can be extracted. The dominant contribution comes from $\gamma \; Z^0$ interference, which is denoted by $xF^{\gamma Z}_3$. The reduced cross sections for the two charged states and the structure function $xF^{NC}_3$ as measured by H1~\cite{H1NCCC89} are presented in Figs.~\ref{f:Hncxf3vsxab}(a,b) for different $Q^2$ intervals. 

\begin{figure}[ht]
\begin{center}
\epsfig{file=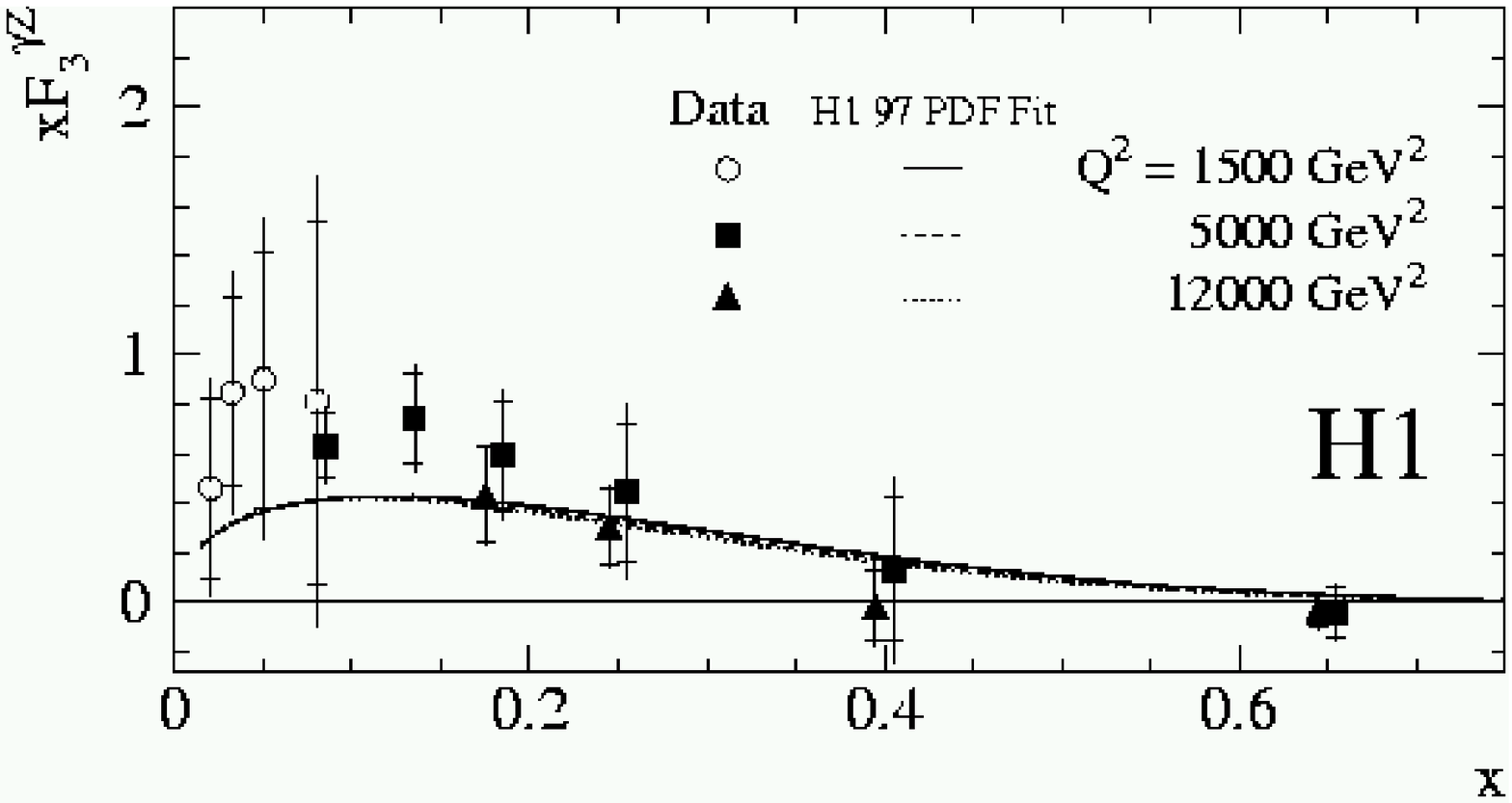,height=5cm}
\end{center}
\caption{\label{f:Hncxf3vsxc}
      { The structure function $xF^{\gamma Z}_3$ for NC scattering 
        as a function of $x$
        for different values of $Q^2$ (from H1). 
        The curves show the predictions from the H1 97 PDF fit.
      }}
\end{figure}

In Fig.~\ref{f:Hncxf3vsxc} $xF^{\gamma Z}_3$ is shown as a function of $x$ for $Q^2$ values of 1500, 5000 and 12000 GeV$^2$. It is remarkable that only little dependence on $Q^2$ is observed despite the large $Q^2$ range. This is in agreement with QCD where a dependence on $Q^2$ is expected only from scaling violations. Note also that $xF^{\gamma Z}_3$ depends on the difference between quark and antiquark densities and is therefore primarily sensitive to the valence quark contribution. The data are compared with the results of a QCD fit performed previously by H1~\cite{HNCe+47} to their NC $e^+p$ data taken in 1994-7 combined with the data from NMC and BCDMS (called H1 97 PDF fit). The fit certainly reproduces the qualitative features of the data.

\subsection{Experimental results: CC scattering}
The recent evaluations of CC cross sections by H1 and ZEUS are based on an order of magnitude more data compared to the previous analyses, H1~\cite{H1NCCC89} ($e^-p$ data from 1998-9 with 16 pb$^{-1}$; $e^+ p$ data from 1994-7 with 36 pb$^{-1}$) and ZEUS~\cite{ZCC99,ZCC00} ($e^-p$ data from 1998-9 with 16 pb$^{-1}$; $e^+ p$ data from 1994-7 with 48 pb$^{-1}$).

\begin{figure}[hpbt]
\begin{center}
\epsfig{file=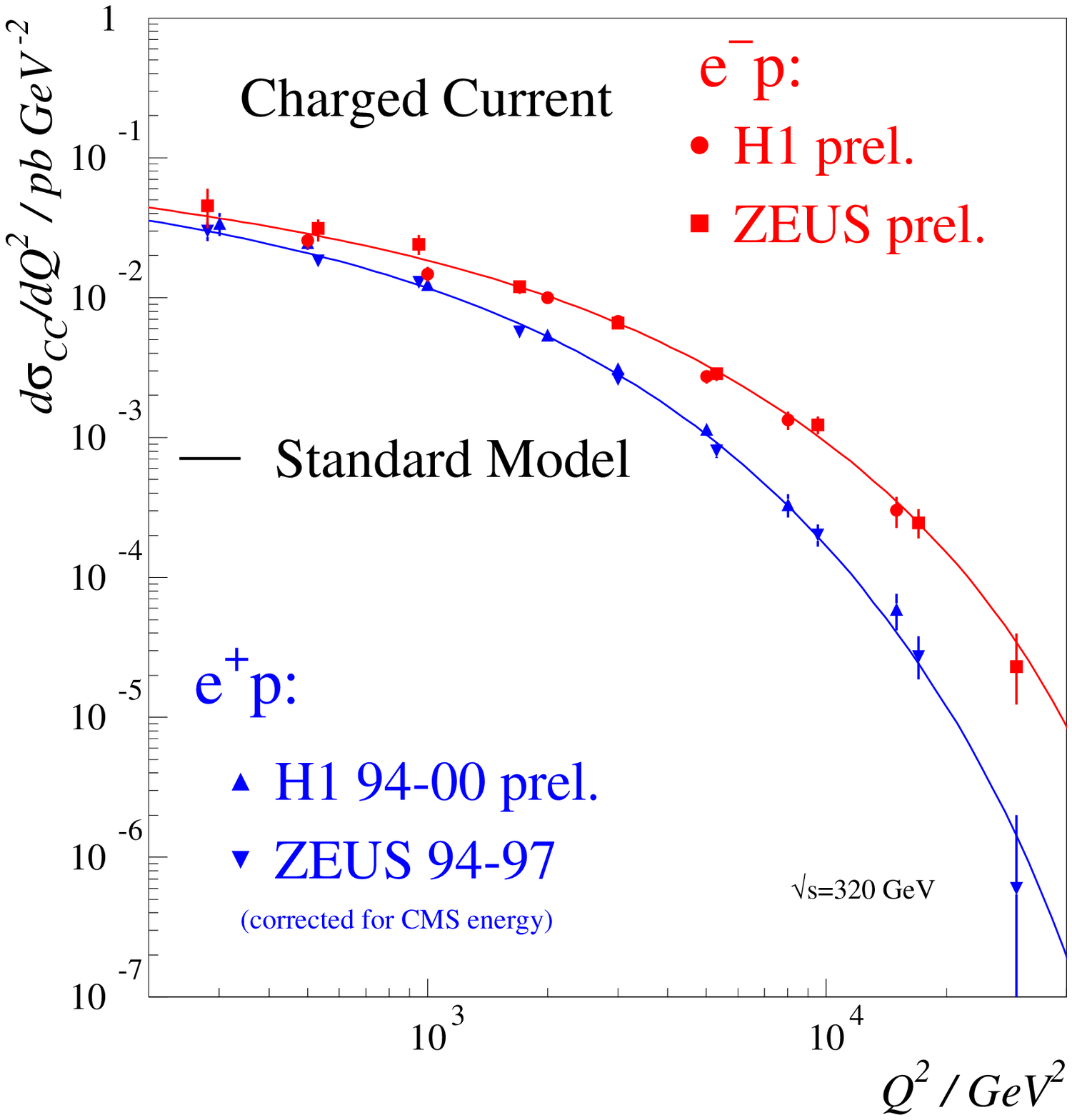,bbllx=50pt,bblly=150pt,bburx=510pt,bbury=650pt,height=11.05cm,width=12.9cm}
\end{center}
\caption{\label{f:HZccvsq00}
      { The CC cross sections for $e^-p$ and $e^+p$ as a function of $Q^2$ 
        as measured by H1 and ZEUS. The curves show QCD NLO predictions. 
      }}
\end{figure}

The combined CC cross section data from the two experiments are shown in Fig.~\ref{f:HZccvsq00} for $e^+p$ and $e^-p$ scattering as a function of $Q^2$. For $Q^2 < M^2_W$ the cross sections show a slow decrease with $Q^2$ which is mainly due to the shrinking phase space in $x$. The rapid fall at $Q^2 > M^2_W$ is mainly driven by the propagator term $~\frac{1}{(M^2_W + Q^2)^2}$. At low $Q^2$, the cross sections for the two charge states are almost the same: the dominant contribution comes from scattering on sea quarks which is approximately flavor symmetric and contributes roughly equally in the two cases. At high $Q^2$ the cross section for $e^-p$ scattering is larger than for $e^+p$ by almost an order of magnitude. Here valence quarks dominate the cross sections and different quark flavors contribute to the two charge states. Furthermore, the $xF_3$ contribution is added to the $e^-p$ cross section and subtracted for $e^+p$.

\begin{figure}[ht]
\begin{center}
\epsfig{file=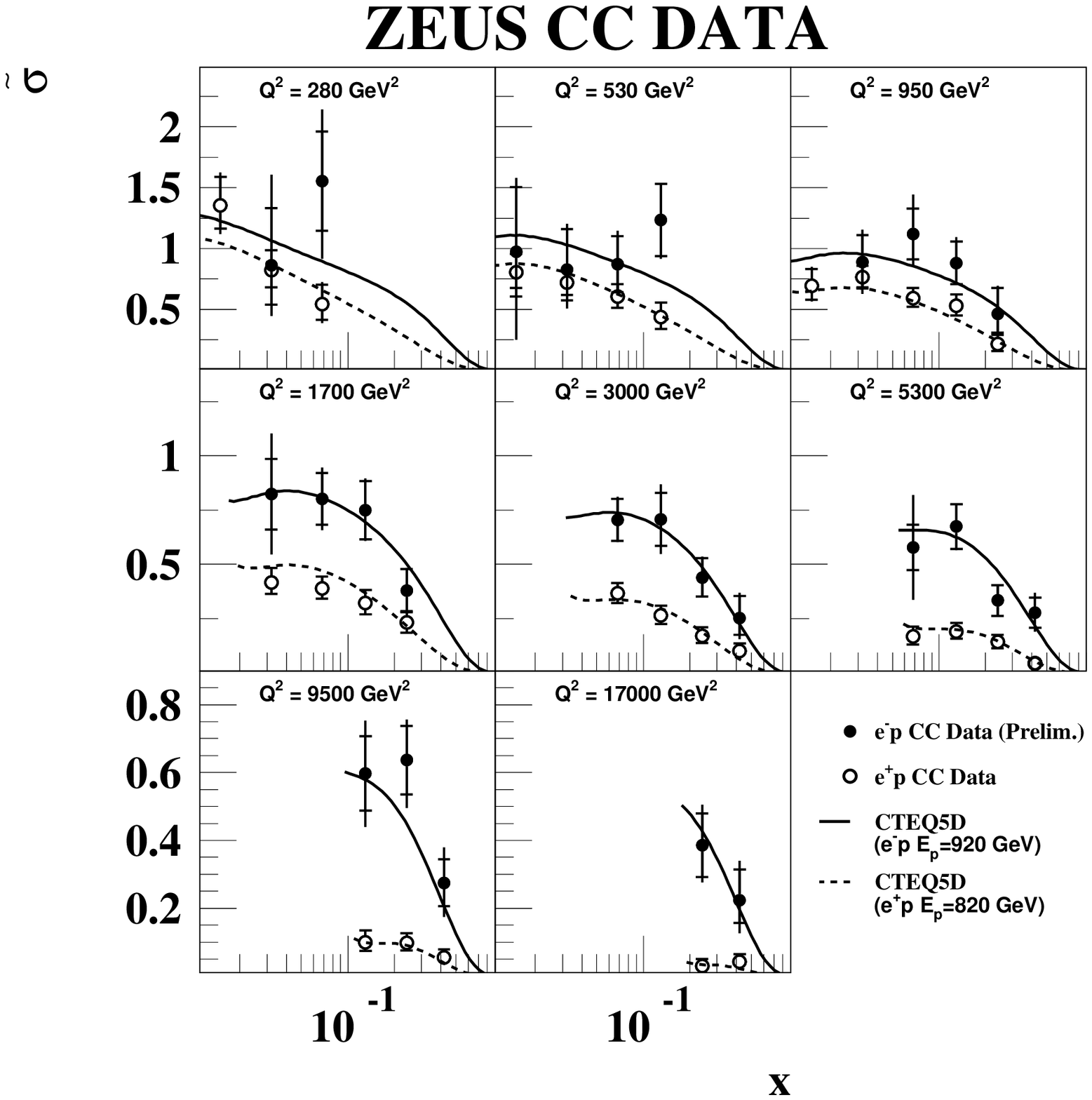,bbllx=50pt,bblly=50pt,bburx=510pt,bbury=601pt,height=12cm}
\end{center}
\caption{\label{f:Zccrvsx00}
      {  The reduced CC cross sections for $e^-p$ and $e^+p$ scattering as
         a function of $x$ for different $Q^2$ values (from ZEUS).
        The curves show the predictions using CTEQ5D.
      }}
\end{figure}

In Fig.~\ref{f:Zccrvsx00} the reduced cross section,
\begin{eqnarray}
\tilde{\sigma}^{\mp}_{CC} = \frac{2\pi x}{G^2_F} \Biggl[\frac{M^2_W+Q^2}{M^2_W} \Biggl]^2 \frac{d^2\sigma_{CC}}{dxdQ^2} 
\end{eqnarray}

as measured by ZEUS, is shown as a function of $x$ for different $Q^2$ intervals. As $Q^2$ increases the reduced cross section for $e^+p$ scattering becomes small with respect to the case of $e^-p$ scattering.

While the charged vector bosons $W^{\pm}$ have been directly observed (CERN 1983) the effect of the $W$ propagator on weak interaction cross sections, before HERA, had not been seen directly. As mentioned above, the $Q^2$ fall-off of the charged current cross sections depends primarily on the $W$ propagator. The new data from HERA allow a significant measurement of $M_W$ thanks to the large $Q^2$ range and the substantial size of the data samples. Using for $G_F$ the measured value, the $W$ mass was found to be:

\begin{eqnarray}
M_W = 79.9 \pm 2.2 (stat) \pm 0.9(syst) \pm 2.1 (theor) {\; \rm GeV} \;\; {\rm (H1)} \nonumber \\
M_W = 81.4^{+2.7}_{-2.6} (stat) \pm 2.0 (syst) ^{+3.3}_{-3.0} (PDF) {\; \rm GeV} \;\; {\rm (ZEUS)}.
\end{eqnarray}
The values are in good agreement with the direct measurement of $M_W = 80.419 \pm 0.056$ GeV~\cite{PDG2000}.

\subsection{Comparison of NC and CC cross sections}
\begin{figure}[ht]
\begin{center}
\epsfig{file=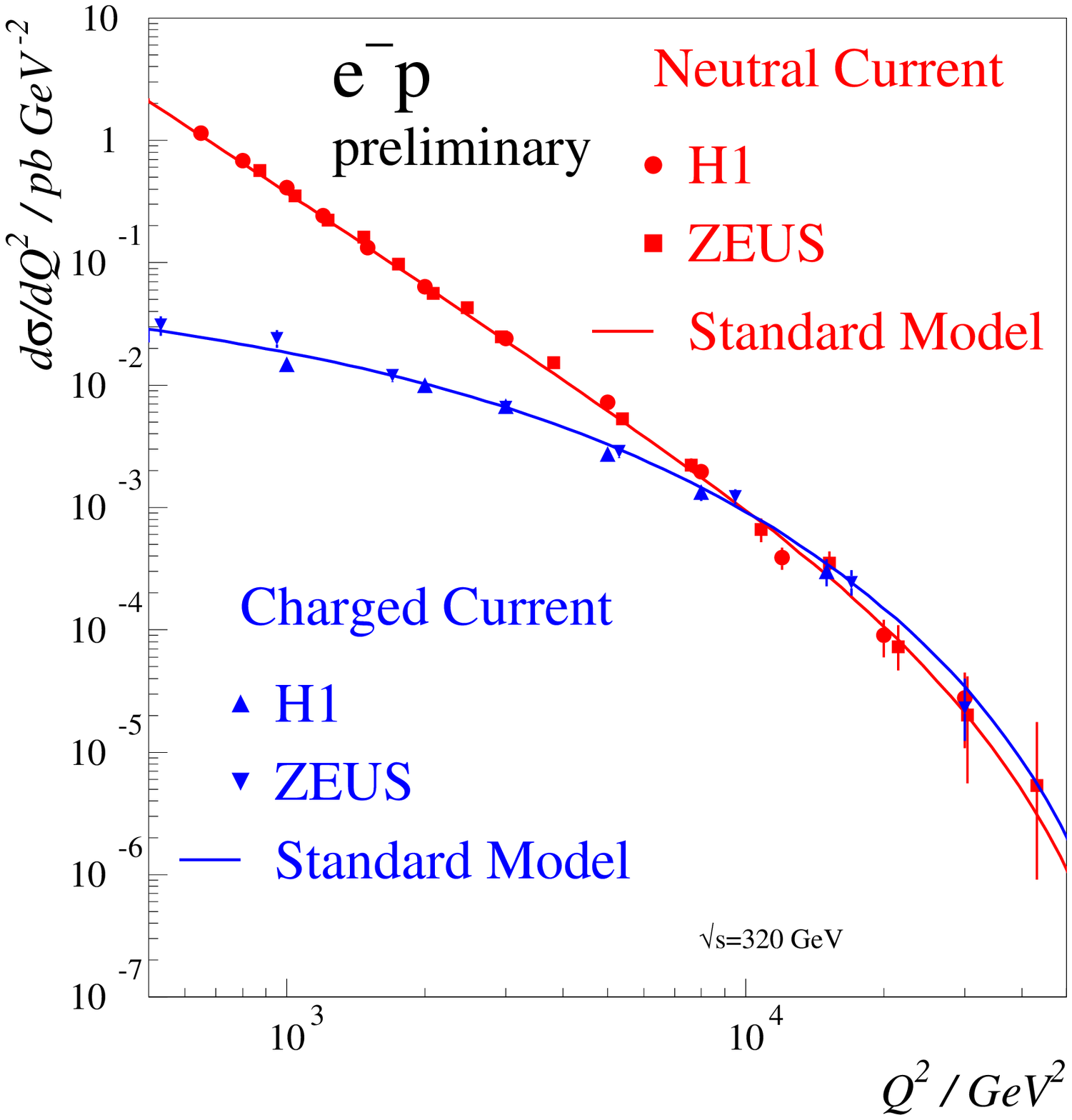,bbllx=50pt,bblly=160pt,bburx=510pt,bbury=740pt,height=12cm}
\end{center}
\caption{\label{f:HZe-ncccvsq00}
      {  Comparison of the cross sections for $e^-p$ scattering by NC and CC
         exchange as a function of $Q^2$, measured by H1 and ZEUS. The curves
         show the predictions of the Standard Model.
      }}
\end{figure}

\begin{figure}[ht]
\begin{center}
\epsfig{file=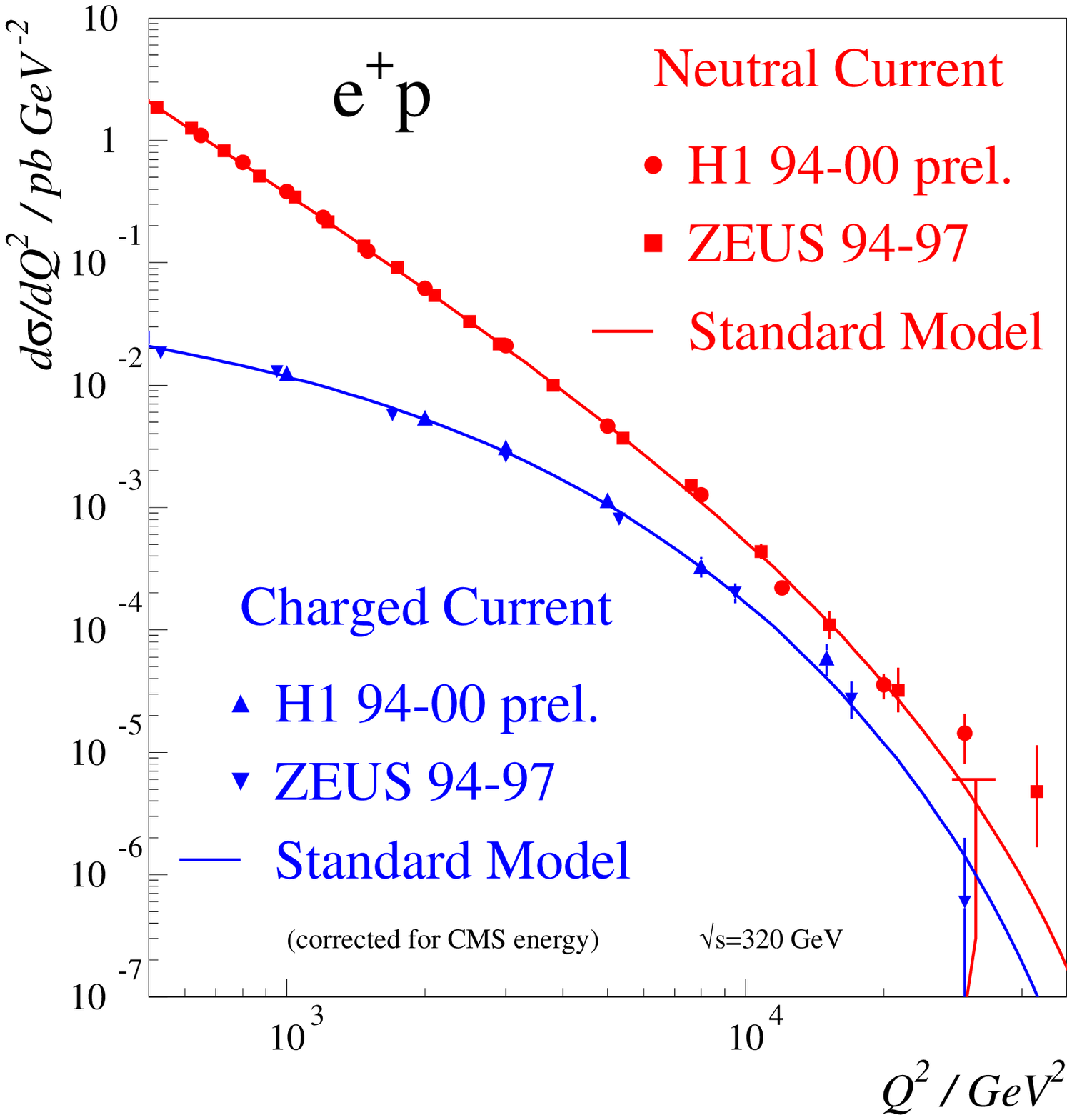,bbllx=50pt,bblly=160pt,bburx=510pt,bbury=740pt,height=12cm}
\end{center}
\caption{\label{f:HZe+ncccvsq00}
      {  Comparison of the cross sections for $e^+p$ scattering by NC and CC
         exchange as a function of $Q^2$, measured by H1 and ZEUS. The curves
         show the predictions of the Standard Model.
      }}
\end{figure}

In Figs.~\ref{f:HZe-ncccvsq00},~\ref{f:HZe+ncccvsq00} the cross sections for NC and CC scattering in $e^-p$ and $e^+p$ interactions are compared. At low $Q^2$, where NC scattering arises predominantly from electromagnetic interactions (= photon exchange), the NC cross section exceeds by far the CC cross section. However, at $Q^2$ values above 10000 GeV$^2$ the two processes have about the same cross sections; at these large values of $Q^2$, the weak force is of similar strength as the electromagnetic one: the HERA measurements show directly the unification of the electromagnetic and weak forces.

It is instructive to compare this result with typical electromagnetic and weak particle decays. There, the weak force is about ten orders of magnitude smaller than the electromagnetic one. For instance, the decay time for the electromagnetic decay $\Sigma^0 \to \Lambda \gamma$ is $7.4 \pm 0.1 \cdot 10^{-20}$ s while it is $4.1 \cdot 10^{-10}$ s for the weak decay $\Lambda \to p \pi^-$, (note, both decays have similar c.m. momentum, viz. 0.10 GeV and 0.074 GeV, respectively, and therefore similar phase space).

\subsection{Testing the Standard Model}
The large space covered in $x$ and $Q^2$ and their high precision make the NC and CC cross section measurements of H1 and ZEUS a powerful testing ground of the Standard Model (SM). Uncertainties in the SM predictions arise from three sources: electroweak parameters, electroweak radiative corrections, and the parton momentum distributions including their higher order QCD corrections. A detailed discussion of these uncertainties can be found in~\cite{H1NCCC89,HNCe+47,ZNCe+47,ZCC99}

For NC scattering, the electroweak parameters have been measured to high precision by other experiments and contribute less than 0.3\% uncertainty for the SM predictions. Higher order corrections for radiative corrections, vertex and propagator corrections and two-boson exchange are expected to be less than 1\%. The primary source of uncertainties stem from the parton momentum distributions which have been determined from data mostly at low $Q^2$ and then extrapolated to higher $Q^2$ using DGLAP QCD evolution. The HERA data included in these fits are mostly from $x < 10^{-2}$ and have little influence on the predictions for the high $Q^2$ regime considered here. The parton densities (PDF) give a total uncertainty on the SM predictions for $d\sigma^{NC}/dQ^2$ of 4\% at $Q^2 = 400$ GeV$^2$ increasing to 8\% at the highest $Q^2$ covered.

For CC scattering, the main uncertainty of the SM predictions comes also from the uncertainties of the PDF's. The resulting uncertainty in $d\sigma^{CC}/dQ^2$ ranges from 4\% at $Q^2 = 400$ GeV$^2$ to 10\% at $Q^2 = 10000$ GeV$^2$, and increases further at higher $Q^2$. The large uncertainty at high $Q^2$ is due to the $d$-quark density which is poorly constrained at high $x$ by the experimental data.

\begin{figure}[ht]
\begin{center}
\epsfig{file=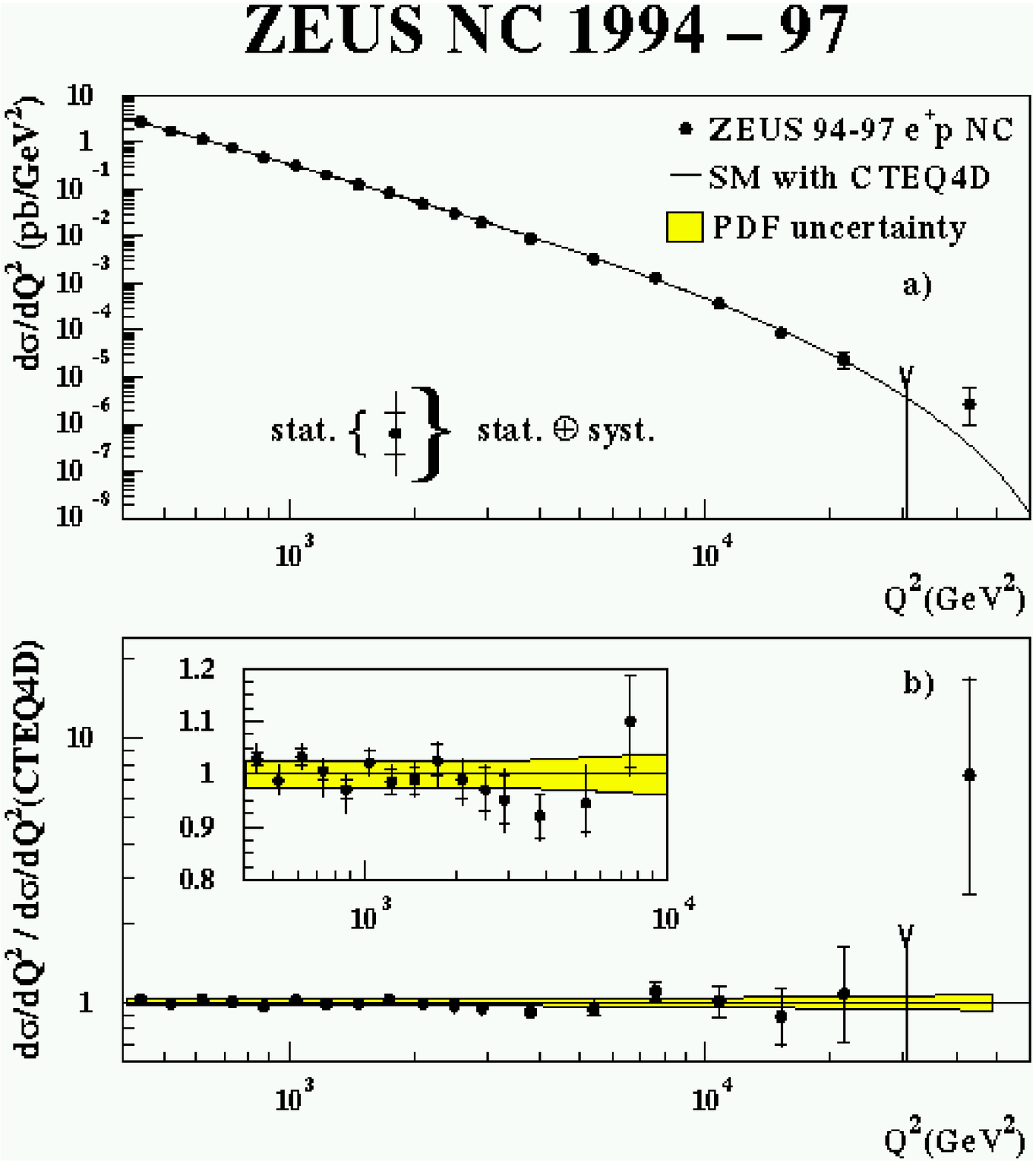,bbllx=50pt,bblly=100pt,bburx=510pt,bbury=740pt,height=12cm}
\end{center}
\caption{\label{f:Ze+pncsm}
      {  The $e^+p$ NC DIS cross section for data and the Standard Model
         (SM) predictions (a) and the ratio of data to the SM prediction 
         (b). The shaded region gives the uncertainty in the SM prediction
         due to the uncertainty in the PDF's.
      }}
\end{figure}

\begin{figure}[ht]
\begin{center}
\epsfig{file=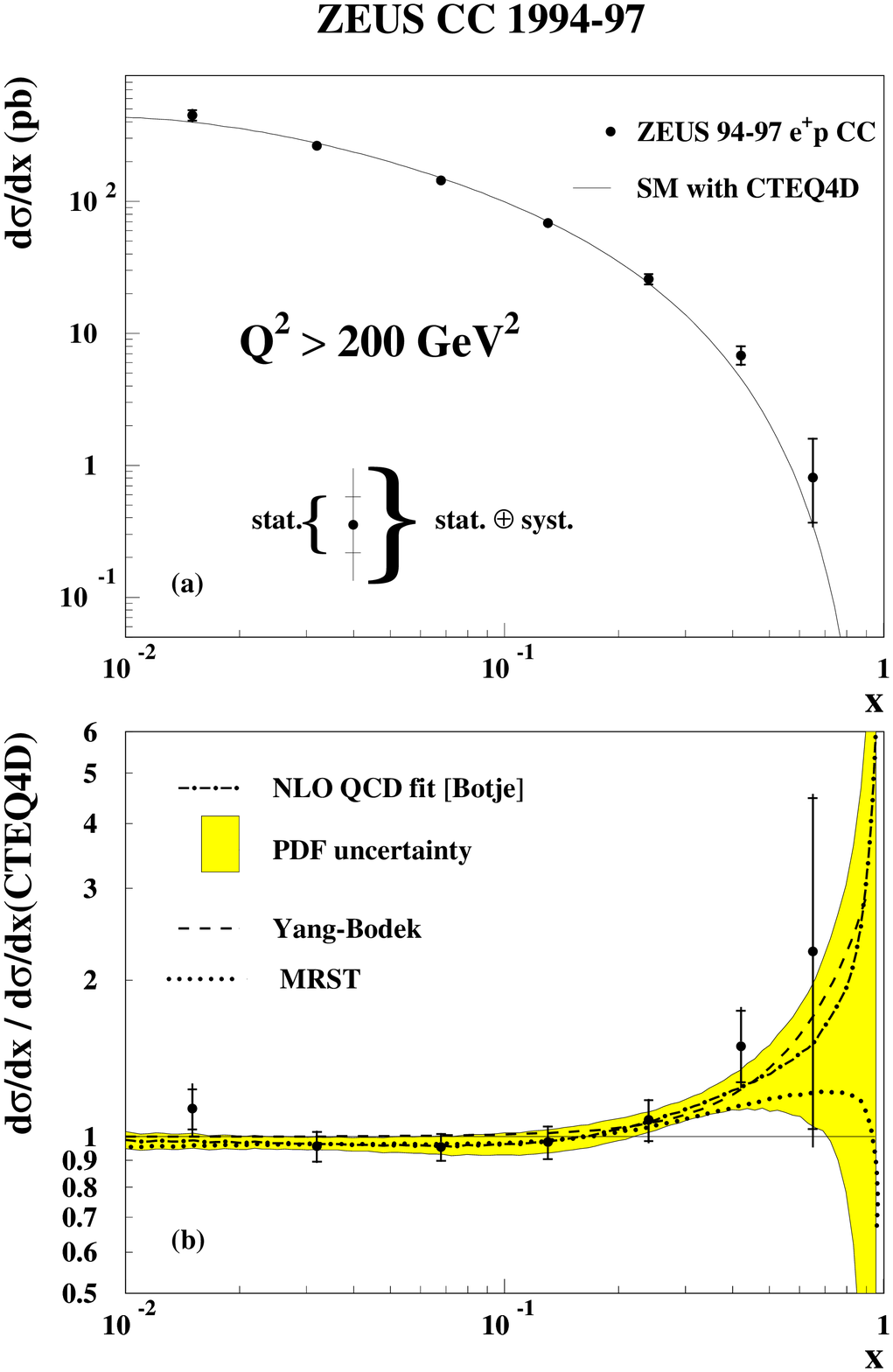,bbllx=0pt,bblly=0pt,bburx=510pt,bbury=840pt,width=11cm,height=11cm}
\end{center}
\caption{\label{f:Ze+pccsm}
      { The $e+p$ CC DIS cross section for data and the Standard Model (SM) predictions (a) and the ratio of data to the SM prediction (b). The shaded region gives the uncertainty in the SM prediction due to the uncertainty in the PDF's.
      }}
\end{figure}

In Fig.~\ref{f:Ze+pncsm} $d\sigma^{NC}/dQ^2$ as measured for $e^+p$ scattering is compared with the SM predictions calculated with the CTEQ4D PDF's. Taking the errors of the data and the uncertainties of the SM into account there is excellent agreement between the SM predictions and the data up to $Q^2 = 36000$ GeV$^2$. At higher $Q^2$ there are 2 events observed compared to 0.27 predicted. The corresponding data for CC scattering are presented in Fig.~\ref{f:Ze+pccsm}. For $Q^2 < 5000$ GeV$^2$ the uncertainties in the PDF's - mainly due to the uncertainty of the d-quark contribution - prevent a precise test of the SM.

The Standard Model (SM) provides a constraint between $G_F$ and $M_W$:
\begin{eqnarray}
G_F = \frac{\pi \alpha}{\sqrt{2}} \frac{M^2_Z}{(M^2_Z-M^2_W)M^2_W} \; \frac{1}{1 - \Delta_r},
\end{eqnarray}
where $M_Z$ is the mass of the $Z$ boson. The term $\Delta_r$ contains the radiative corrections to the lowest order for $G_F$ and is a function of $\alpha_{em}$ and the masses of the fundamental bosons and fermions~\cite{MWradcorr}. Using this constraint, assuming $M_H = 100$ GeV and treating only $M_W$ as a free parameter ZEUS found:

\begin{eqnarray}
M_W = 80.50^{+0.24}_{-0.25} (stat) ^{+0.13}_{-0.16} (syst) \pm 0.31(PDF)^{+0.03}_{-0.06}(\Delta M_t, \Delta M_H, \Delta M_Z) {\rm GeV} \;\;
\end{eqnarray}

The excellent agreement with the directly measured value of $M_W$ indicates that the SM gives a consistent description of a variety of phenomena over a wide range of energy scales.

\section{Search for new physics}
\subsection{Instantons}
The Standard Model contains a class of hard processes which cannot be treated perturbatively. These processes violate conservation of baryon and lepton number in electroweak interactions and of chirality in strong interactions~\cite{tHooft76}. Such processes are induced by instantons. QCD-instantons represent tunneling transitions between topologically different vacua~\cite{Belavin75}. Originally, instanton effects were expected to lead to observable effects only at energies of the order of $10^8$ GeV until it was realized that emission of gauge bosons can bring down the threshold~\cite{Ringwald90,Espinosa90}. 

Up to now, experiments have been unable to demonstrate the presence of instanton effects. Deep inelastic lepton nucleon scattering (DIS) is particularly suited to search for such effects, firstly because DIS provides a hard scale ($Q^2$), secondly, there exist theoretical estimates of the instanton contribution involving quark-gluon (= gauge boson) interactions~\cite{Balitskii93} and thirdly, the contribution is expected to be substantial~\cite{Ringwald94}. 

First studies have been performed by H1~\cite{Hinst96,Carli98,Kuhlen97} which placed limits on a possible instanton contribution. Preliminary results based on a much larger data sample (16 pb$^{-1}$), taken in 1997, were reported recently by the same experiment~\cite{Hinst00}. 

\begin{figure}[ht]
\epsfig{file=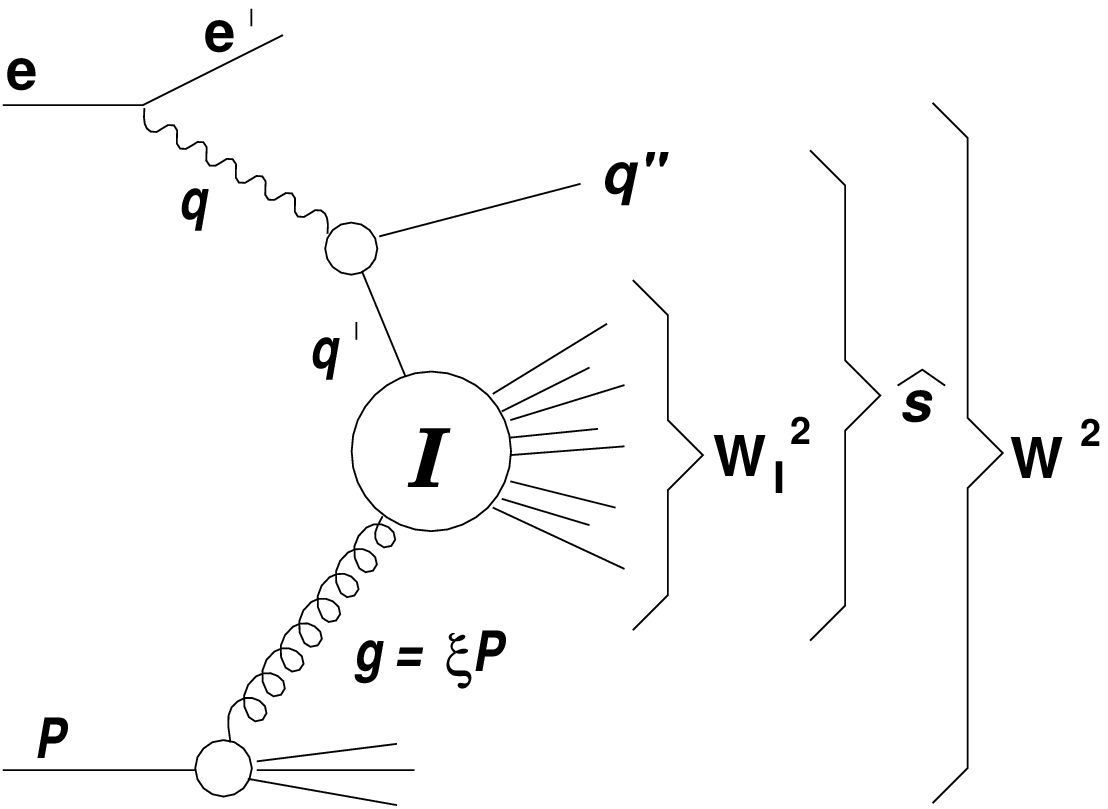,bbllx=-230pt,bblly=0pt,bburx=510pt,bbury=400pt,height=8cm}
\caption{Diagram and kinematic variables for instanton interaction in DIS}
\label{f:diaginstanton}
\end{figure}

\begin{figure}[ht]
\epsfig{file=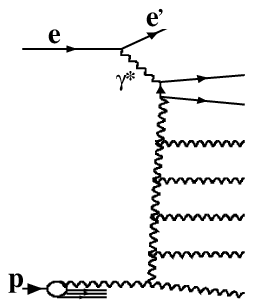,bbllx=-25pt,bblly=50pt,bburx=510pt,bbury=180pt,height=8cm}
\caption{DIS multi-ladder diagram.}
\label{f:diagdismultih}
\end{figure}

The kinematics for DIS interactions involving instanton interactions is sketched in Fig.\ref{f:diaginstanton}. By instanton interaction a quark (momentum $q^{\prime}$, mass squared $Q^{\prime 2}$) emitted by the virtual photon fuses with a gluon (momentum $g = \xi p$) from the proton forming a hadronic system with c.m. energy $W_I$. The c.m. energy of the total hadronic system without the proton remnant is $\hat{s}$. The instanton events have to be isolated from the standard DIS events depicted in Fig.~\ref{f:diagdismultih}. The main difference between the two classes of events lies in the fact that the particles produced by instanton interaction are phase space like distributed whereas those from standard DIS processes result from partons emitted in ladder-type diagrams, see Fig.~\ref{f:diagdismultih}. The challenge of the search for instanton effects lies in the task to devise suitable kinematic variables and find a kinematic region which maximize the instanton signal over the normal DIS background. 

The jet with the highest transverse energy ($Et_{jet})$ is used to estimate the momentum $q^{\prime \prime}$ of the current quark, see~Fig.\ref{f:diaginstanton}. Knowing $q^{\prime \prime}$ the momentum $q^{\prime}$ of the quark entering the instanton subprocess can be determined. The objects belonging to the jet are not used for computing the following variables. The $E_t$-weighted rapidity mean $\overline{\eta} = \sum_h E_{t,h} \eta_h/ \sum_h E_{t,h}$ of all remaining objects is used as an estimator of the instanton band which is defined by $\overline{\eta} = \pm 1$. The total scalar transverse energy $Et_b$ within this band is measured. Furthermore, the axis $\vec{i}$ is found for which the summed projections of the 4-momenta of all hadronic objects are minimal resp. maximal. The relative difference between $E_{in} = min_i \sum_h |\vec{p}_h \cdot \vec{i}|$ and $E_{out} = max_i \sum_h |\vec{p}_h \cdot \vec{i}|$ is called $\Delta_b = (E_{in}-E_{out})/E_{in}$ which measures the $E_t$ weighted azimuthal isotropy of an event, i.e. $\Delta_b$ = small (large) for isotropic (pencil-like) events.

\begin{figure}[ht]
\begin{center}
\epsfig{file=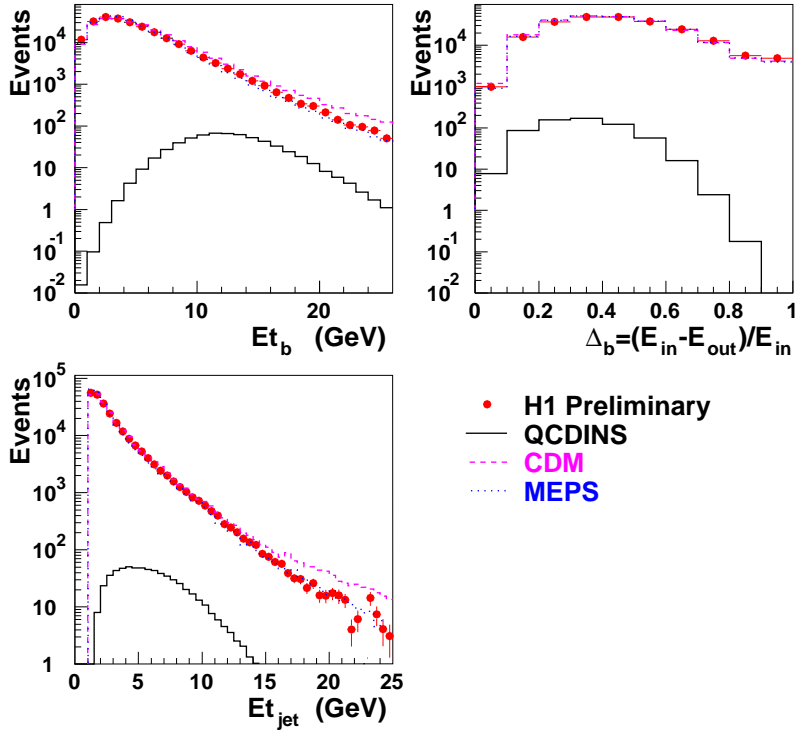,bbllx=50pt,bblly=30pt,bburx=510pt,bbury=640pt,height=12cm}
\end{center}
\caption{\label{f:Hinstet}
      {  Distributions of $Et_b$, $\Delta_b$ and $Et_{jet}$ for the ``instanton band'' $\overline{\eta} \pm 1$. The data are compared with the expectations for standard DIS (MEPS, CDM) and for instanton production (QCDINS). From H1.
      }}
\end{figure}

The search for instanton effects was performed with a sample of 275k DIS events. Figure~\ref{f:Hinstet} shows the distributions of $Et_{jet}, Et_b$ and $\Delta_b$ for the data. They are well reproduced by the expectations for standard DIS as calculated with MEPS which incorporates the QCD matrix elements up to order $\alpha_s$ plus the emission of parton showers to all orders. A second model for standard DIS, CDM, where gluon emission is simulated via the colour dipole model, exceeds the data at large $Et_b$ and large $Et_{jet}$. The expectations for the contributions from instanton processes as calculated with QCDINS are also shown~\cite{QCDINS}. Good discrimination between instanton and standard DIS contributions can be expected from $Et_b$: the instanton contribution is predicted to produce, on average, much higher $Et_b$ values. The predicted amount of instanton induced events is, however, at least a factor of about 20 below the expectations for standard DIS production determined with MEPS. In addition, at large $Et_b$ there is a substantial uncertainty in the predictions for standard DIS contributions as shown by the difference between CDM and MEPS predictions.

\begin{figure}[ht]
\begin{center}
\epsfig{file=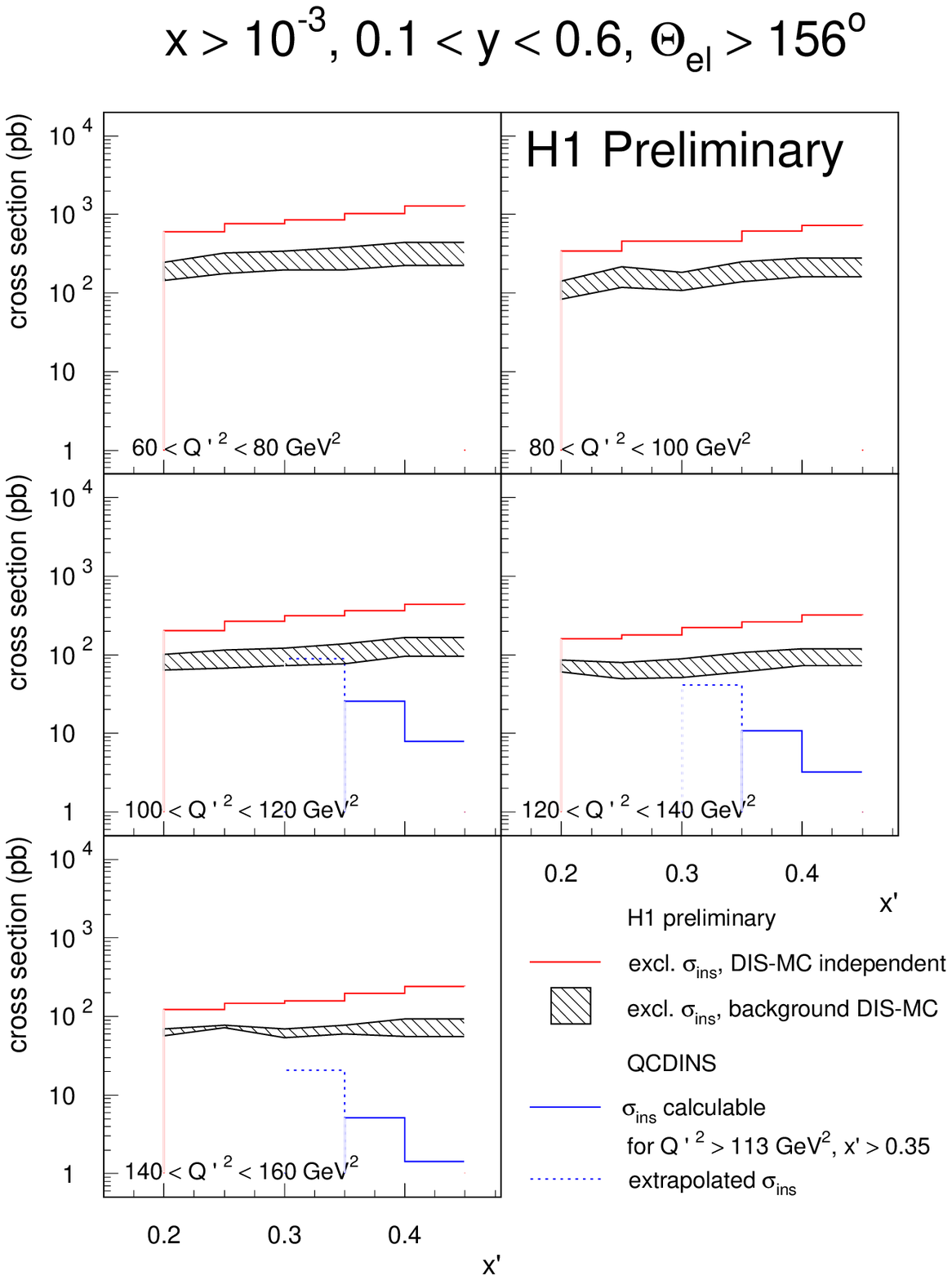,bbllx=-100pt,bblly=30pt,bburx=510pt,bbury=640pt,height=13cm}
\end{center}
\caption{\label{f:Hinstx}
      {  Measured upper limit and expected cross section for instanton induced events as predicted by QCDINS as a function of $x^{\prime}$ and $Q^{\prime 2}$. 
      }}
\end{figure}

A summary of upper limits on instanton production from the H1 analysis and the theoretical expectations for instanton production is presented in Fig.~\ref{f:Hinstx}. Instanton cross sections above 100 - 1000 nb are excluded. The instanton cross section predicted for $\x^{\prime} > 0.35$ and $Q^{\prime 2} > 113$ GeV$^2$ is below the upper limit derived from the data. In conclusion, an additional factor of 10 - 100 in discrimination against standard DIS is required in order to establish an instanton contribution of the predicted size.

\subsection{Deep inelastic scattering at high $Q^2$ and high $x$}
During 1994-6 H1 and ZEUS had collected sufficient data (14 pb$^{-1}$ and 20 pb$^{-1}$, respectively) for a first look at deep inelastic $e^+p$ scattering beyond $Q^2$ values of 10 000 GeV$^2$. Both experiments reported good agreement with the SM predictions for $Q^2 <$ 15 000 GeV$^2$ and an excess of events in the region of $Q^2 >$ 15 000 GeV$^2$, $x >$ 0.4~\cite{Hhixync946,Zhixync946}.

\begin{figure}[ht]
\begin{center}
\epsfig{file=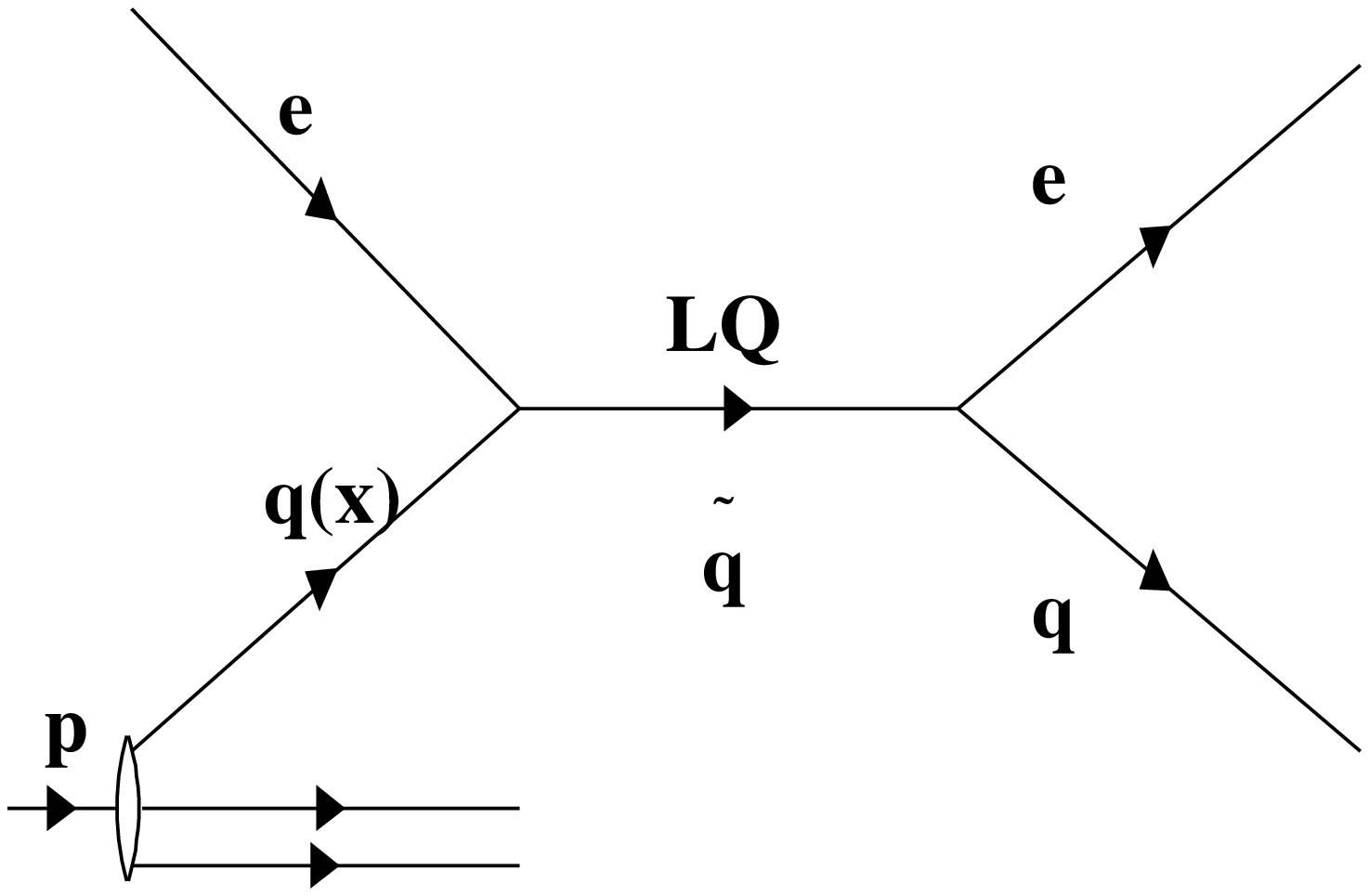,bbllx=0pt,bblly=100pt,bburx=610pt,bbury=501pt,height=6cm,clip=}
\end{center}
\caption{Diagram for leptoquark (or R-parity violating squark) production by $ep$ scattering.}
\label{f:diaglq}
\end{figure}

H1 observed 12 events with $Q^2 >$ 15 000 GeV$^2$ where only $4.7\pm 0.76$ were expected from SM calculations. Assuming the SM to be correct, the probability for such an excess was found to be about 1\%. Further, these high $Q^2$ events tended to cluster around $x$ values of 0.4 - 0.5 corresponding to a positron-quark mass of $M \approx \sqrt{xs} \approx 200$ GeV which was suggestive for the production of a leptoquark or a R-parity violating SUSY state, see Fig.~\ref{f:diaglq}. In the region $187.5 < M < 212.5$ GeV, $y = Q^2/(xs) > 0.4$, 7 events were observed where $0.95 \pm 0.18$ were expected from SM processes. 

The ZEUS experiment found two events with $Q^2 >$ 35 000 GeV$^2$ while $0.145 \pm 0.013$ were expected from SM. For $x > 0.55$ and $y > 0.25$ four events were observed compared to $0.91 \pm 0.08$ expected. In the SM the probability for such a fluctuation to occur in this kinematic region was 0.7\%, and 8\% for the entire region of $Q^2 >$ 5000 GeV$^2$.

In 2000, both experiments presented preliminary analyses of the high $Q^2$, high $x$ region based on a factor 4 - 6 more $e^+p$ data and first high statistics data from $e^-p$ scattering. 

\begin{figure}[hpbt]
\begin{center}
\epsfig{file=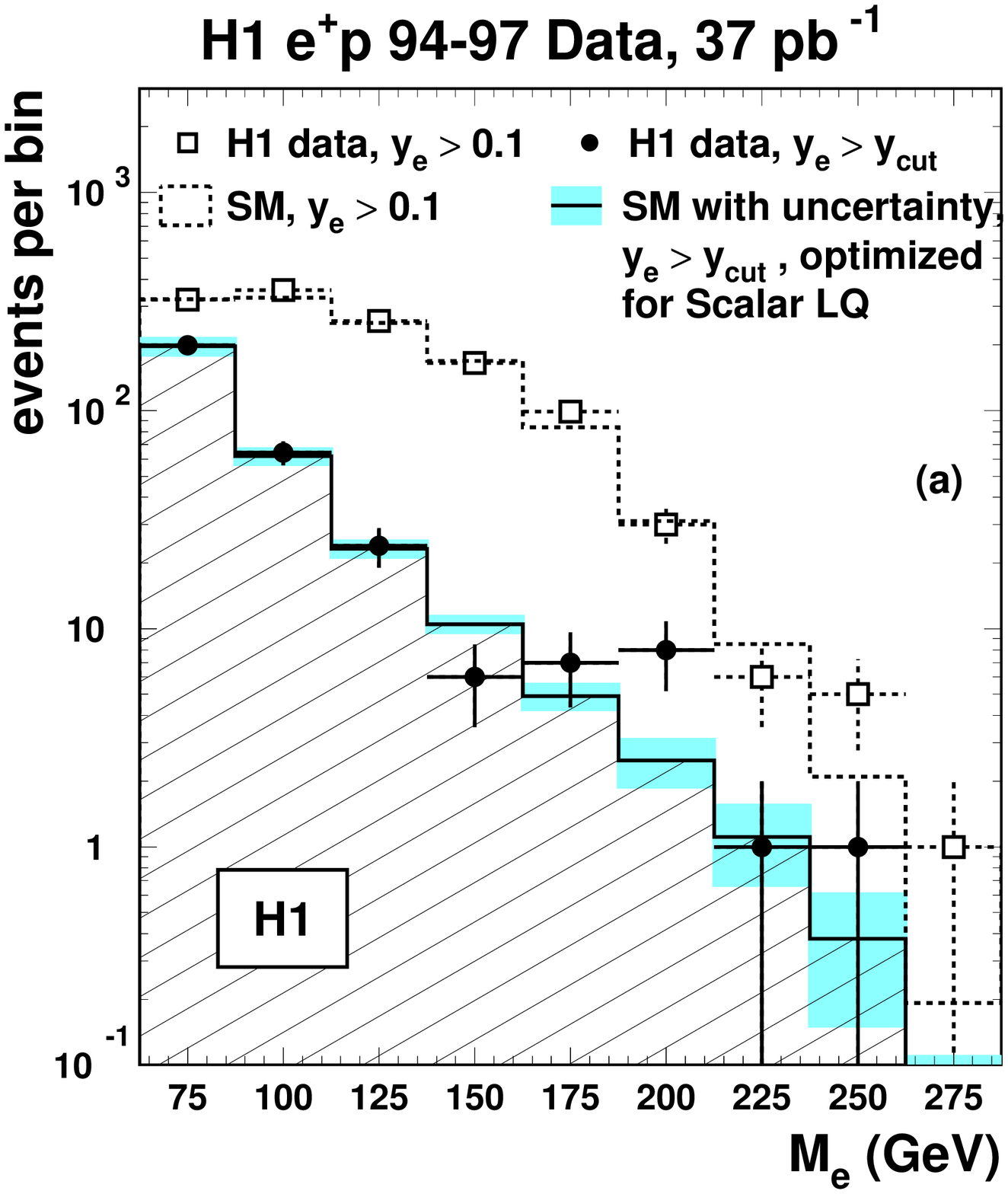,height=8cm}
\epsfig{file=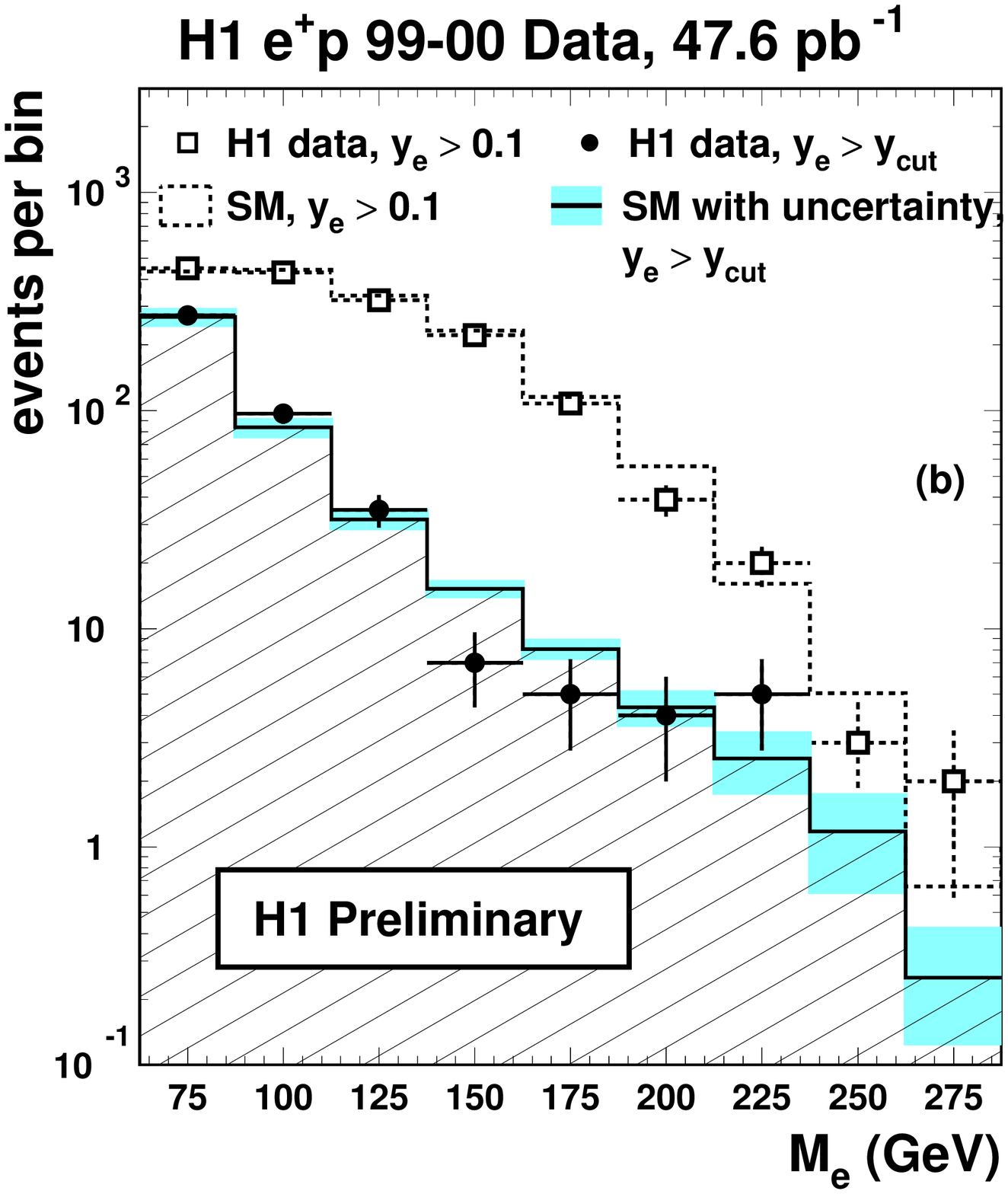,height=8cm}
\end{center}
\caption{\label{f:He+hixync940}
      { Mass spectra for NC DIS-like final states (points with error bars) and DIS SM predictions (histograms) for $y > 0.1$ and for a $y$ cut which maximizes a possible signal from leptoquarks (squarks). Limits from HERA, LEP and TEVATRON.
      }}
\end{figure}

\begin{figure}[hpbt]
\begin{center}
\epsfig{file=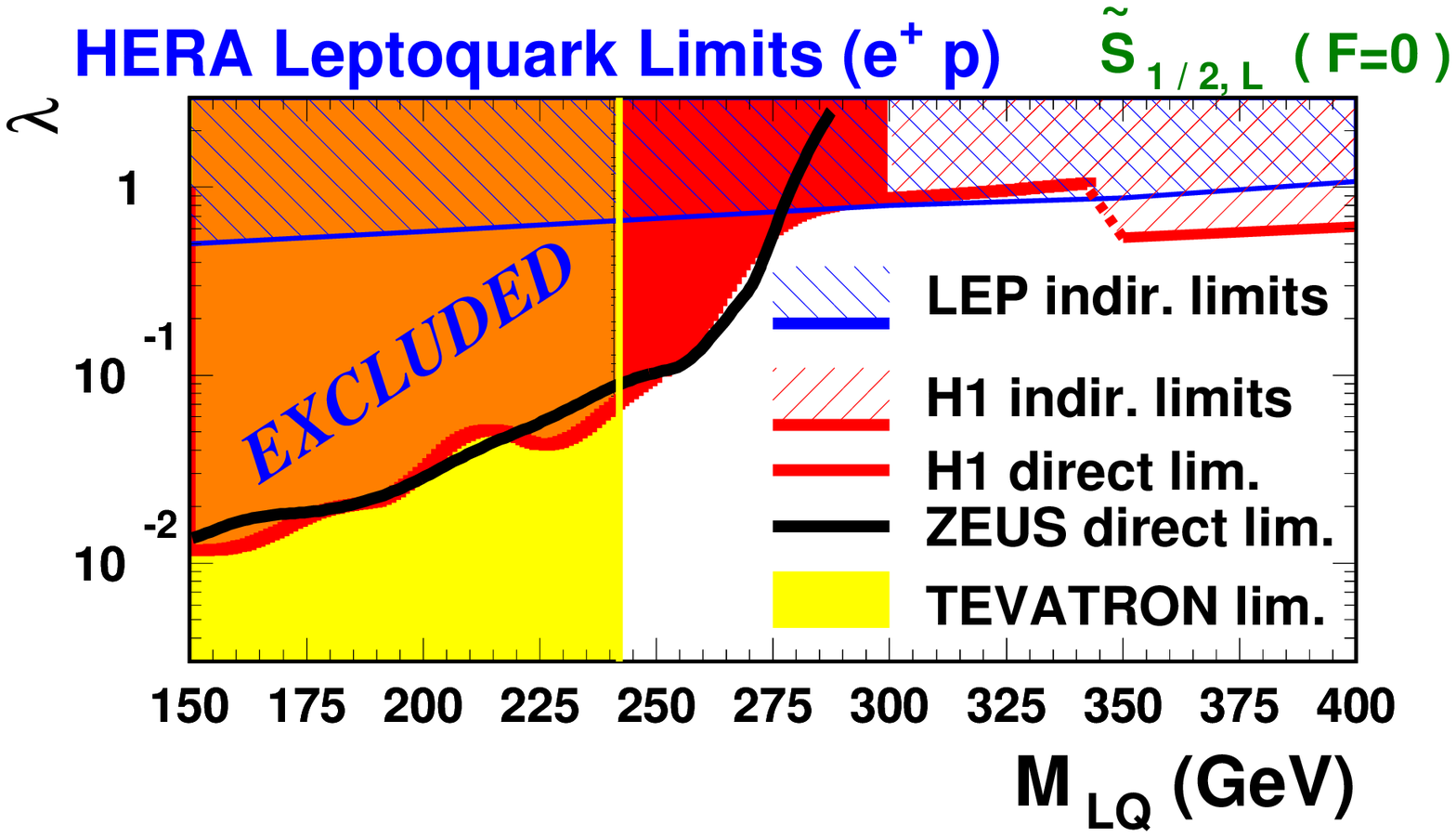,height=8cm}
\epsfig{file=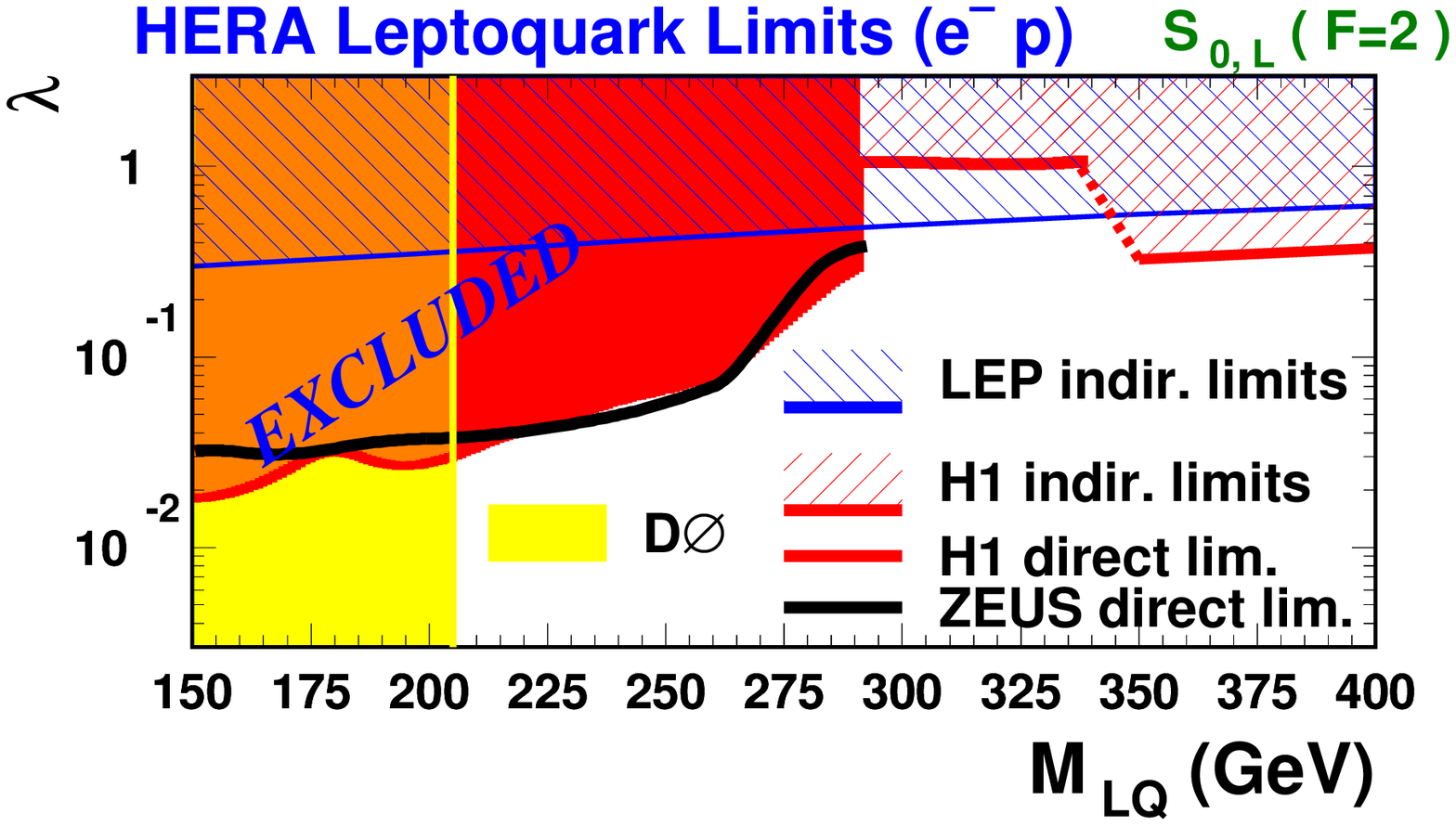,height=8cm}
\end{center}
\caption{\label{f:ZHelqlim}
      { Exclusion limits from HERA, LEP and TEVATRON at 95\% CL on the Yukawa coupling $\lambda$ as a function of the LQ mass for a scalar LQ with $F = 0$ (top) and $F = 2$ (bottom) described by the BRW model. The grey and hatched domains are excluded. 
      }}
\end{figure}

The new results from H1~\cite{Hhixync40} are presented in Fig.~\ref{f:He+hixync940} in terms of the electron-quark mass $M$, separately for the data taken in 1994-7 and 1999-0. While the earlier data show still some excess of events above the SM predictions (see histograms) near $M = 200$ GeV the data taken later agree well with the SM. Based on the data from 1994-7 which correspond to an integrated luminosity of 37 pb$^{-1}$ H1 has placed tight constraints on the properties of first generation leptoquarks~\cite{He+lqlimit99}. These studies were extended using the additional data taken during 1998-2000. The limits derived on the Yukawa coupling $\lambda$ for leptoquark (or R-parity violating squark) positron quark vertices based on the BRW model~\cite{BRW97} are summarized in Fig.~\ref{f:ZHelqlim}. For masses between 205 and 300 GeV these limits are more stringent than those from the TEVATRON or LEP.

\begin{figure}[ht]
\begin{center}
\epsfig{file=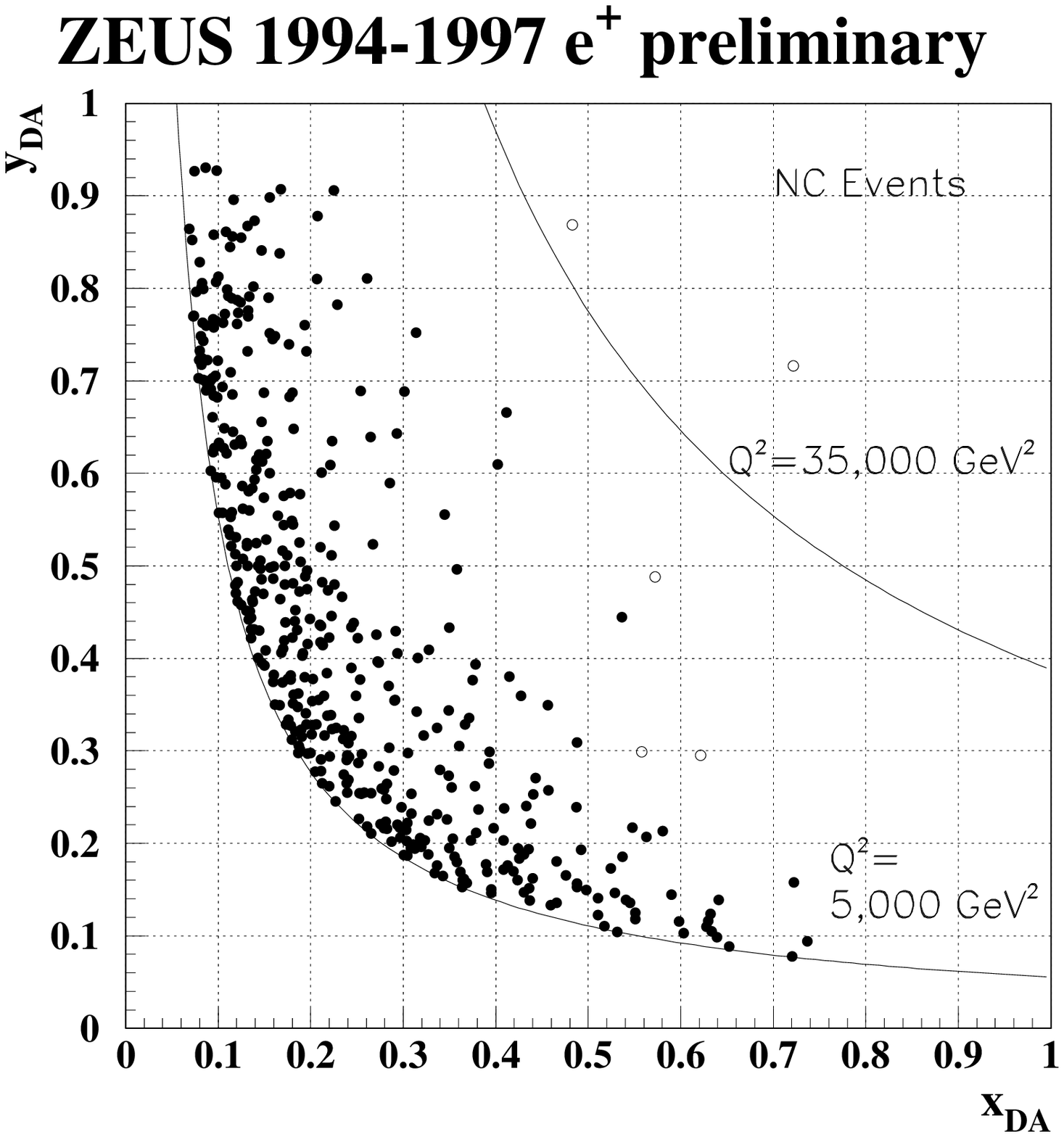,bbllx=50pt,bblly=0pt,bburx=510pt,bbury=740pt,height=12cm}
\epsfig{file=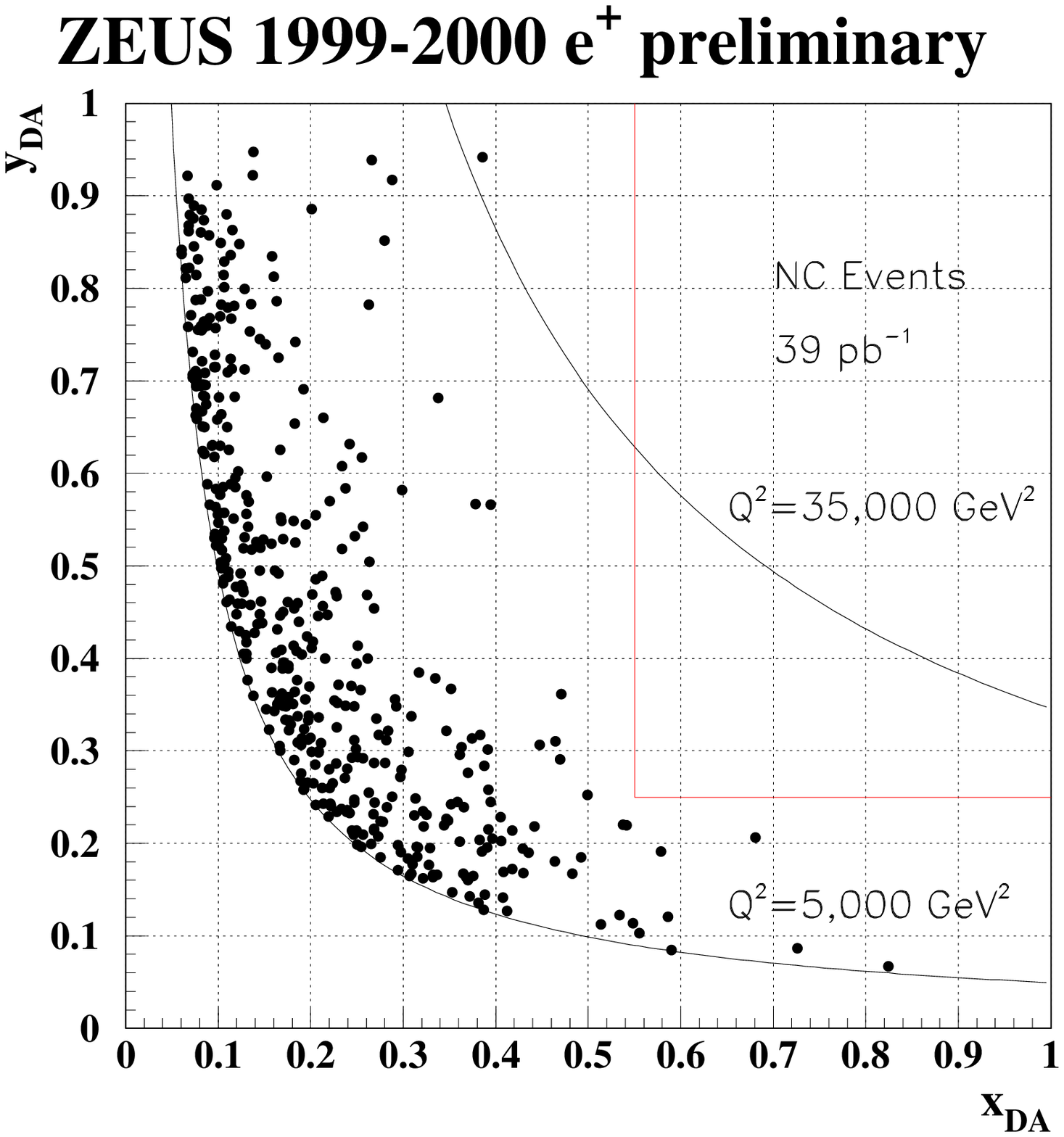,bbllx=50pt,bblly=0pt,bburx=510pt,bbury=740pt,height=12cm}
\end{center}
\caption{\label{f:Ze+hixync947}
      { The distribution in $x_{DA}$ and $y_{DA}$ of $e^+p$ data taken in 1994-7 (left) and 1999-0 (right). The curves indicate constant values of $Q^2_{DA} = x_{DA}y_{DA}s$ for $Q^2_{DA} =$ 5000 and 35 000 GeV$^2$. In the left distribution the open circles mark the five ``high-x high-y'' events from the 1994-6 data set. From ZEUS. 
      }}
\end{figure}

Figure.~\ref{f:Ze+hixync947} shows the distribution of events in the $x$ - $y$ plane as obtained by ZEUS~\cite{Zhixync40} for $e^+p$ data taken in 1994-7 and 1999-2000. Table~\ref{t:Zhiq} summarizes the number of events observed and expected at high $Q^2$, and at high $x$, high $y$, including the results from $e^-p$ data taken in 1998-9. The combined data sets are in good agreement with the SM expectations and exclude a statistically significant excess.

\begin{table}[hpbt]
\centering
\caption{The observed and expected numbers of events with $Q^2 >$ 35 000 GeV$^2$ and $x_{DA} > 0.55, y_{DA}> 0.25$, respectively, from different years of HERA running, from ZEUS.}
\vspace{0.5cm}
\begin{tabular}{|l|c|c|c|c|c|c|c|}
\hline
      &     &     &   & $Q^2>$ &       & $x_{DA}>0.55$ & \\ 
      &     &     &   & 35000 GeV$^2$ & & $y_{DA}>0.25$ & \\
Year  & $\int L dt$  & $E_e$   &  $E_p$   & data & SM   & data & SM \\
      & (pb$^{-1}$)  & (GeV)   &  (GeV)   & Nevt & Nevt & Nevt & Nevt \\
\hline
1994-7 $e^+p$        & 47.7  & 27.5 & 820 & 2    & 0.34 & 4    & 1.9 \\
1999-0 $e^+p$        & 39.2  & 27.6 & 920 & 1    & 0.53 & 0    & 1.6 \\
part of:             &       &      &     &      &      &      &     \\
1998-9 $e^-p$        & 16.2  & 27.6 & 920 & 2    & 1.02 & 1    & 1.3 \\
\hline
\end{tabular}
\label{t:Zhiq}
\end{table}

\begin{figure}[hpbt]
\begin{center}
\epsfig{file=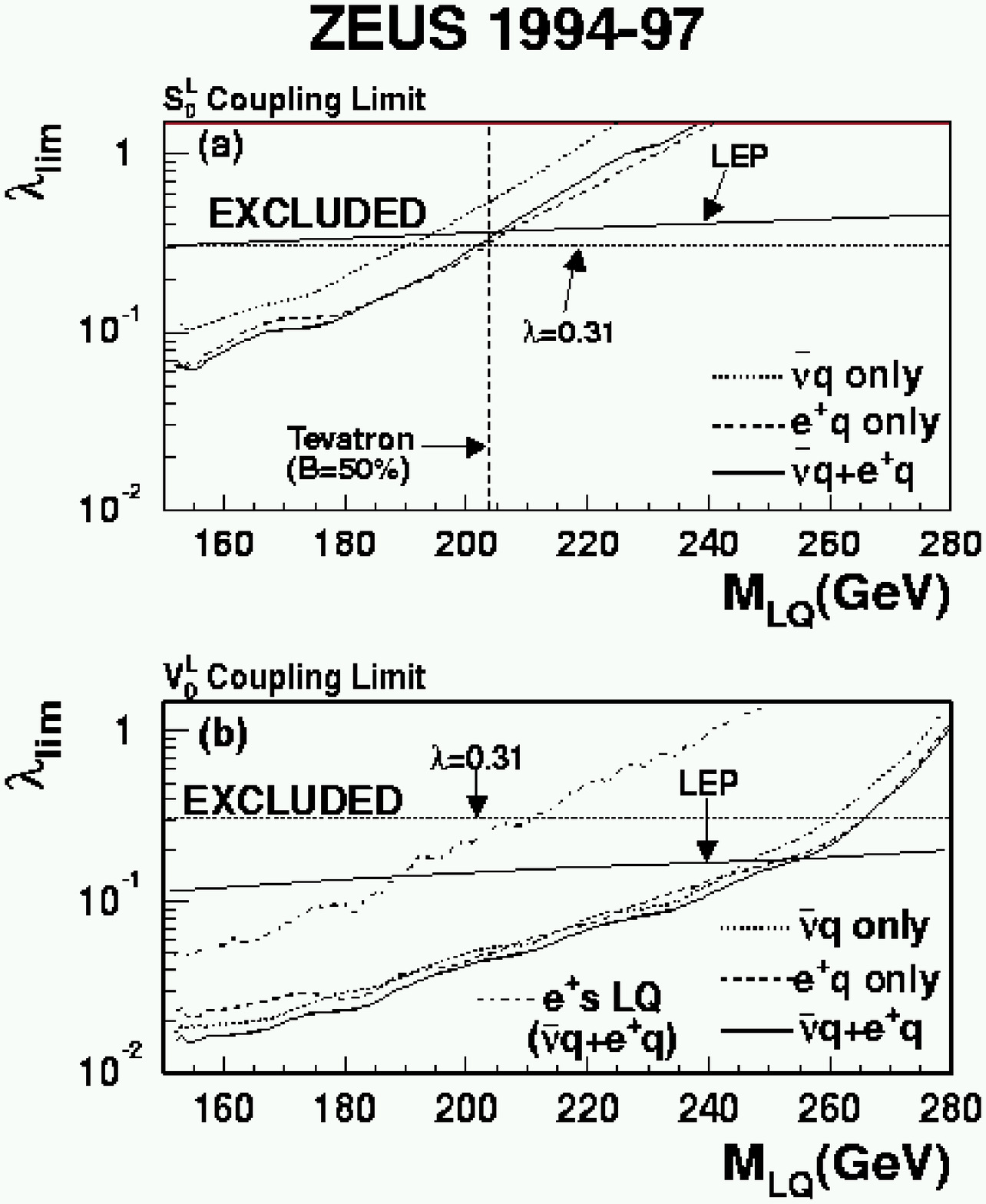,height=14cm}
\end{center}
\caption{\label{f:Ze+1jetbranch}
      { Exclusion limits at 95\% CL on the Yukawa coupling $\lambda$ as a function of the LQ (squark) mass for scalar and vector states decaying into $e^+q$ and $\overline{\nu} q$ channels, derived from $e^+jet$ and $\overline{\nu}jet$ final states. The horizontal lines marked $\lambda=0.31$ indicate an electromagnetic coupling strength $\lambda = \sqrt{4\pi \alpha} = 0.31$. From ZEUS. 
      }}
\end{figure}

ZEUS has searched $e^+ jet$ and $\overline{\nu} +jet$ final states produced in $e^+p$ interactions for leptoquarks and squarks~\cite{Ze+ejet00,Ze+nujet00}. No signal has been observed. The limits derived for decays into $e^+ jet$ or $\overline{\nu} +jet$  on the Yukawa couplings for scalar and vector LQ-lepton-quark and squark-lepton-quark couplings are summarized in Figs.~\ref{f:ZHelqlim},~\ref{f:Ze+1jetbranch}. For vector states these limits are tighter than those obtained at the TEVATRON~\cite{D0lq98} and at LEP~\cite{OPALlq99}. First generation scalar (vector) leptoquarks with an electromagnetic-type coupling strength, $\lambda = \sqrt{4\pi \alpha} = 0.31$, are excluded for masses below 204 (265) GeV.  

\subsection{Search for lepton-flavor violation} 
\begin{figure}[hpbt]
\begin{center}
\epsfig{file=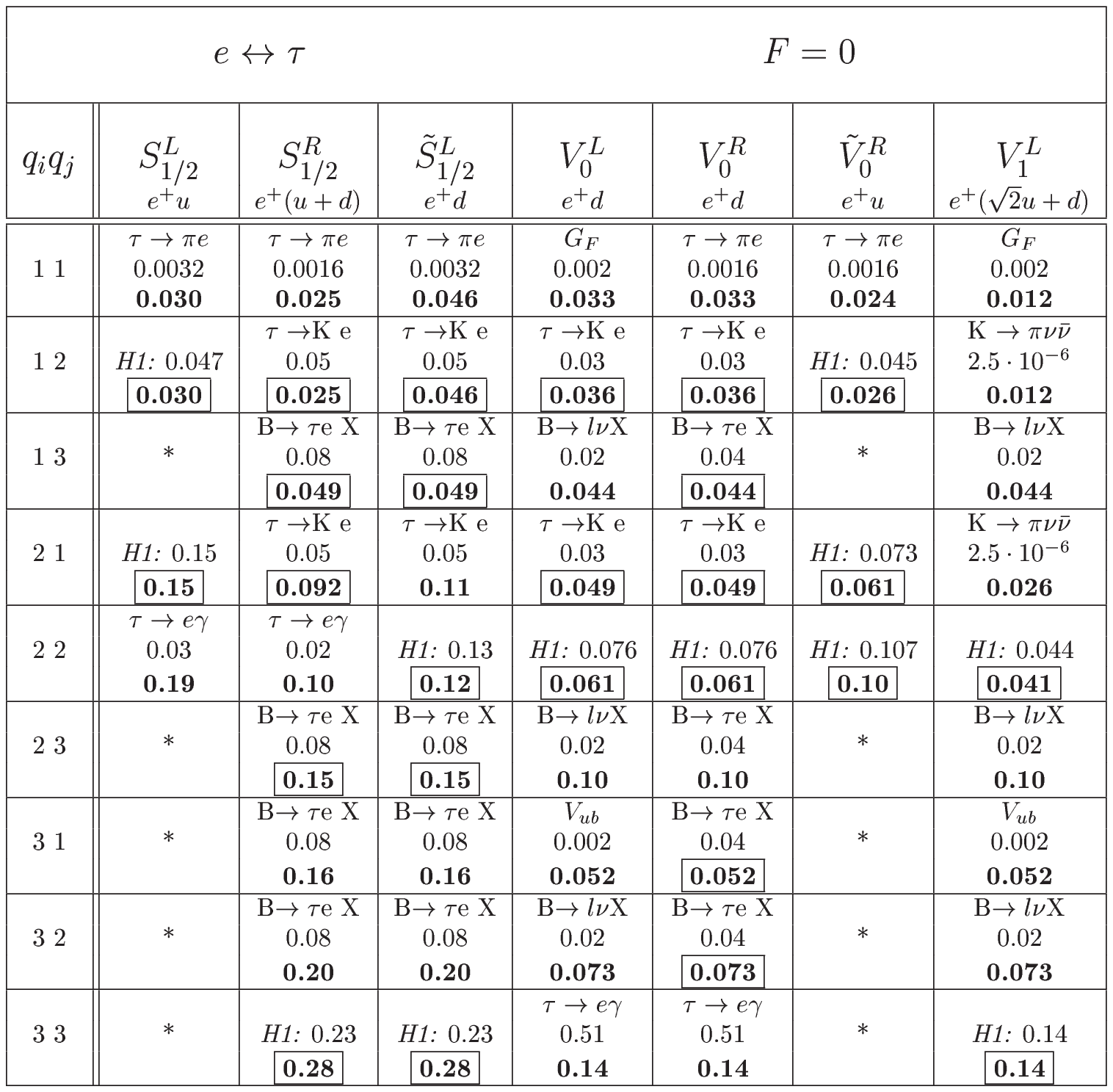,bbllx=80pt,bblly=270pt,bburx=510pt,bbury=750pt,height=16cm}
\end{center}
\caption{\label{f:Zlfv00tab4}
    95\% CL rejection limits on $\frac{{\lambda_{eq_i}}{\lambda_{\tau q_j}}}{M^2_{LQ}}$ for $F=0$ $LQ$s (in units of $10^{-4}$ GeV$^{-2}$). The first column indicates the quark generations coupling to $LQ-e$ and $LQ-\tau$ respectively. The second row indicates the different $F=0$ $LQ$s. The ZEUS preliminary results are shown in the third line (bold) for each case. The process providing the most stringent limit on the given $LQ$ is shown in the first line of each case together with its value (second line). If there is no constraint from low energy experiments, the ZEUS result is compared to the H1 result. ZEUS limits which are comparable (within a factor of two) or better than the low energy results are enclosed in a box. The $*$ indicates the cases where only a top quark would be involved. From ZEUS.
      }
\end{figure}

H1~\cite{He+lqlimit99} and ZEUS~\cite{Zlfv00} have searched for events of the type $e^+p \to l X$ where $l$ is a final-state higher-generation lepton $\mu$ or $\tau$ of high transverse momentum. Since no evidence was found for lepton-flavor violation the data were used to set constraints on couplings of leptoquarks (LQ) mediating lepton-flavor-violating $ep$ reactions. As an example, Table~\ref{f:Zlfv00tab4} shows the constraints derived for $e \leftrightarrow \tau$ transitions via fermion number $F=0$ LQ's. The data from HERA are shown together with the tightest limits from other experiments. For many transitions H1 and ZEUS provide either the only or the smallest limit.

\subsection{Substructure of quarks and electrons}
The SM assumes quarks and leptons to be pointlike. If, instead, they are extended objects the DIS cross sections at large $Q^2$ will be reduced in comparison to the SM prediction. The modification resulting from formfactors of electron and quarks can be written as 

\begin{eqnarray}
\frac{d\sigma}{dQ^2} = \frac{d\sigma^{SM}}{dQ^2} F^2_e(Q^2) F^2_q(Q^2)
\end{eqnarray}

where $\frac{d\sigma^{SM}}{dQ^2}$ represents the SM cross section. Possible modifications of the $q\overline{q}-gluon$ vertex from extended quarks are expected to be small and are neglected. 

With the assumption that electrons are pointlike H1~\cite{He+compos00} has derived an upper limit for the quark radius of $R_q < 1.7 \cdot 10^{-16}$ cm.
Assuming the electron and quark formfactors to be the same and of the form $F_e(Q^2) = F_q(Q^2) = 1/(1+\frac{1}{6} R^2 Q^2)$ the data presented by ZEUS~\cite{ZNCe+47} for the ratio $\frac{d\sigma/dQ^2 (data)}{d\sigma/dQ^2 (SM)}$ in $e^+p$ NC scattering lead~\footnote{Extracted by the author.} to a limit of $R_{e,q} < 1.5 \cdot 10^{-16}$ cm for the radii of electrons and quarks. The HERA limits can be compared with the result $R_q < 1 \cdot 10^{-16}$ cm obtained at the TEVATRON~\cite{CDFD0979} from Drell-Yan pair production assuming pointlike leptons. An analysis of the contributions of anomalous magnetic dipole moments to the $Zq\overline{q}$ vertex yielded $R_q < 1.2 \cdot 10^{-16}$ cm for $u$ and $d$ quarks~\cite{Koepp95}.  

\subsection{Extra dimensions}
Starting from the Kaluza-Klein model~\cite{Kaluza1921} it has recently been suggested that the gravitational scale in $4+n$ dimensions may be as low as the electroweak scale~\cite{Arkani99,Perez-Lorenzana00}. While in conventional string theory the extra dimensions are assumed to be compactified, in models with large extra dimensions the spin 2 graviton propagtaes into the extra spatial dimensions. As a result the gravitational force maybe strong in $4+n$ dimensions and much reduced in our four dimensional world. In lepton-quark scattering this mechanism leads effectively to a contact interaction with a coupling coefficient of $\eta_G = \lambda/M^4_S$~\cite{Giudice99}. 

\begin{figure}[hpbt]
\begin{center}
\epsfig{file=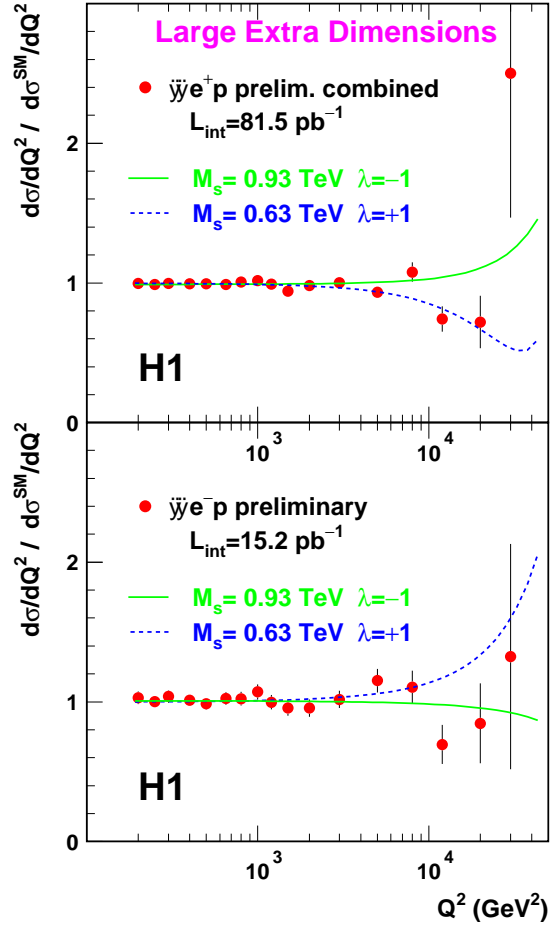,bbllx=0pt,bblly=0pt,bburx=510pt,bbury=750pt,height=14cm}
\end{center}
\caption{\label{f:Hepncextradim00}
NC cross sections $d\sigma/dQ^2$ normalised to the SM expectation for $e^+p$ and $e^-p$ data. The curves show the effects of graviton exchange in large extra dimensions given by a common fit to the scale $M_S$ with couplings $\lambda = +1$ and $\lambda = -1$. The errors represent statistics and uncorrelated experimental systematics. The values shown for $M_S$ represent 95\% C.L. lower limits.
      }
\end{figure}

H1~\cite{Hepnccontact00} has searched for the presence of such a contact interaction in $e^+p$ and $e^-p$ NC interactions by studying the ratio  $\frac{d\sigma/dQ^2 (data)}{d\sigma/dQ^2 (SM)}$, see Fig.~\ref{f:Hepncextradim00}. No evidence for contact interactions was observed. The 95\% C.L. lower limits are $M_S = 0.63$ TeV ($\lambda = +1$) and $M_S = 0.93$ TeV ($\lambda = -1$). Similar limits have been derived by LEP experiments from $e^+e^-$ annihilation~\cite{LEPextradim00}. 

\subsection{FCNC and single top production}
In the SM, flavour-changing neutral current (FCNC) transitions are forbidden at the tree level, and a transition between a top quark and a charm or up quark can only occur via loops which renders the cross section extremely small. However, some models beyond the SM predict anomalous effective couplings of the top-quark of the type $tuV$ and $tcV$ with $V=\gamma, Z^0$, which could give rise to experimentally detectable signals~\cite{Han95}. In $ep$ collisions they can lead to single top production. Since the top decays almost exclusively into $bW$  a promising event signature are topologies with an isolated lepton, missing transverse momentum and a large hadronic transverse momentum.

Both HERA experiments have searched for such events. No candidate was observed by ZEUS in $e^{\pm}p$ data from a total of 82 pb$^{-1}$. ZEUS~\cite{Zisolmispt00} set a preliminary upper limit on the anomalous magnetic coupling of $\kappa_{tu\gamma} < 0.22$ at 95\%C.L. 

H1~\cite{Hisolmispt00} performed a search for single top production in $e^+p$ (82 pb$^{-1}$) and $e^-p$ data (14 pb$^{-1}$) looking at events with either an isolated lepton plus missing transverse momentum, or hadronic events with three jets of large transverse momentum. Un upper limit of $\kappa_{tu\gamma} < 0.25$ at 95\%C.L. was derived. 

\begin{figure}[hpbt]
\begin{center}
\epsfig{file=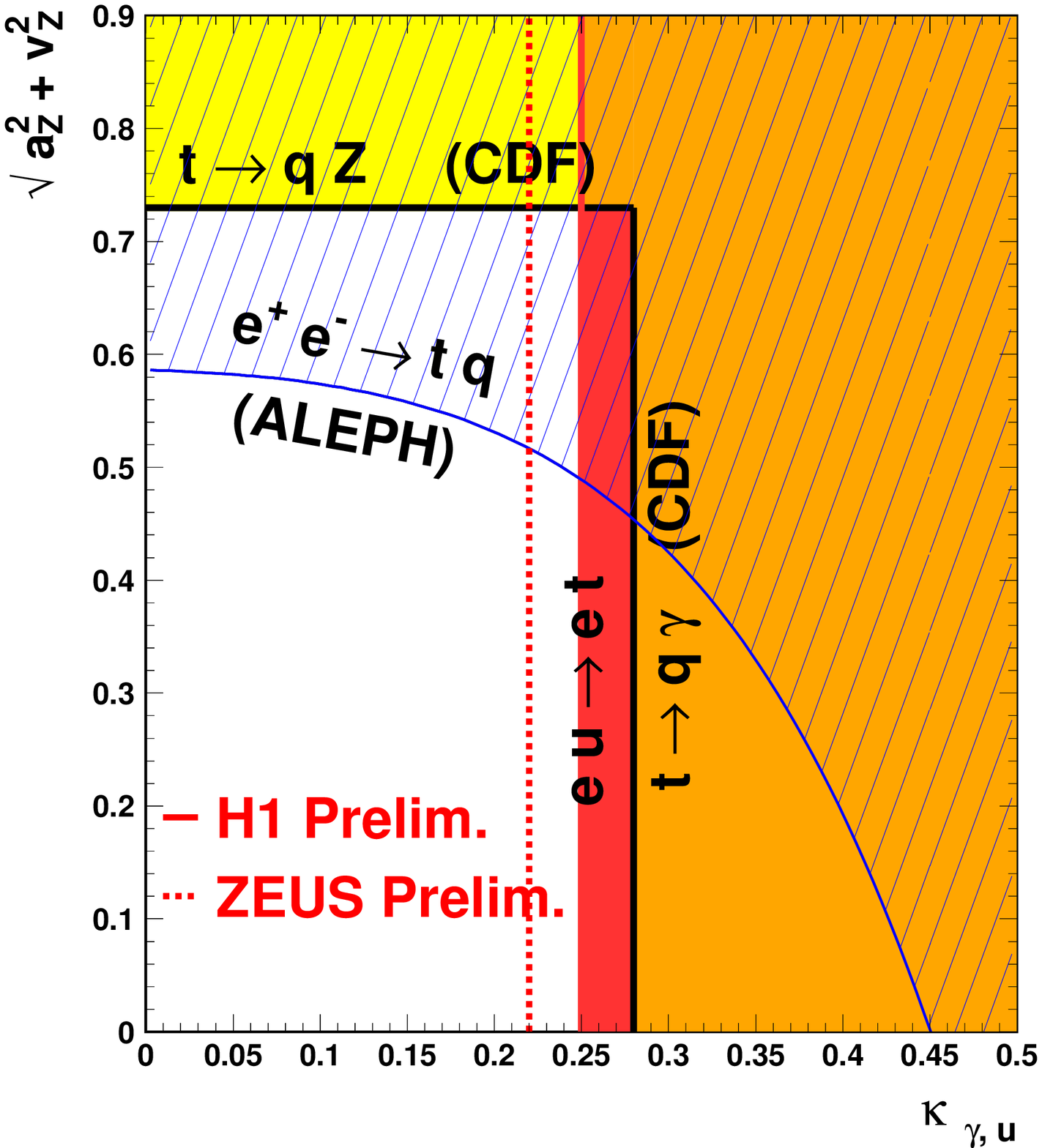,bbllx=0pt,bblly=50pt,bburx=510pt,bbury=570pt,height=8cm}
\end{center}
\caption{\label{f:tuvanomcoup}
Limits on the anomalous $tu\gamma$ (magnetic) coupling $\kappa_{\gamma,u}$ and on the anomalous $tuZ$ (vector) coupling $\sqrt{a^2_Z+v^2_Z}$ obtained at HERA, LEP and the TEVATRON.
      }
\end{figure}

The HERA limits restrict strongly the allowed space for anomalous couplings of the top as shown in Fig.~\ref{f:tuvanomcoup} where the limits on the anomalous magnetic and vector couplings $tuZ$: $\sqrt{a^2_Z + v^2_Z}$ are shown from HERA, LEP and the TEVATRON.

\subsection{Events with isolated leptons and large missing transverse momentum}
 Events produced in $ep$ collisions with high energy leptons and large missing transverse momentum are promising candidates for new physics such as production of squarks resulting from R-parity violating interactions. In the SM, events of this topology are predicted from $W$ production where the $W$ decays leptonically. However, since the cross section for $W$ production is only about 1 pb, this contribution may constitute a small and manageable background.
 
From a study of $e^+p$ interactions with 37 pb$^{-1}$ H1 has reported six events with an isolated, energetic lepton plus large missing transverse momentum~\cite{Hhilmisspt98}. For one of these events the lepton is an electron, for the remaining five it is a muon. The number of muons is significantly above the SM expectations. A similar search performed by ZEUS found agreement with the SM~\cite{Zhilmisspt00}. 

Preliminary results obtained from substantially larger data sets (H1: $e^+p$ with 82 pb$^{-1}$, $e^-p$ with 14 pb$^{-1}$; ZEUS: $e^+p$ with 66 pb$^{-1}$,  $e^-p$ with 16 pb$^{-1}$) have been presented in~\cite{Hhilmissptosaka00,Zhilmissptosaka00}. 

\begin{figure}[hpbt]
\begin{center}
\epsfig{file=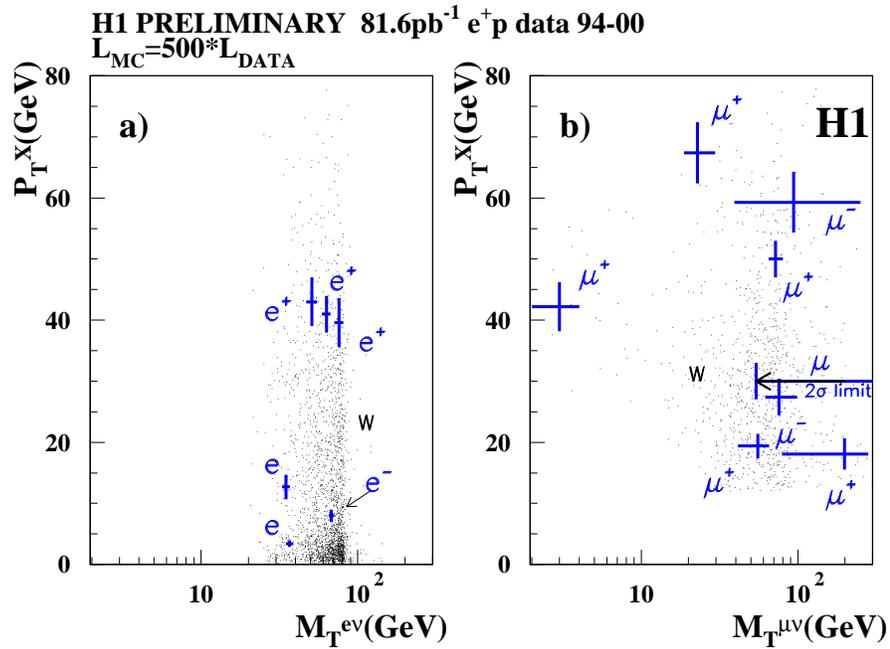,bbllx=50pt,bblly=30pt,bburx=510pt,bbury=380pt,height=8cm}
\end{center}
\caption{\label{f:Hhilmisptptxmt}
      { Distribution of events with an isolated lepton and large missing transverse momentum as a function of the transverse momentum $p^X_T$ of the hadronic system, and of the mass $M^{e\nu}_T$ of the system formed by the lepton and the missing 4-momentum vector. The small dots represent the SM prediction for $W$ production assuming a luminosity 500 times of the data sample. From H1.
      }}
\end{figure}

For H1, the increased data set showed again a substantial excess over the SM prediction while ZEUS still observed good agreement with SM. For H1 events the transverse momentum, $p^X_T$, of the hadron system is plotted against the mass $M^{e\nu}_T$ of the system formed by the lepton and the missing 4-momentum vector in Fig.~\ref{f:Hhilmisptptxmt}. The points with error bars show the observed events while the dots indicate the expected event distribution from $W$ production which was calculated for a luminosity which is 500 times larger than that for the data. In the muon sample several events stand out in a region where the contribution expected from $W$ production is very small. 

In an effort to compare the H1 and ZEUS data in a more transparent way, both experiments made an attempt to apply the same or at least similar selection criteria when selecting the topology of an isolated, energetic lepton plus large missing transverse momentum. In particular, H1 restricted the track angular region to $0.3 < \theta < 2.0$ which is the range used by ZEUS. ZEUS required for the transverse momentum of the hadron system $p^X_t > 25$ GeV. For the muon sample the missing transverse momentum was required to be $p_t(miss) > 12$ GeV. Furthermore, events with a second $\mu$ of $p_t > 1$ GeV were rejected. This requirement suppresses $\mu$ pair production by two-photon scattering, $e p \to eN \mu \mu$, where $N$ is a proton or a low mass nucleonic state. None of the $\mu$ candidate events from ZEUS survived the latter cut. 

The number of observed and predicted events obtained with these criteria by the two experiments are presented in Table~\ref{t:emispt}~\cite{Kuze00}:

\begin{table}[hbt]
\centering
\caption{Observed and predicted (total and for $W$-production alone) numbers for events with a lepton, transverse momentum $p^X_t$ of the hadron system above the cut values indicated and missing transverse momentum $p_t(miss) > 12$ GeV. From H1 and ZEUS.}
\vspace{0.5cm}
\label{t:emispt}
\begin{tabular}{|c|c|c|c|c|c|c|}
\hline 
$p^X_t >25$ GeV & N-e & N-e & N-e & N-$\mu$ & N-$\mu$ & N-$\mu$ \\
            & data   & SM     & SM(W)  & data  & SM    & SM(W) \\
\hline
H1          & 3      & 0.84   & 0.67   & 6     & 0.94  & 0.78  \\
ZEUS        & 1      & 0.78   & 0.74   & 0     & 0.82  & 0.66  \\
\hline
\hline
$p^X_t >40$ GeV & N-e & N-e & N-e & N-$\mu$ & N-$\mu$ & N-$\mu$ \\
            & data   & SM     & SM(W)  & data  & SM    & SM(W) \\
\hline
H1          & 2      & 0.27   & 0.26   & 4     & 0.35  & 0.33  \\
ZEUS        & 0      & 0.27   & 0.27   & 0     & 0.32  & 0.28  \\
\hline
\end{tabular}
\end{table}
The numbers of events expected from SM processes as calculated by the two experiments agree well with each other. However, the numbers of events observed differ substantially: while H1 sees a substantial excess over SM expectations, viz. for $p^X_t > 25$ GeV  9 events are observed compared to 1.8 events predicted, the ZEUS event rates are in good agreement with SM: 1 event observed and 1.6 events expected. 

An increase in the data samples by factors 4 - 10 seems to be required before a definitive conclusion can be drawn.

\section{Outlook}
The year-2000 upgrade of HERA is expected to raise the luminosity by a factor of 3 - 5 which promises for H1 and ZEUS each data samples as large as 1 fb$^{-1}$. At the same time the interaction regions for the two collider experiments are being equiped with spin rotators in order to turn the transverse polarisation of electrons and positrons into a longitudinal polarisation. After the upgrade, the collider detectors will have improved vertexing capabilities for identifying $c$ and $b$ quarks with high efficiency.  The combination of these improvements will increase substantially the physics capabilities~\cite{HERA95} of the experiments.

The charm and beauty part of the proton structure function can then be measured with high precision. A special feature of $c$ and $b$ quarks is their heavy mass which by itself provides a hard scale and which should suppress in deep-inelastic scattering possible contributions from nonperturbative processes.

The structure functions $F_2, F_L, xF_3, F^c_2, F^b_2$ can be measured between $Q^2 = 10$ and 40000 GeV$^2$ in a single experiment.  These measurements and the analyses of jet production in deep inelastic scattering are expected to determine the strong coupling with a precision of $\Delta\alpha_s = 0.0015$, provided the necessary theoretical calculations are available in at least NNLO.

With 1 fb$^{-1}$ the study of diffractive processes in CC processes becomes a possibility.

A luminosity of 1 fb$^{-1}$ will allow to test quarks and electrons for substructure down to $4.10^{-17}$cm. The measured NC and CC cross sections will be sensitive to deviations from the Standard Model predictions down to the level of 30 - 50 MeV in $M_W$. Additional $W's$ and $Z's$ can be detected up to 600 - 800 GeV. Perhaps most exciting will be the measurement of the charged current cross section for right (left) handed beam electrons (positrons). Any nonzero cross section contribution will be a sign for a new right handed charged current.

\section{Acknowledgements}
I want to thank Prof. Herrera and his staff for their kind hospitality at Metetpec, Mexico, and Prof. A. Zichichi and the scientific directors for stimulating discussions in Erice. I am grateful to Dr. R. Yoshida for help with the data, and to Dr. M. Kuze and Prof. E. Lohrmann for a critical reading of the manuscript.

%

\end{document}